\overfullrule=0pt

\newcount\mgnf  
\mgnf=0

\ifnum\mgnf=0
\def\openone{\leavevmode\hbox{\ninerm 1\kern-3.3pt\tenrm1}}%
\def\*{\vglue0.3truecm}\fi
\ifnum\mgnf=1
\def\openone{\leavevmode\hbox{\ninerm 1\kern-3.63pt\tenrm1}}%
\def\*{\vglue0.5truecm}\fi

\ifnum\mgnf=0
   \magnification=\magstep0
   \hsize=15truecm\vsize=23.truecm
   \parindent=4.pt\baselineskip=0.45cm
\font\titolo=cmbx12 \font\titolone=cmbx10 scaled\magstep 2
\font\cs=cmcsc10  
 \font\euftw=eufm10 \font\msytw=msbm10
  \font\indbf=cmbx10
scaled\magstep1  \fi \ifnum\mgnf=1
   \magnification=\magstep1\hoffset=0.truecm
   \hsize=14truecm\vsize=24.truecm
   \baselineskip=18truept plus0.1pt minus0.1pt \parindent=0.9truecm
   \lineskip=0.5truecm\lineskiplimit=0.1pt      \parskip=0.1pt plus1pt
\font\titolo=cmbx12 scaled\magstep 1 \font\titolone=cmbx10
scaled\magstep 3 \font\cs=cmcsc10 scaled\magstep 1
 1 
scaled\magstep1 \font\euftw=eufm10 scaled\magstep1
\font\msytw=msbm10 scaled\magstep1 
scaled\magstep1 
 \font\indbf=cmbx10
scaled\magstep2 \fi

\global\newcount\numsec\global\newcount\numapp
\global\newcount\numfor\global\newcount\numfig\global\newcount\numsub
\global\newcount\numlemma\global\newcount\numtheorem\global\newcount\numdef
\global\newcount\appflag \numsec=0\numapp=0\numfig=1
\def\veroparagrafo{\number\numsec}\def\veraformula{\number\numfor}
\def\veraappendice{\number\numapp}\def\verasub{\number\numsub}
\def\verafigura{\number\numfig}
\def\verolemma{\number\numlemma}
\def\verotheorem{\number\numtheorem}
\def\veradef{\number\numdef}

\def\section(#1,#2){\advance\numsec by 1\numfor=1\numsub=1%
\numlemma=1\numtheorem=1\numdef=1\appflag=0%
\SIA p,#1,{\veroparagrafo} %
\write15{\string\Fp (#1){\secc(#1)}}%
\write16{ sec. #1 ==> \secc(#1)  }%
\hbox to \hsize{\titolo\hfill \number\numsec. #2\hfill%
\expandafter{\alato(sec. #1)}}\*}

\def\appendix(#1,#2){\advance\numapp by 1\numfor=1\numsub=1%
\numlemma=1\numtheorem=1\numdef=1\appflag=1%
\SIA p,#1,{A\veraappendice} %
\write15{\string\Fp (#1){\secc(#1)}}%
\write16{ app. #1 ==> \secc(#1)  }%
\hbox to \hsize{\titolo\hfill Appendix A\number\numapp. #2\hfill%
\expandafter{\alato(app. #1)}}\*}

\def\senondefinito#1{\expandafter\ifx\csname#1\endcsname\relax}

\def\SIA #1,#2,#3 {\senondefinito{#1#2}%
\expandafter\xdef\csname #1#2\endcsname{#3}\else \write16{???? ma
#1#2 e' gia' stato definito !!!!} \fi}

\def \Fe(#1)#2{\SIA fe,#1,#2 }
\def \Fp(#1)#2{\SIA fp,#1,#2 }
\def \Fg(#1)#2{\SIA fg,#1,#2 }
\def \Fl(#1)#2{\SIA fl,#1,#2 }
\def \Ft(#1)#2{\SIA ft,#1,#2 }
\def \Fd(#1)#2{\SIA fd,#1,#2 }

\def\etichetta(#1){%
\ifnum\appflag=0(\veroparagrafo.\veraformula)%
\SIA e,#1,(\veroparagrafo.\veraformula) \fi%
\ifnum\appflag=1(A\veraappendice.\veraformula)%
\SIA e,#1,(A\veraappendice.\veraformula) \fi%
\global\advance\numfor by 1%
\write15{\string\Fe (#1){\equ(#1)}}%
\write16{ EQ #1 ==> \equ(#1)  }}

\def\getichetta(#1){Fig. \verafigura%
\SIA g,#1,{\verafigura} %
\global\advance\numfig by 1%
\write15{\string\Fg (#1){\graf(#1)}}%
\write16{ Fig. #1 ==> \graf(#1) }}

\def\etichettap(#1){%
\ifnum\appflag=0{\veroparagrafo.\verasub}%
\SIA p,#1,{\veroparagrafo.\verasub} \fi%
\ifnum\appflag=1{A\veraappendice.\verasub}%
\SIA p,#1,{A\veraappendice.\verasub} \fi%
\global\advance\numsub by 1%
\write15{\string\Fp (#1){\secc(#1)}}%
\write16{ par #1 ==> \secc(#1)  }}

\def\etichettal(#1){%
\ifnum\appflag=0{\veroparagrafo.\verolemma}%
\SIA l,#1,{\veroparagrafo.\verolemma} \fi%
\ifnum\appflag=1{A\veraappendice.\verolemma}%
\SIA l,#1,{A\veraappendice.\verolemma} \fi%
\global\advance\numlemma by 1%
\write15{\string\Fl (#1){\lm(#1)}}%
\write16{ lemma #1 ==> \lm(#1)  }}

\def\etichettat(#1){%
\ifnum\appflag=0{\veroparagrafo.\verotheorem}%
\SIA t,#1,{\veroparagrafo.\verotheorem} \fi%
\ifnum\appflag=1{A\veraappendice.\verotheorem}%
\SIA t,#1,{A\veraappendice.\verotheorem} \fi%
\global\advance\numtheorem by 1%
\write15{\string\Ft (#1){\thm(#1)}}%
\write16{ th. #1 ==> \thm(#1)  }}

\def\etichettad(#1){%
\inum\appflag=0{\veroparagrafo.\veradef}%
\SIA d,#1,{\veroparagrafo.\veradef} \fi%
\inum\appflag=1{A\veraappendice.\veradef}%
\SIA d,#1,{A\veraappendice.\veradef} \fi%
\global\advance\numdef by 1%
\write15{\string\Fd (#1){\defz(#1)}}%
\write16{ def. #1 ==> \defz(#1)  }}

\def\Eq(#1){\eqno{\etichetta(#1)\alato(#1)}}
\def\eq(#1){\etichetta(#1)\alato(#1)}
\def\eqg(#1){\getichetta(#1)\alato(fig #1)}
\def\sub(#1){\0\palato(p. #1){\bf \etichettap(#1)\hskip.3truecm}}
\def\lemma(#1){\0\palato(lm #1){\cs Lemma \etichettal(#1)\hskip.3truecm}}
\def\theorem(#1){\0\palato(th #1){\cs Theorem \etichettat(#1)%
\hskip.3truecm}}
\def\definition(#1){\0\palato(df #1){\cs Definition \etichettad(#1)%
\hskip.3truecm}}

\def\equv(#1){\senondefinito{fe#1}$\clubsuit$#1%
\write16{eq. #1 non e' (ancora) definita}%
\else\csname fe#1\endcsname\fi}
\def\grafv(#1){\senondefinito{fg#1}$\clubsuit$#1%
\write16{fig. #1 non e' (ancora) definito}%
\else\csname fg#1\endcsname\fi}

\def\secv(#1){\senondefinito{fp#1}$\clubsuit$#1%
\write16{par. #1 non e' (ancora) definito}%
\else\csname fp#1\endcsname\fi}

\def\lmv(#1){\senondefinito{fl#1}$\clubsuit$#1%
\write16{lemma #1 non e' (ancora) definito}%
\else\csname fl#1\endcsname\fi}

\def\thmv(#1){\senondefinito{ft#1}$\clubsuit$#1%
\write16{th. #1 non e' (ancora) definito}%
\else\csname ft#1\endcsname\fi}

\def\defzv(#1){\senondefinito{fd#1}$\clubsuit$#1%
\write16{def. #1 non e' (ancora) definito}%
\else\csname fd#1\endcsname\fi}

\def\equ(#1){\senondefinito{e#1}\equv(#1)\else\csname e#1\endcsname\fi}
\def\graf(#1){\senondefinito{g#1}\grafv(#1)\else\csname g#1\endcsname\fi}
\def\secc(#1){\senondefinito{p#1}\secv(#1)\else\csname p#1\endcsname\fi}
\def\lm(#1){\senondefinito{l#1}\lmv(#1)\else\csname l#1\endcsname\fi}
\def\thm(#1){\senondefinito{t#1}\thmv(#1)\else\csname t#1\endcsname\fi}
\def\defz(#1){\senondefinito{d#1}\defzv(#1)\else\csname d#1\endcsname\fi}
\def\sec(#1){{\S\secc(#1)}}

\def\BOZZA{
\def\alato(##1){\rlap{\kern-\hsize\kern-1.2truecm{$\scriptstyle##1$}}}
\def\palato(##1){\rlap{\kern-1.2truecm{$\scriptstyle##1$}}}
}

\def\alato(#1){}
\def\galato(#1){}
\def\palato(#1){}

{\count255=\time\divide\count255 by 60
\xdef\hourmin{\number\count255}
        \multiply\count255 by-60\advance\count255 by\time
   \xdef\hourmin{\hourmin:\ifnum\count255<10 0\fi\the\count255}}

\def\oramin{\hourmin }

\def\data{\number\day/\ifcase\month\or gennaio \or febbraio \or marzo \or
aprile \or maggio \or giugno \or luglio \or agosto \or settembre
\or ottobre \or novembre \or dicembre \fi/\number\year;\ \oramin}
\setbox200\hbox{$\scriptscriptstyle \data $}
\footline={\rlap{\hbox{\copy200}}\tenrm\hss \number\pageno\hss}

\let\a=\alpha \let\b=\beta  \let\g=\gamma     \let\d=\delta  \let\e=\varepsilon
\let\z=\zeta  \let\h=\eta    \def\th{\theta}
   \let\l=\lambda
\let\m=\mu    \let\n=\nu            \let\p=\pi      \let\r=\rho
\let\s=\sigma \let\t=\tau   \let\f=\varphi     \let\c=\chi
   \let\o=\omega 
 \let\D=\Delta     \let\L=\Lambda

\def\\{\hfill\break} \let\==\equiv

\let\io=\infty 

\let\0=\noindent

\def\ie{\hbox{\it i.e.\ }}

\let\dpr=\partial 
\let\bs=\backslash

\def\defin{{\buildrel def\over=}}
\def\tende#1{\,\vtop{\ialign{##\crcr\rightarrowfill\crcr
 \noalign{\kern-1pt\nointerlineskip}
 \hskip3.pt${\scriptstyle #1}$\hskip3.pt\crcr}}\,}
\def\otto{\,{\kern-1.truept\leftarrow\kern-5.truept\to\kern-1.truept}\,}
\def\fra#1#2{{#1\over#2}}

\def\VV{{\cal V}}
\def\CC{{\cal C}}\def\WW{{\cal W}}
\def\TT{{\cal T}}\def\NN{{\cal N}}\def\BB{{\cal B}}
\def\RR{{\cal R}}\def\LL{{\cal L}}
\def\DD{{\cal D}}

\def\T#1{{#1_{\kern-3pt\lower7pt\hbox{$\widetilde{}$}}\kern3pt}}
\def\VVV#1{{\underline #1}_{\kern-3pt
\lower7pt\hbox{$\widetilde{}$}}\kern3pt\,}
\def\W#1{#1_{\kern-3pt\lower7.5pt\hbox{$\widetilde{}$}}\kern2pt\,}

\def\indica{\leaders \hbox to 0.5cm{\hss.\hss}\hfill}
\def\guida{\leaders\hbox to 1em{\hss.\hss}\hfill}
\mathchardef\oo= "0521

\def\pp{{\bf p}}\def\qq{{\bf q}}\def\xx{{\bf x}}
\def\yy{{\bf y}}\def\kk{{\bf k}}
\def\zz{{\bf z}}

\def\bT{{\bf T}}

\def\Halmos{\hfill\vrule height6pt width4pt depth2pt \par\hbox to \hsize{}}
\def\virg{\quad,\quad}
\def\proof{\0{\cs Proof - } }

\def\oo{{\underline \omega}}

\def\un{{\underline \nu}}
 \def\ux{{\underline x}}
\def\xxx{{\underline\xx}}

\def\qed{\raise1pt\hbox{\vrule height5pt width5pt depth0pt}}

\def\indic{\hbox{\raise-2pt \hbox{\indbf 1}}}
\def\bk#1#2{\bar\kk_{#1#2}}

\def\MMM{\hbox{\euftw M}} \def\BBB{\hbox{\euftw B}}
\def\RRR{\hbox{\msytw R}}

%
%
%
\def\ins#1#2#3{\vbox to0pt{\kern-#2 \hbox{\kern#1 #3}\vss}\nointerlineskip}
%
%
%
\newdimen\xshift \newdimen\xwidth \newdimen\yshift

\def\insertplot#1#2#3#4#5{\par%
\xwidth=#1 \xshift=\hsize \advance\xshift by-\xwidth \divide\xshift by 2%
\yshift=#2 \divide\yshift by 2%
\line{\hskip\xshift \vbox to #2{\vfil%
#3 \includegraphics{#4.pst}}\hfill \raise\yshift\hbox{#5} }}

\openin14=\jobname.aux \ifeof14 \relax \else
\input \jobname.aux \closein14 \fi
\openout15=\jobname.aux


\centerline{\titolone Ward identities and Chiral anomaly}
\centerline{\titolone in the Luttinger liquid} \vskip.5cm
\centerline{{\titolo G. Benfatto, V. Mastropietro}}
\centerline{Dipartimento di Matematica, Universit\`a di Roma ``Tor
Vergata''} \centerline{Via della Ricerca Scientifica, I-00133,
Roma}

\vskip1cm \0{\cs Abstract.} {\it Systems of interacting non relativistic
fermions in $d=1$, as well as spin chains or interacting bidimensional Ising
models, verify an hidden approximate Gauge invariance which can be used to
derive suitable Ward identities. Despite the presence of corrections and
anomalies, such Ward identities can be implemented in a Renormalization Group
approach and used to exploit nontrivial cancellations which allow to control
the flow of the running coupling constants; in particular this is achieved
combining Ward identities, Dyson equations and suitable correction identities
for the extra terms appearing in the Ward identities, due to the presence of
cutoffs breaking the local gauge symmetry. The correlations can be computed
and show a Luttinger liquid behavior characterized by non universal critical
indices, so that the general Luttinger liquid construction for one
dimensional systems is completed without any use of exact solutions. The
ultraviolet cutoff can be removed and a Quantum Field Theory corresponding to
the Thirring model is also constructed.}

\vskip1cm
\section(0, Introduction)

\* \sub(1.2) {\it Luttinger liquids.}

A key notion in solid state physics is the one of {\it Fermi
liquids}, used to describe systems of {\it interacting} electrons
which, in spite of the interaction, have a physical behavior
qualitatively similar to the one of the {\it free Fermi gas}. In
analogy with Fermi liquids, the notion of {\it Luttinger liquids}
has been more recently introduced, to describe systems behaving
qualitatively as the Luttinger model, see
for instance [A] or [Af]; their correlations have an anomalous
behavior described in terms of non universal (\ie nontrivial
functions of the coupling) critical indices.

A large number of models, for which an exact solution is lacking
(at least for the correlation functions) are indeed believed to be
in the same class of universality of the Luttinger model or of
its massive version, and indeed in the last decade, starting from [BG], it has been
possible to substantiate this assumption on a large class of
models, by a quantitative analysis based on Renormalization Group
techniques, which at the end allow us to write the correlations and
the critical indices as convergent series in the coupling. We
mention the Schwinger functions of interacting non relativistic
fermions in $d=1$ (modelling the electronic properties of metals
so anisotropic to be considered as one dimensional), in the
spinless [BGPS], in the spinning case with repulsive interaction [BoM],
or with external periodic or quasi-periodic potentials [M]; the
spin-spin correlations of the {\it Heisenberg $XYZ$ spin chain}
[BM1], [BM2]; the thermodynamic functions of classical Ising
systems on a bidimensional lattice with quartic interactions like
the {\it Eight-vertex} or the {\it Ashkin-Teller} models [M1]; and
many others, see for instance the review [GM].

In all such models the observables are written as Grassmannian
integrals, and a naive evaluation of them in terms of a series
expansion in the perturbative parameter does not work; it is however
possible, by a multiscale analysis based on Renormalization Group,
to write the Grassmannian integrals as series of suitable
finitely many parameters, called {\it running coupling constants}, and this
expansion is convergent if the running coupling constants are
small. The running coupling constants obey a complicated set of
recursive equations, whose right hand side is called, as usual,
the {\it Beta function}.
The Beta function can be written as sum
of two terms; one, which we call
{\it principal part} of the Beta function,
is common to all such models
while the other one is model dependent.
It turns out
that, {\it if} the principal part of
the beta function is asymptotically
vanishing, than the flow of the running coupling constants in all
such models can be controlled just by dimensional bounds, and the
expansion is really convergent; the observables are then expressed by
explicit convergent series from which the physical information
can be extracted. On the other hand the principal part of the Beta
function coincides with the Beta function of a model, which we call
{\it reference model}, describing two kind of fermions
with a linear "relativistic" dispersion relation
and with momenta restricted by infrared and ultraviolet cutoffs,
interacting by a local quartic potential.

In order to prove the vanishing of the Beta function of the reference model
(in the form of the bound \equ(2.1) below), one can use an
indirect argument, based on the fact that the reference model
is close to the Luttinger model, and then use the
Luttinger model exact solution of [ML], see [BGPS], [BM3].

{\it In this paper we show that indeed the vanishing of the principal
part of the Beta
function can be proved without any use of
the exact solution, by using only Ward identities based on the
approximate chiral gauge invariance of the reference model.}

As exact solutions are quite rare and generally peculiar to $d=1$, while RG
analysis and Ward identities are general methods working in any dimension,
our results might become relevant for the theory of non-Fermi liquids in
$d>1$. We remember that there is experimental evidence for non Fermi liquid
(probably Luttinger) behavior in high $T_c$ superconductors [A], which are
essentially planar systems.

Ward identities play a crucial role in Quantum Field Theory and
Statistical Mechanics, as they allow to prove cancellations in a
non perturbative way. The advantage of reducing the analysis of
quantum spin chains or interacting Ising models to the reference model is that
such model, formally neglecting the cutoffs, verifies many
symmetries which were not verified by the spin chain or spin
models; in particular it verifies a {\it total gauge invariance}
symmetry $\psi_{\xx,\o}^\pm \to e^{\pm i\a_\xx} \psi_{\xx,\o}^\pm$
and {\it chiral gauge invariance} $\psi_{\xx,\o}^\pm\to e^{\pm
i\a_{\xx,\o}}\psi_{\xx,\o}^\pm$. Such symmetries are {\it hidden}
in Ising or spin chain models, as they do not verify chiral gauge
invariance, even if they are ``close'', in an RG sense, to a model
formally verifying them.
\*
\* \sub(1.1) {\it The reference model.}

The reference model is not Hamiltonian and is defined in terms of {\it
Grassmannian variables}. It describes a system of two kinds of fermions with
linear dispersion relation interacting with a local potential; the presence
of an ultraviolet and an infrared cutoff makes the model {\it not} solvable.
Given the interval $[0,L]$, the inverse temperature $\b$ and the (large)
integer $N$, we introduce in $\L=[0,L]\times[0,\b]$ a lattice $\L_N$, whose
sites are given by the {\it space-time points} $\xx=(x,x_0)=(n a,n_0 a_0)$,
$a=L/N$, $a_0=\b/N$, $n,n_0=0,1,\ldots,N-1$. We also consider the set $\DD$
of {\it space-time momenta} $\kk=(k,k_0)$, with $k={2\pi\over L}(n+{1\over
2})$ and $k_0={2\pi\over \b}(n_0+{1\over 2})$, $n,n_0=0,1,\ldots,N-1$. With
each $\kk\in\DD$ we associate four Grassmannian variables
$\hat\psi^{[h,0]\s}_{\kk,\o}$, $\s,\o\in\{+,-\}$. Then we define the
functional integration $\int \DD\psi^{[h,0]}$ as the linear functional on the
Grassmann algebra generated by the variables $\hat\psi^{[h,0]\s}_{\kk,\o}$,
such that, given a monomial $Q(\hat\psi)$ in the variables
$\hat\psi^{[h,0]\s}_{\kk,\o}$, its value is $0$, except in the case
$Q(\hat\psi)= \prod_{\kk\in\DD,\o=\pm} \hat\psi^{[h,0]-}_{\kk,\o}
\hat\psi^{[h,0]+}_{\kk,\o}$, up to a permutation of the variables. In this
case the value of the functional is determined, by using the anticommuting
properties of the variables, by $\int\;\DD\psi^{[h,0]} Q(\hat\psi)=1\;$.

The lattice $\L_N$ is introduced only to allow us to perform a non formal
treatment of the Grassmannian integrals, as the number of Grassmannian
variables is finite, and eventually the limit $N\to\io$ is taken, see [BM1].

We also define the {\it Grassmannian field} on the lattice $\L_N$ as
$$\psi^{[h,0]\s}_{\xx,\o}={1\over L\b}\sum_{\kk\in \DD}e^{i\s\kk\xx}
\hat\psi^{[h,0]\s}_{\kk,\o}\;,\quad \xx\in\L_N\;.\Eq(1.1)$$
Note that $\psi^{[h,0]\s}_{\xx,\o}$ is antiperiodic
both in time and space variables.

We define
$$V(\psi^{[h,0]})=\l \int d\xx\; \psi^{[h,0]+}_{\xx,+} \psi^{[h,0]-}_{\xx,+}
\psi^{[h,0]+}_{\xx,-} \psi^{[h,0]-}_{\xx,-}\Eq(1.2)$$
and
$$P(d\psi^{[h,0]}) = \NN^{-1} \DD\psi^{[h,0]} \cdot\;\exp \left\{-{1\over L\b}
\sum_{\o=\pm 1} \sum_{\kk\in \DD} C_{h,0}(\kk)(-i k_0+\o k)
\hat\psi^{[h,0]+}_{\kk,\o} \hat\psi^{[h,0]-}_{\kk,\o}\right\}\;,\Eq(1.3)$$
with $\NN=\prod_{\kk\in
\DD}[(L\b)^{-2}(-k_0^2-k^2)C_{h,0}(\kk)^2]$ and $\int d\xx$ is a
shorthand for ``$a\ a_0\ \sum_{\xx\in\L_N}$''. The function
$C_{h,0}(\kk)$ acts as an ultraviolet and infrared cutoff and it
is defined in the following way. We introduce a positive number
$\g>1$ and a positive function $\c_0(t) \in C^{\io}(\RRR_+)$ such
that
$$\c_0(t) = \cases{ 1 & if
$0\le t \le 1\;,$ \cr 0 & if $t \ge \g_0\;,\quad 1<\g_0\le\g\;,$\cr}\Eq(1.4)$$
and we define, for any integer $j\le 0$,
$f_j(\kk)=\c_0(\g^{-j}|\kk|)-\c_0(\g^{-j+1}|\kk|)$.
Finally  we define
$\c_{h,0}(\kk) =[C_{h,0}(\kk)]^{-1} =
\sum_{j=h}^0 f_j(\kk)$
so that $[C_{h,0}(\kk)]^{-1}$ is a smooth function with support in
the interval $\{\g^{h-1} \le |\kk| \le \g\}$, equal to $1$ in the
interval $\{\g^h \le |\kk| \le 1\}$.

We call $\psi^{[h,0]}$ simply $\psi$ and we introduce the {\it
generating functional}
$$\WW(\phi,J)= \log \int P(d\psi) e^{-V(\psi)+ \sum_\o \int d\xx
\left[J_{\xx,\o}\psi^{+}_{\xx,\o}\psi^{-}_{\xx,\o}+
\phi^+_{\xx,\o}\psi^{-}_{\xx,\o}+ \psi^{+}_{\xx,\o}\phi^-_{\xx,\o}\right]}
\;.\Eq(1.7)$$
\insertplot{300pt}{90pt}%
{\ins{30pt}{60pt}{$+$}
\ins{60pt}{60pt}{$+$}
\ins{20pt}{40pt}{$-$}
\ins{70pt}{40pt}{$-$}

\ins{120pt}{60pt}{$\o$}
\ins{170pt}{60pt}{$\o$}

}%
{vertici}{}
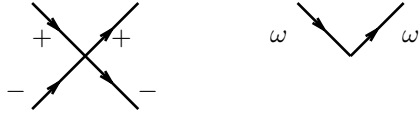
\centerline{\eqg(1aa): Graphical representation of the interaction $V(\psi)$
and the density $\psi^{+}_{\xx,\o}\psi^{-}_{\xx,\o}$} \*

The Grassmannian variables $\phi^\s_{\xx,\o}$ are antiperiodic in $x_0$ and
$x$ and anticommuting with themselves and $\psi^{\s}_{\xx,\o}$, while the
variables $J_{\xx,\o}$ are periodic and commuting with themselves and all the
other variables. The Schwinger functions can be obtained by functional
derivatives of \equ(1.7); for instance
$$G^{2,1}_\o(\xx;\yy,\zz)={\dpr\over\dpr J_{\xx,\o}}
{\dpr^2\over\dpr\phi^+_{\yy,+}\dpr\phi^-_{\zz,+}}
\WW(\phi,J)|_{\phi=J=0}\;,\Eq(1.8)$$
$$G_{\o}^{4,1}(\xx;\xx_1,\xx_2,\xx_3,\xx_4)={\dpr\over\dpr J_{\xx,\o}}
{\dpr^2\over\dpr\phi^+_{\xx_1,\o}\dpr\phi^-_{\xx_2,\o}}
{\dpr^2\over\dpr\phi^+_{\xx_3,-\o}\dpr\phi^-_{\xx_4,-\o}}
\WW(\phi,J)|_{\phi=J=0}\;,\Eq(1.9)$$
$$G_{\o}^{4}(\xx_1,\xx_2,\xx_3,\xx_4)=
{\dpr^2\over\dpr\phi^+_{\xx_1,\o}\dpr\phi^-_{\xx_2,\o}}
{\dpr^2\over\dpr\phi^+_{\xx_3,-\o}\dpr\phi^-_{\xx_4,-\o}}
\WW(\phi,J)|_{\phi=J=0}\;,\Eq(1.10)$$

$$G_{\o}^{2}(\yy,\zz)= {\dpr^2\over\dpr\phi^+_{\yy,\o}\dpr\phi^-_{\zz,\o}}
\WW(\phi,J)|_{\phi=J=0}\;.\Eq(1.11a)$$

\insertplot{300pt}{150pt}%
{\ins{25pt}{82pt}{$G^{2,1}_\o$}
\ins{28pt}{32pt}{$\xx$}
\ins{-5pt}{110pt}{$\yy$}
\ins{58pt}{110pt}{$\zz$}
\ins{10pt}{97pt}{$\o$}
\ins{45pt}{97pt}{$\o$}
\ins{20pt}{50pt}{$\o$}
\ins{35pt}{50pt}{$\o$}
\ins{105pt}{82pt}{$G^{4,1}_\o$}
\ins{88pt}{32pt}{$\xx$}
\ins{128pt}{32pt}{$\xx_4$}
\ins{80pt}{120pt}{$\xx_1$}
\ins{108pt}{122pt}{$\xx_2$}
\ins{130pt}{120pt}{$\xx_3$}
\ins{80pt}{105pt}{$\o$}
\ins{102pt}{107pt}{$\o$}
\ins{118pt}{110pt}{$-\o$}
\ins{112pt}{45pt}{$-\o$}
\ins{83pt}{50pt}{$\o$}
\ins{98pt}{50pt}{$\o$}
\ins{195pt}{82pt}{$G^4_\o$}
\ins{165pt}{120pt}{$\xx_1$}
\ins{228pt}{120pt}{$\xx_2$}
\ins{168pt}{38pt}{$\xx_3$}
\ins{228pt}{38pt}{$\xx_4$}
\ins{165pt}{105pt}{$\o$}
\ins{218pt}{108pt}{$\o$}
\ins{162pt}{55pt}{$-\o$}
\ins{210pt}{45pt}{$-\o$}
\ins{275pt}{82pt}{$G^2_\o$}
\ins{279pt}{114pt}{$\zz$}
\ins{279pt}{38pt}{$\yy$}
\ins{272pt}{105pt}{$\o$}
\ins{272pt}{45pt}{$\o$}
}%
{schw}{}

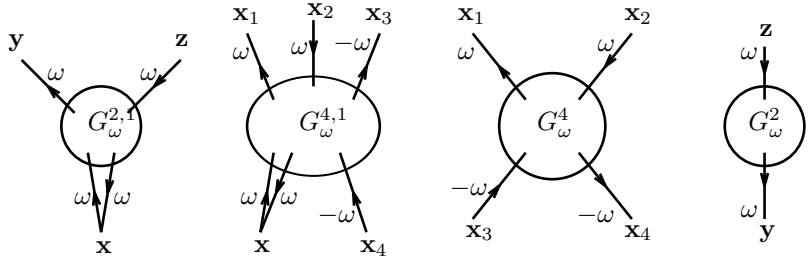
\* \centerline{\eqg(1a): Graphical representation of the Schwinger functions
$G^{2,1}_\o, G^{4,1}_\o, G^{4}_\o, G^2_\o$}

\*

If $Q(\psi)$ is a monomial in the Grassmannian variables, it is
easy to see that $\int P(d\psi)$ is given by the anticommutative
{\it Wick rule}; the corresponding propagator
$$\hat g_\o(\kk)={\chi_{h,0}(\kk)\over -i k_0+\o k}\Eq(0)$$
is singular as $L,\b\to\io$ at $\kk=0$. The simplest way of
computing the Grassmannian integral
\equ(1.7) is to expand in power series of $\l$ the exponential, obtaining
many Grassmannian integrals of monomials, which can be computed by
the anticommutative Wick rule. This procedure makes it possible to
write series expansions for the Schwinger functions, for example
those defined in \equ(1.8)-\equ(1.11a), and to prove that they are
absolutely convergent, in the limit $L,\b\to\io$, for $|\l|\le
\e_h$, with $\e_h\to 0$ as $h\to-\io$;  {\it in other words, the
estimated radius of convergence vanishes as the infrared cutoff is
removed}. In this paper we will show that it is possible to modify
the expansions, through a {\it resummation} of the power series in
$\l$, so that it is possible to prove that they are well defined
and convergent even when the infrared cutoff is removed
($h\to-\io$); as a corollary we prove
vanishing of the Beta function in the form of the bound \equ(2.1) below.

It is also possible to remove the ultraviolet cutoff, so constructing a
relativistic quantum field theory, the massless {\it Thirring model} [T];
this is shown in the Appendix. Note finally that the reference model, if both
the infrared and ultraviolet cutoffs are removed and the local potential is
replaced by a short-ranged one, coincides with the {\it Luttinger model},
which was solved in [ML]; the presence of cutoffs makes however the model
\equ(1.7) {\it not solvable}.

\* \sub(1.3) {\it Sketch of the proof: the Dyson equation.}

This paper is the conclusion of our construction of $d=1$
Luttinger liquids {\it with no use of exact solutions}, started in
[BM1], [BM2], [BM3], whose results will be used here.

The analysis starts by expressing the Grassmann integration in \equ(1.7) as
the product of many independent integrations, each of them "describing the
theory at a certain momentum scale $\g^j$", with $j$ an integer such that
$h\le j\le 0$ and $\g>1$. This allows us to perform the overall functional
integration by iteratively integrating the Grassmanian variables of
decreasing momentum scale. After $|j|$ integration steps one gets a
Grassmannian integral still similar to \equ(1.7), the main differences being
that the Grassmannian fields acquire a {\it wave function renormalization}
$Z_j$, the local terms quartic in the $\psi$ fields in the interaction have
coupling $\l_j$ (the effective interaction strength at momentum scale $\g^j$)
and the local terms $J \psi^+_\o \psi^-_\o$ have coupling $Z_j^{(2)}$ (the
density renormalization). This iterative procedure allows us to get an
expansion, resumed in \S\secc(1), for the Schwinger functions; they are
written as series in the set of parameters $\l_j$, $j=0,-1,...h$, called {\it
running coupling constants}. It was proved in [BM1] that, if the running
coupling constants are small enough, such expansions are convergent, as a
consequence of suitable cancellations due to the anticommutativity of
fermions.

However, the property that the running coupling constants are small is not
trivial at all; it is related to very complex and intricate cancellations at
all orders in the perturbative series, eventually implying that the effective
interaction strength $\l_j$ stays close to its initial value $\l$ for any
$j$; while one can easily check that cancellations are present at lowest
orders by an explicit computation, to prove directly that the cancellations
are present at every order seems essentially impossible. In order to prove
that $\l_j$ remains close to $ \l$ for any $j$, we use the fact that the
Schwinger functions, even if they are expressed by apparently very different
series expansion, are indeed related by remarkable identities; on the other
hand the Fourier transform of Schwinger functions computed at the cut-off
scale are related to the running coupling constants or to the renormalization
constants at the cutoff scale (see Theorem \thm(th2)), so that identities
between Schwinger functions imply identities between coupling or
renormalization constants.

As $\hat G^4$ computed at the cutoff scale is proportional to $\l_h$, see
\equ(1.47) below, it is natural to write (see \S\secc(2)) a {\it Dyson}
equation for $\hat G^4$:
$$\eqalign{
-\hat G^{4}_+(\kk_1,\kk_2,\kk_3,\kk_4) &= \l \hat g_-(\kk_4) \Big[
\hat G^2_-(\kk_3) \hat G^{2,1}_+(\kk_1-\kk_2,\kk_1,\kk_2)+\cr
&+ {1\over L\b} \sum_\pp
G^{4,1}_+(\pp;\kk_1,\kk_2,\kk_3,\kk_4-\pp)\Big]\;,}\Eq(2.11a)$$
relating the correlations in
\equ(1.8),\equ(1.9),\equ(1.10),\equ(1.11a); see Fig. \graf(3).

\insertplot{300pt}{150pt}%
{\ins{45pt}{82pt}{$\hat G^{4}_+$}
\ins{30pt}{110pt}{$\kk_1$}
\ins{15pt}{110pt}{$+$}
\ins{65pt}{115pt}{$\kk_2$}
\ins{80pt}{105pt}{$+$}
\ins{65pt}{45pt}{$\kk_3$}
\ins{20pt}{55pt}{$-$}
\ins{30pt}{45pt}{$\kk_4$}
\ins{75pt}{55pt}{$-$}
\ins{95pt}{75pt}{$=$}
\ins{143pt}{123pt}{$\hat G^{2,1}_+$}
\ins{147pt}{50pt}{$\hat G^{2}_-$}
\ins{130pt}{140pt}{$\kk_1$}
\ins{120pt}{132pt}{$+$}
\ins{165pt}{145pt}{$\kk_2$}
\ins{180pt}{135pt}{$+$}
\ins{130pt}{85pt}{$\kk_4$}
\ins{120pt}{75pt}{$-$}
\ins{155pt}{90pt}{$\kk_1-\kk_2$}
\ins{140pt}{95pt}{$+$}
\ins{155pt}{97pt}{$+$}
\ins{155pt}{70pt}{$\kk_3$}
\ins{143pt}{68pt}{$-$}
\ins{155pt}{20pt}{$\kk_3$}
\ins{143pt}{28pt}{$-$}
\ins{200pt}{75pt}{$+$}
\ins{245pt}{108pt}{$\hat G^{4,1}_+$}
\ins{220pt}{130pt}{$\kk_1$}
\ins{220pt}{142pt}{$+$}
\ins{240pt}{135pt}{$\kk_2$}
\ins{252pt}{138pt}{$+$}
\ins{275pt}{125pt}{$\kk_3$}
\ins{275pt}{142pt}{$-$}
\ins{255pt}{50pt}{$\kk_4$}
\ins{240pt}{40pt}{$-$}
\ins{260pt}{70pt}{$\kk_4-\pp$}
\ins{260pt}{80pt}{$-$}
\ins{235pt}{70pt}{$\pp$}
\ins{235pt}{80pt}{$+$}
}%
{dyson}{}

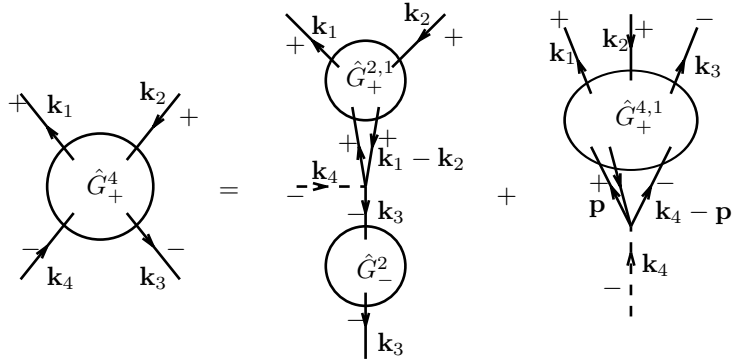
\*\* \centerline{\eqg(3): Graphical representation of the Dyson equation
\equ(2.11a);} \centerline{the dotted line represents the ``bare'' propagator
$g(\kk_4)$}

\*
The l.h.s. of the Dyson equation computed at the cutoff scale
is indeed proportional to the
effective interaction $\l_h$ (see \equ(1.47)
below), while the r.h.s. is proportional to
$\l$. {\it If one does not take into account cancellations in
\equ(2.11a)}, this equation only allows us to prove that
$|\l_h|\le C_h |\l|$, with $C_h$ diverging as $h\to-\io$. However,
inspired by the analysis in the physical literature, see [DL], [S], [MD],
we can try to express $\hat G^{2,1}_\o$ and $\hat G^{4,1}_\o$, in the
r.h.s. of \equ(2.11a), in terms of $\hat G^{2}_\o$ and $\hat
G^{4}_\o$ by suitable {\it Ward identities} and {\it correction identities}.

\* \sub(1.4) {\it Ward identities and the first addend of \equ(2.11a).}

To begin with, we consider the first addend in the r.h.s. of the Dyson
equation \equ(2.11a). A remarkable identity relating $\hat G^{2,1}_+$ to
$\hat G^2_+$ can be obtained by the chiral Gauge transformation
$\psi^\pm_{\xx,+}\to e^{\pm i\a_{\xx}} \psi^\pm_{\xx,+}$,
$\psi^\pm_{\xx,-}\to \psi^\pm_{\xx,-}$ in the generating functional
\equ(1.7); one obtains the identity \equ(2.12) below, represented pictorially
in Fig. \graf(4), with $D_\o(\pp)=-i p_0+\o p$.

\insertplot{300pt}{150pt}%
{\ins{53pt}{83pt}{$\hat G^{2,1}_+$}
\ins{5pt}{83pt}{$D_+(\pp)$}
\ins{45pt}{110pt}{$\kk$}
\ins{70pt}{110pt}{$\qq$}
\ins{42pt}{25pt}{$\pp=\kk-\qq$}
\ins{97pt}{80pt}{$=$}
\ins{125pt}{81pt}{$\hat G^2_+$}
\ins{135pt}{50pt}{$\qq$}
\ins{135pt}{105pt}{$\qq$}
\ins{155pt}{80pt}{$-$}
\ins{185pt}{81pt}{$\hat G^2_+$}
\ins{195pt}{50pt}{$\kk$}
\ins{195pt}{105pt}{$\kk$}
\ins{215pt}{80pt}{$+$}
\ins{248pt}{83pt}{$\hat \D^{2,1}_+$}
\ins{240pt}{110pt}{$\kk$}
\ins{265pt}{110pt}{$\qq$}
\ins{253pt}{25pt}{$\pp$}
}%
{ward1}{}

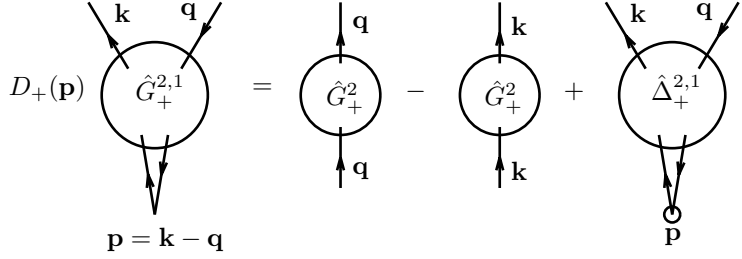
\* \centerline{\eqg(4): Graphical representation of the Ward identity
\equ(2.12);} \centerline{the small circle in $\hat \D^{2,1}_+$ represents the
function $C_+$ of \equ(2.14)}

\*

The above Ward identity provides a relation between $\hat G^{2,1}_+$, $\hat
G^{2}_+$ and a ${\it correction \ term}$ $\hat\D^{2,1}_+$, which can be
obtained, through the analogue of \equ(1.8), from a functional integral very
similar to \equ(1.7), with the difference that $\sum_\o \int d\pp d\kk \hat
J_{\pp,\o} \psi^+_{\kk,\o} \psi^-_{\kk-\pp,\o}$ is replaced by $\int d\pp
d\kk C_+(\kk,\kk-\pp) \hat J_{\pp} \psi^+_{\kk,+}  \psi^-_{\kk-\pp,+}$; the
function $C_+(\kk,\kk-\pp)$, defined in \equ(2.14) below and represented by
the small circle in Fig. \graf(4), is vanishing (like the term
$\hat\D^{2,1}_+$ itself) if $C_{h,0}^{-1}=1$, that is the correction term
would vanish if no cutoffs (which break Gauge invariance) were present in the
model.

{\it Remark.} The above Ward identity is usually stated in the physical
literature by neglecting the correction term $\hat \D^{2,1}_+$; we shall call
in general {\it formal} Ward identities the Ward identities one obtains by
putting equal to $0$ the correction terms. The formal Ward identities are
generally derived, see [DL], [S], [MD], by neglecting the cutoffs, so that
the propagator becomes simply $D_\o(\kk)^{-1}$ and one can use the following
relation
$$D_\o(\pp)^{-1}(D_\o(\kk)^{-1}-D_\o(\kk+\pp)^{-1})=
D_\o(\kk)^{-1}D_\o(\kk+\pp)^{-1}\;.\Eq(7.44)$$
After the derivation of the formal Ward identities, the cutoffs are
introduced in order to have non diverging quantities; this approximation
leads however to some well known inconsistencies, see [G].

\*

The use of Ward identities is to provide relations between Schwinger
functions, but the correction terms (due to the cutoffs) substantially affect
the Ward identities and apparently spoil them of their utility. However there
are other remarkable relations connecting the correction terms to the
Schwinger functions; such {\it correction identities} can be proved by
performing a careful analysis of the renormalized expansion for the
correction terms, and come out of the peculiar properties of the function
$C_+(\kk,\kk-\pp)$, see [BM2] and section \secc(3). For example, the analysis
of [BM2] implies that $\hat \D_+^{2,1}$ verifies the following {\it
correction identity}, see Fig.\graf(5)
$$\hat \D_+^{2,1}(\pp,\kk,\qq)=\n_+  D_+(\pp)
\hat G_+^{2,1}(\pp,\kk,\qq)+\n_- D_-(\pp)
\hat G_-^{2,1}(\pp,\kk,\qq)+
\hat H_+^{2,1}(\pp,\kk,\qq)\Eq(1.3b)$$
where $\n_+,\n_-$ are $O(\l)$ and weakly dependent on $h$, once we prove that
$\l_j$ is small enough for $j\ge h$, and $\hat H_+^{2,1}(\pp,\kk,\qq)$ can be
obtained through the analogue of \equ(1.8), from a functional integral very
similar to \equ(1.7), with the difference that $\sum_\o \int d\pp d\kk \hat
J_{\pp,\o} \psi^+_{\kk,\o} \psi^-_{\kk-\pp,\o}$ is replaced by
$$\int d\pp d\kk C_+(\kk,\kk-\pp)
\hat J_{\pp} \psi^+_{\kk,+}  \psi^-_{\kk-\pp,+} -\sum_\o  \n_\o \int d\pp
d\kk \hat J_{\pp} D_\o(\pp) \psi^+_{\kk,\o} \psi^+_{\kk-\pp,\o}\;.\Eq(ri)$$
The crucial point is that $\hat H_+^{2,1}$, when computed for momenta at the
cut-off scale, {\it is $O(\g^{\th h})$ smaller, with $0<\th<1$ a positive
constant, with respect to the first two addends of the r.h.s. of \equ(1.3b)}.
In other words the correction identity \equ(1.3b) says that the correction
term $\hat \D_+^{2,1}$, which is usually neglected, can be written in terms
of the Schwinger functions $\hat G_+^{2,1}$ and $\hat G_-^{2,1}$ up to the
exponentially smaller term $\hat H_+^{2,1}$.

\insertplot{300pt}{140pt}%
{\ins{33pt}{130pt}{$\hat \D^{2,1}_+$}
\ins{67pt}{80pt}{$=$}
\ins{105pt}{130pt}{$\n_+ D_+\hat G^{2,1}_+$}
\ins{155pt}{80pt}{$+$}
\ins{185pt}{130pt}{$\n_- D_-\hat G^{2,1}_-$}
\ins{235pt}{80pt}{$+$}
\ins{258pt}{130pt}{$\hat H^{2,1}_+$}
}%
{ward3}{}

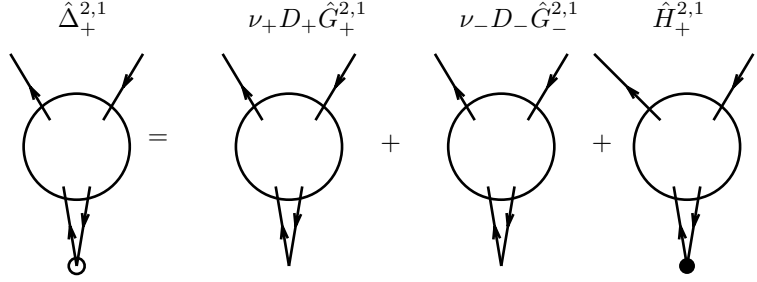
\centerline{\eqg(5): Graphical representation of the correction identity
\equ(1.3b);} \centerline{the filled point in the last term represents
\equ(ri)} \*

Note that \equ(1.3b) was not explicitly stated in [BM2], but its proof,
which we omit here too, is implicitly contained in the proof of Theorem 4 of
that paper. One has to use a strategy similar (but much simpler) to that used
in \S\secc(3.3) below.

Inserting the correction identity \equ(1.3b) in the Ward identity \equ(2.12),
we obtain the new identity
$$(1-\n_+)D_+(\pp)\hat G^{2,1}_{+}(\pp,\kk,\qq)-
\n_- D_-(\pp)\hat G^{2,1}_{-}(\pp,\kk,\qq)=\hat G^{2}_{+}(\qq)- \hat
G^{2}_{+}(\kk)+\hat H^{2,1}_{+}(\pp,\kk,\qq)\;.\Eq(2.12b)$$
In the same way one can show that the formal Ward identity $D_-(\pp)\hat
G^{2,1}_{-}(\pp,\kk,\qq)=0$ becomes, if the cutoffs are taken into account
$$(1-\n'_-)D_-(\pp)\hat G^{2,1}_-(\pp,\kk,\qq)-
\n'_+ D_+(\pp)\hat G^{2,1}_+(\pp,\kk,\qq)= \hat
H^{2,1}_-(\pp,\kk,\qq)\;,\Eq(2.12c)$$
where, by symmetry reasons, $\n'_\pm=\n_\mp$ and $H^{2,1}_-$ satisfying a
bound similar to that of $H^{2,1}_+$, when computed for momenta at the cutoff
scale.

{\it Remark.} By removing both the ultraviolet and the infrared cutoff, see
the Appendix, {\it the functions $\hat H_\pm^{2,1}$ vanishes}, if we fix the
momenta to some cutoff independent values. Hence the formal Ward identities
are not true, even after the removal of the infrared and ultraviolet cutoff,
but must be replaced by \equ(2.12b) and \equ(2.12c), with $\hat
H_\pm^{2,1}=0$. In other words, the presence of cutoffs produces
modifications to the formal Ward identities, which persist when the cutoffs
are removed, a phenomenon known as {\it anomaly}; in the case of Ward
identities based on chiral gauge transformations, one speaks of {\it chiral
anomaly}, see [Z].

\*

The identities \equ(2.12b) and \equ(2.12c) allow us to write $\hat
G^{2,1}_{+}$ in terms of $\hat G_2$ and $\hat H_\pm^{2,1}$. If we put the
expression so obtained in the first addend of \equ(2.11a), we can prove that
it is indeed proportional to $\l$ with the ``right'' proportionality
constant, see \equ(2.23a), \equ(2.23) below. Note that in \equ(2.23) we make reference to
a bound explicitly proved in [BM2], which is expressed directly in terms of
the function $\D_+^{2,1}$. In the following section we shall explain how a
similar strategy can be applied to the second addend of
\equ(2.11a).

\* \sub(1.5) {\it Ward identities and the second addend of \equ(2.11a).}

The analysis of the second addend of \equ(2.11a) is more complex, the reason
being that $\pp$ is integrated instead of being fixed at the infrared cutoff
scale, as it was the case for the first addend. If we simply compute $\hat
G^{4,1}_\o$ by our series expansion and we insert it in the second addend of
\equ(2.11a), we get a "bad" bound, just by dimensional reasons. We can
however derive a Ward identity for $\hat G^{4,1}$, in the form of \equ(2.14a)
below, see Fig. \graf(6).

\insertplot{300pt}{150pt}%
{\ins{53pt}{83pt}{$\hat G^{4,1}_+$}
\ins{5pt}{83pt}{$D_+(\pp)$}
\ins{15pt}{110pt}{$\kk_1 +$}
\ins{40pt}{120pt}{$\kk_2 +$}
\ins{66pt}{120pt}{$\kk_3 -$}
\ins{75pt}{110pt}{$\kk_4-\pp$}
\ins{90pt}{100pt}{$-$}
\ins{68pt}{40pt}{$\pp$}
\ins{97pt}{80pt}{$=$}
\ins{125pt}{81pt}{$\hat G^{4}_+$}
\ins{110pt}{115pt}{$\kk_1-\pp$}
\ins{120pt}{102pt}{$+$}
\ins{150pt}{115pt}{$\kk_2$}
\ins{145pt}{95pt}{$+$}
\ins{140pt}{50pt}{$\kk_3$}
\ins{145pt}{60pt}{$-$}
\ins{99pt}{45pt}{$\kk_4-\pp$}
\ins{112pt}{63pt}{$-$}
\ins{165pt}{80pt}{$-$}
\ins{195pt}{81pt}{$\hat G^{4}_+$}
\ins{175pt}{115pt}{$\kk_1$}
\ins{190pt}{102pt}{$+$}
\ins{200pt}{115pt}{$\kk_2+\pp$}
\ins{215pt}{95pt}{$+$}
\ins{210pt}{50pt}{$\kk_3$}
\ins{215pt}{60pt}{$-$}
\ins{175pt}{45pt}{$\kk_4-\pp$}
\ins{180pt}{60pt}{$-$}
\ins{235pt}{80pt}{$+$}
\ins{268pt}{83pt}{$\hat \D^{4,1}_+$}
\ins{235pt}{110pt}{$\kk_1 +$}
\ins{255pt}{120pt}{$\kk_2 +$}
\ins{281pt}{120pt}{$\kk_3 -$}
\ins{292pt}{110pt}{$\kk_4-\pp$}
\ins{303pt}{100pt}{$-$}
\ins{283pt}{40pt}{$\pp$}
\ins{100pt}{20pt}{$0=\kk_1+\kk_3-\kk_2-\kk_4$}
}%
{ward2}{}

\* \centerline{\eqg(6): Graphical representation of the Ward identity
\equ(2.14a)}

\*

If we insert the above identity in the second addend of the Dyson equation,
we get three terms (all multiplied by $\l$); two of them, the ones involving
the $\hat G^4_+$ functions (which are of the form $g(\kk_4)(L\b)^{-1}
\sum_\pp \hat G^{4}_+ D_+(\pp)^{-1}$), admit good bounds, see \equ(2.24)
below. On the contrary, the third term, which is of the form
$g(\kk_4)(L\b)^{-1} \sum_\pp \hat \D^{4,1}_+ D_+(\pp)^{-1}$, has a "bad"
bound, see [BM3]. This is not surprising, as also $\hat\D^{4,1}_+$ verifies a
{\it correction identity}, which represents it in terms of $\hat G^{4,1}_+$
and $\hat G^{4,1}_-$; see \equ(2.17a) below and Fig. \graf(7).

\insertplot{300pt}{150pt}%
{\ins{33pt}{130pt}{$\hat \D^{4,1}_+$}
\ins{67pt}{80pt}{$=$}
\ins{100pt}{135pt}{$\n_+ D_+\hat G^{4,1}_+$}
\ins{155pt}{80pt}{$+$}
\ins{180pt}{135pt}{$\n_- D_-\hat G^{4,1}_-$}
\ins{235pt}{80pt}{$+$}
\ins{258pt}{130pt}{$\hat H^{4,1}_+$}
}%
{ward5}{}

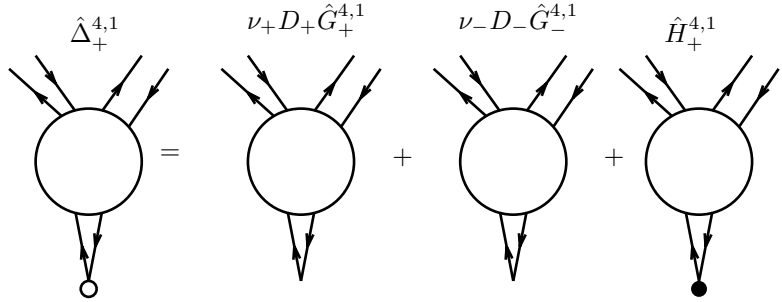
\centerline{\eqg(7): The correction identity for $\hat\D^{4,1}_\o$; the
filled point in the last term represents \equ(ri)} \*

By combining the above equation and the Ward identities for $\hat G^{4,1}_+$
and $\hat G^{4,1}_-$ we obtain, after some algebra, an equation relating
$\hat G^{4,1}_+$ to $\hat G^{4}_+$ and functions $\hat H^{4,1}_+$ and $\hat
H^{4,1}_-$; see \equ(2.18) below. Inserting this expression in the second
addend of the r.h.s. of the Dyson equation, we get our final expression, see
\equ(2.20) below, for the Dyson equation, containing several terms; among
them the ones still requiring a further analysis are the ones involving the
functions $\hat H^{4,1}_\pm$, namely $g(\kk_4) (L\b)^{-1}\sum_\pp \hat
H^{4,1}_\pm D_+(\pp)^{-1}$, represented as in Fig. 8.

\insertplot{300pt}{150pt}%
{\ins{145pt}{108pt}{$\hat H^{4,1}_+$}
\ins{118pt}{130pt}{$\kk_1$}
\ins{120pt}{142pt}{$+$}
\ins{139pt}{135pt}{$\kk_2$}
\ins{152pt}{138pt}{$+$}
\ins{175pt}{125pt}{$\kk_3$}
\ins{175pt}{142pt}{$-$}
\ins{155pt}{50pt}{$\kk_4$}
\ins{140pt}{40pt}{$-$}
\ins{160pt}{70pt}{$\kk_4-\pp$}
\ins{160pt}{80pt}{$-$}
\ins{135pt}{70pt}{$\pp$}
\ins{135pt}{80pt}{$+$}
}%
{dyson2}{}

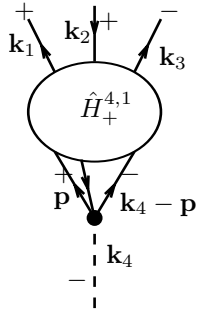
\centerline{\eqg(8): Graphical representation of the term containing $\hat
H^{4,1}_\pm$} \*\*

The analysis of such terms is done in \S 4; we again start by writing for
them a Dyson equation similar to \equ(2.11a), in which the analogous of the
first addend in the l.h.s. vanishes; this Dyson equation allows to write this
term in terms of a function $\tilde G_4$ similar to $G_4$.

\insertplot{300pt}{150pt}%
{\ins{45pt}{82pt}{$\tilde G^{4}_+$}
\ins{30pt}{110pt}{$\kk_1$}
\ins{15pt}{110pt}{$+$}
\ins{60pt}{115pt}{$\kk_2$}
\ins{80pt}{105pt}{$+$}
\ins{65pt}{45pt}{$\kk_3$}
\ins{20pt}{55pt}{$-$}
\ins{30pt}{45pt}{$\kk_4$}
\ins{75pt}{55pt}{$-$}
\ins{125pt}{75pt}{$=$}
\ins{245pt}{108pt}{$\hat H^{4,1}_+$}
\ins{220pt}{130pt}{$\kk_1$}
\ins{220pt}{142pt}{$+$}
\ins{238pt}{135pt}{$\kk_2$}
\ins{252pt}{138pt}{$+$}
\ins{275pt}{125pt}{$\kk_3$}
\ins{275pt}{142pt}{$-$}
\ins{255pt}{50pt}{$\kk_4$}
\ins{240pt}{40pt}{$-$}
\ins{260pt}{70pt}{$\kk_4-\pp$}
\ins{260pt}{80pt}{$-$}
\ins{235pt}{70pt}{$\pp$}
\ins{235pt}{80pt}{$+$}
}%
{dyson1}{}

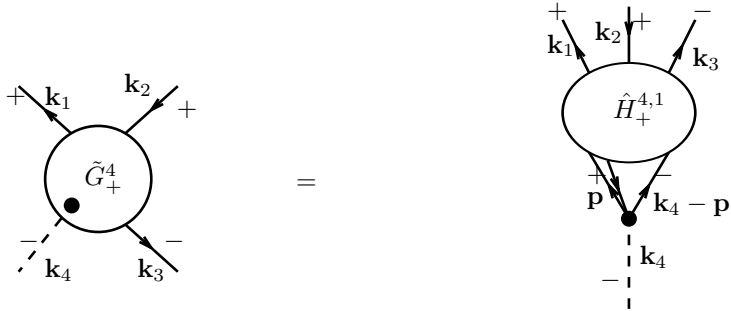
\centerline{\eqg(9): Graphical representation of the Dyson equation for the
correction.}

\*

We can study $\tilde G_4$ by a multiscale analysis very similar to the one
for $\hat G^4$; the presence of a "special" vertex (the one associated to the
filled point in Fig \graf(9)) has however the effect that a new running
coupling appears, associated with the local part of the terms with four
external lines among which one is the dotted line in Fig. \graf(9), to which
the bare propagator $\hat g(\kk_4)$ is associated; we will call this new
running coupling constant $\tilde\l_j$. It would seem that we have a problem
more difficult than our initial one, since we have now to control the flow of
two running coupling constants, $\l_j$ and $\tilde\l_j$, instead of one.
However, it turns out, see Lemma 4.2, that $\l_j$ and $\tilde\l_j$ are not
independent but are essentially {\it proportional}; this follows from a
careful analysis of the expansion for $\tilde\l_j$, based on the properties
of the function $C_\o(\kk, \kk-\pp)$. One then gets for $\tilde G_4$ a bound
very similar to the one for $\hat G_4$, except that $\l_h$ is replaced by
$\tilde\l_h$ (but $\tilde\l_h$ and $\l_h$ are proportional) and there is no
wave function renormalization associated to the external line with momentum
$\kk_4$ (to such line is associated a "bare" instead of a "dressed"
propagator, like in $\hat G_4$). We can however identify two class of terms
in the expansion of $\tilde G_4$, see Fig. \graf(5vd), and summing them has
the effect that also the external line with momentum $\kk_4$ is dressed by
the interaction, and this allows us to complete the proof that
$\l_h=\l+O(\l^2)$ for any $h$.

\insertplot{300pt}{100pt}%
{\ins{120pt}{60pt}{$-$}
\ins{72pt}{53pt}{$\tilde\l_{j^*}$}
\ins{222pt}{53pt}{$\l_{j^*}$}
\ins{235pt}{70pt}{$\hat g^{(h)}$}
}%
{lambda}{}

\centerline{\eqg(5vd): The last resummation}

\*

Finally we show in the Appendix that a simple extension of our analysis
implies that also the ultraviolet cutoff can be removed, that is we can
construct a QFT corresponding to the Thirring model.

\*\*

\section(1, Renormalization Group analysis)

\* \sub(2.1) {\it Multiscale integration.}

We resume the Renormalization Group analysis in [BM1] for the
generating function \equ(1.7). The functional integration of
\equ(1.7) can be performed iteratively in the following way. We
prove by induction that, for any negative $j$, there are a
constant $E_j$, a positive function $\tilde Z_j(\kk)$ and
functionals $\VV^{(j)}$ and $\BB^{(j)}$ such that
$$e^{\WW(\phi,J)}=e^{-L\b E_j}
\int P_{\tilde Z_j,C_{h,j}} (d\psi^{[h,j]})
e^{-\VV^{(j)}(\sqrt{Z_j}\psi^{[h,j]})+\BB^{(j)}
(\sqrt{Z_j}\psi^{[h,j]},\phi,J)}\,,\Eq(1.11)$$
where:

1) $P_{\tilde Z_j,C_{h,j}}(d\psi^{[h,j]})$ is the {\it effective
Grassmannian measure at scale $j$}, equal to, if $Z_j=\max_{\kk}
\tilde Z_j({\kk})$,
$$\eqalign{
P_{\tilde Z_j,C_{h,j}}(d\psi^{[h,j]}) &= \prod_{\kk:C_{h,j}(\kk)>0}
\prod_{\o=\pm1}
{d\hat\psi^{[h,j])+}_{\kk,\o}d\hat\psi^{[h,j]-}_{\kk,\o}\over \NN_j(\kk)}
\cdot\cr &\cdot\; \exp \left\{-{1\over L\b} \sum_{\kk} \, C_{h,j}(\kk)
\tilde Z_j(\kk)\sum_{\o\pm1} \hat\psi^{[h,j]+}_{\o} D_\o(\kk)
\hat\psi^{[h,j]-}_{\kk,\o}\right\}\;,\cr} \Eq(1.12)$$
$$\NN_j(\kk)=(L\b)^{-1} C_{h,j}(\kk) \tilde Z_j(\kk)
[-k_0^2-k^2]^{1/2}\;,\Eq(1.13)$$
$$C_{h,j}(\kk)^{-1}=\sum_{r=h}^j f_r(\kk) \=\c_{h,j}(\kk) \virg
D_\o(\kk)=-ik_0+\o k\;;\Eq(1.14)$$

2) the {\it effective potential on scale $j$}, $\VV^{(j)}(\psi)$,
is a sum of monomial of Grassmannian variables multiplied by
suitable kernels. \ie it is of the form
$$\VV^{(j)}(\psi) = \sum_{n=1}^\io {1\over (L\b)^{2n}}
\sum_{\kk_1,\ldots,\kk_{2n} \atop \o_1,\ldots,\o_{2n}}
\prod_{i=1}^{2n} \hat\psi^{\s_i}_{\kk_i,\o_i}
\hat W_{2n,\oo}^{(j)}(\kk_1,...,\kk_{2n-1})
\d\left(\sum_{i=1}^{2n}\s_i\kk_i\right)\;,\Eq(1.15)$$
where $\s_i=+$ for $i=1,\ldots,n$, $\s_i=-$ for $i=n+1,\ldots,2n$
and $\oo=(\o_1,\ldots,\o_{2n})$;

3) the {\it effective source term at scale $j$},
$\BB^{(j)}(\sqrt{Z_j}\psi, \phi,J)$, is a sum of monomials of
Grassmannian variables and $\phi^\pm,J$ field, with at least one
$\phi^\pm$ or one $J$ field; we shall write it in the form
$$\BB^{(j)}(\sqrt{Z_j}\psi, \phi,J) = \BB_\phi^{(j)}(\sqrt{Z_j}\psi)+
\BB_J^{(j)}(\sqrt{Z_j}\psi) + W_R^{(j)}(\sqrt{Z_j}\psi,\phi,J)\;,\Eq(1.16)$$
where $\BB_\phi^{(j)}(\psi)$ and $\BB_J^{(j)}(\psi)$ denote the sums over
the terms containing only one $\phi$ or $J$ field, respectively.

Of course \equ(1.11) is true for $j=0$, with
$$\eqalign{
&\tilde Z_0(\kk)=1,\qquad E_0=0,\qquad \VV^{(0)}(\psi)=V(\psi),\qquad
W_R^{(0)}=0,\cr
&\BB_\phi^{(0)}(\psi)=\sum_\o \int d\xx [
\phi^+_{\xx,\o}\psi^{-}_{\xx,\o}+ \psi^{+}_{\xx,\o}\phi^-_{\xx,\o}],\qquad
\BB_J^{(0)}(\psi)=\sum_\o \int d\xx
J_{\xx,\o}\psi^{+}_{\xx,\o}\psi^{-}_{\xx,\o}\;.\cr}\Eq(1.17)$$
Let us now assume that \equ(1.11) is satisfied for a certain $j\le 0$ and
let us show that it holds also with $j-1$ in place of $j$.

In order to perform the integration corresponding to $\psi^{(j)}$,
we write the effective potential and the effective source
as sum of two terms, according to the following rules.

We split the effective potential
$\VV^{(j)}$ as $\LL \VV^{(j)}+\RR \VV^{(j)}$, where
$\RR=1-\LL$ and $\LL$, the {\it localization operator}, is a linear operator
on functions of the form \equ(1.15), defined in the following way by its action
on the kernels $\hat W_{2n,\oo}^{(j)}$.
\*
1) If $2n=4$, then
$$\LL \hat W_{4,\oo}^{(j)}(\kk_1,\kk_2,\kk_3)=
\hat W_{4,\oo}^{(j)}(\bk++,\bk++,\bk++)\;,\Eq(1.18)$$
where
$\bk\h{\h'} = (\h\p L^{-1},\h'\p\b^{-1})$.
Note that $\LL \hat W_{4,\oo}^{(j)}(\kk_1,\kk_2,\kk_3)=0$, if $\sum_{i=1}^4
\o_i\not=0$, by simple symmetry considerations.
\*
2) If $2n=2$ (in this case there is a non zero contribution only
if $\o_1=\o_2$)
$$\LL \hat W_{2,\oo}^{(j)}(\kk)= \fra14 \sum_{\h,\h'=\pm 1}
\hat W_{2,\oo}^{(j)}(\bk\h{\h'})\left\{ 1+ \h {L\over \p} +
\h'{\b\over \p} k_0 \right\}\;,\Eq(1.19)$$
\*
3) In all the other cases
$$\LL \hat W_{2n,\oo}^{(j)}(\kk_1,\ldots,\kk_{2n-1})=0\;.\Eq(1.20)$$

\*
These definitions are such that $\LL^2=\LL$, a property which plays an
important role in the analysis of [BM1]. Moreover, by using the symmetries
of the model, it is easy to see that
$$\LL\VV^{(j)}(\psi^{[h,j]})=z_j F_\z^{[h,j]}+a_j
F_\a^{[h,j]}+l_j F_\l^{[h,j]}\;,\Eq(1.21) $$
where $z_j$, $a_j$ and $l_j$ are real numbers and
$$\eqalignno{
F_\a^{[h,j]}&=\sum_\o {\o\over (L\b)}
\sum_{\kk:C^{-1}_{h,j}(\kk)>0} k\hat\psi^{[h,j]+}_{\kk,\o}
\hat\psi^{[h,j]-}_{\kk,\o}= \sum_\o i\o \int_\L d\xx
\psi^{[h,j]+}_{\xx,\o}
\dpr_x\psi^{[h,j]-}_{\xx,\o}\;,&\eq(1.22)\cr F_\z^{[h,j]}&=\sum_\o
{1\over (L\b)} \sum_{\kk:C^{-1}_{h,j}(\kk)>0} (-i k_0)
\hat\psi^{[h,j]+}_{\kk,\o}\hat\psi^{[h,j]-}_{\kk',\o} = -\sum_\o
\int_\L d\xx \psi^{[h,j]+}_{\xx,\o} \dpr_0\psi^{[h,j]-}_{\xx,\o}
\;,&\eq(1.23)\cr F_\l^{[h,j]}&={1\over
(L\b)^4}\sum_{\kk_1,...,\kk_4:C^{-1}_{h,j}(\kk_i)>0}
\hat\psi^{[h,j]+}_{\kk_1,+} \hat\psi^{[h,j]-}_{\kk_2,+}
\hat\psi^{[h,j]+}_{\kk_3,-}
\hat\psi^{[h,j]-}_{\kk_4,-}\d(\kk_1-\kk_2+\kk_3-\kk_4).
&\eq(1.24)\cr}$$
$\dpr_x$ and $\dpr_0$ are defined in an obvious way, so that the second
equality in \equ(1.22) and \equ(1.23) is satisfied; if $N=\io$ they are simply
the partial derivative with respect to $x$ and $x_0$.
Note that $\LL\VV^{(0)}=\VV^{(0)}$, hence $l_{0}=\l$, $a_0=z_0=0$.

In the limit $L=\b=\io$ one has $a_j=z_j$ as a trivial
consequence of the symmetries of the propagator. If
$L,\b$ are finite this is not true and
by dimensional arguments it follows that $z_j-a_j$ is of
order $\g^{-j}\max\{L^{-1},\b^{-1}\}$.

Analogously we write $\BB^{(j)}=\LL \BB^{(j)}+\RR \BB^{(j)}$,
$\RR=1-\LL$, according to the following definition. First of all,
we put $\LL W_R^{(j)}=W_R^{(j)}$. Let us consider now
$\BB_J^{(j)}(\sqrt{Z_j}\psi)$. It is easy to see that the field
$J$ is equivalent, from the point of view of dimensional
considerations, to two $\psi$ fields. Hence, the only terms which
need to be renormalized are those of second order in $\psi$, which
are indeed marginal. We shall use for them the definition
$$\eqalign{
\BB_J^{(j,2)}(\sqrt{Z_j}\psi) &= \sum_{\o,\tilde\o} \int d\xx d\yy d\zz
B_{\o,\tilde\o}(\xx,\yy,\zz) J_{\xx,\o} (\sqrt{Z_j}\psi^{+}_{\yy,\tilde\o})
(\sqrt{Z_j}\psi^{-}_{\zz,\tilde\o}) =\cr
&= \sum_{\o,\tilde\o} \int {d\pp\over (2\p)^2} {d\kk\over (2\p)^2}
\hat B_{\o,\tilde\o}(\pp,\kk) \hat J(\pp)
(\sqrt{Z_j}\hat\psi^{+}_{\pp+\kk,\tilde\o}) (\sqrt{Z_j}
\hat\psi^{-}_{\kk,\tilde\o})\;.\cr}\Eq(1.25)$$

We regularize $\BB_J^{(j,2)}(\sqrt{Z_j}\psi)$, in analogy to what we did
for the effective potential, by decomposing it as the sum of
$\LL\BB_J^{(j,2)}(\sqrt{Z_j}\psi)$ and
$\RR\BB_J^{(j,2)}(\sqrt{Z_j}\psi)$, where $\LL$ is defined through its
action on $\hat B_\o(\pp,\kk)$ in the following way:
$$\LL \hat B_{\o,\tilde\o}(\pp,\kk) ={1\over 4}
\sum_{\h,\h'=\pm 1}\hat B_{\o,\tilde\o}(\bar p_\h,\bar k_{\h,\h'})
\;,\Eq(1.26)$$
where $\bar k_{\h,\h'}$ was defined above and
$\bar p_\h=(0,2\pi\h'/\b)$. In the limit $L=\b=\io$
it reduces simply to
$\LL \hat B_{\o,\tilde\o}(\pp,\kk)=\hat B_{\o,\tilde\o}(0,0)$.

This definition apparently implies that we have to introduce two new
renormalization constants. However, one can easily show that, in the
limit $L,\b\to \io$, $\hat B_{\o,-\o}(0,0)=0$, while, at finite $L$
and $\b$, $\LL B_{\o,-\o}$ behaves as an irrelevant term, see [BM1].

The previous considerations imply that we can write
$$\LL\BB_J^{(j,2)}(\sqrt{Z_j}\psi)=
\sum_\o {Z_j^{(2)}\over Z_j} \int d\xx J_{\xx,\o}
(\sqrt{Z_j}\psi^{+}_{\xx,\o}) (\sqrt{Z_j}\psi^{-}_{\xx,\o})\;,\Eq(1.27)$$
which defines the renormalization constant $Z_j^{(2)}$.

Finally we have to define $\LL$ for
$\BB_\phi^{(j)}(\sqrt{Z_j}\psi)$; we want to show that,
by a suitable choice of the localization procedure,
if $j\le -1$, it
can be written in the form
$$\eqalign{
&\qquad\qquad \BB_\phi^{(j)}(\sqrt{Z_j}\psi) =
\sum_\o \sum_{i=j+1}^0 \int d\xx d\yy\;\cdot\cr &\cdot\;\left[
\phi^+_{\xx,\o} g^{Q,(i)}_{\o}(\xx-\yy){\dpr\over \dpr\psi^+_{\yy\o}}
\VV^{(j)}(\sqrt{Z_j}\psi) + {\dpr\over \dpr\psi^-_{\yy,\o}}
\VV^{(j)}(\sqrt{Z_{j}}\psi) g^{Q,(i)}_{\o}(\yy-\xx)\phi^-_{\xx,\o} \right]+\cr
&+ \sum_\o\int {d\kk\over (2\p)^2} \left[ \hat\psi^{[h,j]+}_{\kk,\o}
\hat Q^{(j+1)}_{\o}(\kk)
\hat\phi_{\kk,\o}^- +\hat\phi^+_{\kk,\o} \hat Q^{(j+1)}_{\o}(\kk)
\hat\psi^{[h,j]-}_{\kk,\o} \right]\;,\cr}\Eq(1.28)$$
where $\hat g^{Q,(i)}_{\o}(\kk)= \hat g^{(i)}_{\o}(\kk) \hat
Q^{(i)}_{\o}(\kk)$, with
$$\hat g_\o^{(j)}(\kk)={1\over Z_{j-1}}{\tilde f_j(\kk) \over D_\o(\kk)}\;,
\Eq(1.35)$$
$\tilde f_j(\kk)=f_j(\kk) Z_{j-1} [\tilde Z_{j-1}(\kk)]^{-1}$ and
$Q^{(j)}_{\o}(\kk)$ defined inductively by the relations
$$\hat Q^{(j)}_{\o}(\kk)=\hat Q^{(j+1)}_{\o}(\kk) - z_j Z_j
D_\o(\kk) \sum_{i=j+1}^0 \hat g^{Q,(i)}_{\o}(\kk)\;,
\quad \hat Q^{(0)}_{\o}(\kk)=1\;.\Eq(1.29)$$
Note that $\hat g_\o^{(j)}(\kk)$ does not depend on the infrared
cutoff for $j>h$ and that (even for $j=h$) $\hat g^{(j)}(\kk)$ is
of size $Z_{j-1}^{-1}\g^{-j}$, see discussion in \S3 of [BM2],
after eq. (60). Moreover the propagator $\hat g^{Q,(i)}_{\o}(\kk)$
is equivalent to $\hat g^{(i)}_{\o}(\kk)$, as concerns the
dimensional bounds.

The $\LL$ operation for $\BB^{(j)}_\phi$
is defined by decomposing $\VV^{(j)}$ in the r.h.s. of \equ(1.28)
as $\LL \VV^{(j)}+\RR \VV^{(j)}$, $\LL
\VV^{(j)}$ being defined by \equ(1.21).

After writing $\VV^{(j)}=\LL \VV^{(j)}+\RR\VV^{(j)}$
and $\BB^{(j)}=\LL \BB^{(j)}+\RR\BB^{(j)}$, the next step is
to {\sl renormalize} the free measure $P_{\tilde Z_j,C_{h,j}}
(d\psi^{[h,j]})$, by adding to it part of the r.h.s. of \equ(1.21). We get
$$\eqalign{
& \int P_{\tilde Z_j,C_{h,j}}(d\psi^{[h,j]}) \, e^{-\VV^{(j)}
(\sqrt{Z_j}\psi^{[h,j]})+\BB^{(j)}
(\sqrt{Z_j}\psi^{[h,j]})}=\cr
&\quad =e^{-L\b t_j}\int P_{\tilde Z_{j-1},C_{h,j}}(d\psi^{[h,j]}) \,
e^{-\tilde\VV^{(j)}(\sqrt{Z_j}\psi^{[h,j]})+\tilde\BB^{(j)}
(\sqrt{Z_j}\psi^{[h,j]})}\;,\cr}\Eq(1.30)$$
where
$$\tilde Z_{j-1}(\kk)=Z_j(\kk)[1+\c_{h,j}(\kk) z_j]\;,\Eq(1.31)$$
$$\tilde\VV^{(j)}(\sqrt{Z_j}\psi^{[h,j]}) = \VV^{(j)}(\sqrt{Z_j}\psi^{[h,j]})
-z_j Z_j[F_\z^{[h,j]}+F_\a^{[h,j]}]\,,\Eq(1.32)$$
and the factor $\exp(-L\b t_j)$ in \equ(1.30) takes into account the different
normalization of the two measures. Moreover
$$\tilde\BB^{(j)}(\sqrt{Z_j}\psi^{[h,j]})=
\tilde\BB^{(j)}_\phi(\sqrt{Z_j}\psi^{[h,j]})
+\BB^{(j)}_J(\sqrt{Z_j}\psi^{[h,j]})+W^{(j)}_R\;,\Eq(1.33)$$
where $\tilde\BB^{(j)}_\phi$ is obtained from $\BB^{(j)}_\phi$ by inserting
\equ(1.32) in the second line of \equ(1.28) and by absorbing the terms
proportional to $z_j$ in the terms in the third line of \equ(1.28).

If $j>h$, the r.h.s of \equ(1.30) can be written as
$$e^{-L\b t_j} \int P_{\tilde Z_{j-1},C_{h,j-1}} (d\psi^{[h,j-1]})
\int P_{Z_{j-1},\tilde f_j^{-1}}(d\psi^{(j)}) \, $$
$$e^{-\tilde \VV^{(j)}\big(\sqrt{Z_j}[\psi^{[h,j-1]} + \psi^{(j)}]\big)+
\tilde \BB^{(j)} \big(\sqrt{Z_j}[\psi^{[h,j-1]} + \psi^{(j)}]
\big)}\;,\Eq(1.34)$$
where $P_{Z_{j-1},\tilde f_j^{-1}}(d\psi^{(j)})$ is the
integration with propagator $\hat g_\o^{(j)}(\kk)$.

We now {\it rescale} the field so that
$$\tilde\VV^{(j)}(\sqrt{Z_j}\psi^{[h,j]})= \hat\VV^{(j)}(\sqrt{Z_{j-1}}
\psi^{[h,j]})\quad,\quad \tilde\BB^{(j)}(\sqrt{Z_j}\psi^{[h,j]})=
\hat\BB^{(j)}(\sqrt{Z_{j-1}}
\psi^{[h,j]})\;;\Eq(1.36)$$
it follows that (in the limit $L,\b=\io$, so that $a_j=z_j$, see above)
$$\LL\hat\VV^{(j)}(\psi^{[h,j]})=\l_j F_\l^{[h,j]}\;,
\Eq(1.37)$$
where $\l_j=(Z_j Z_{j-1}^{-1})^2 l_j$. If we now define
$$\eqalign{
&e^{-\VV^{(j-1)}\big(\sqrt{Z_{j-1}} \psi^{[h,j-1]} \big)+
\BB^{(j-1)}\big(\sqrt{Z_{j-1}}\psi^{[h,j-1]}\big)-L\b E_j}=\cr
&=\int P_{Z_{j-1},\tilde f_j^{-1}}(d\psi^{(j)}) \, e^{-\hat
\VV^{(j)} \big(\sqrt{Z_{j-1}}[\psi^{[h,j-1]} + \psi^{(j)}]\big)+
\hat \BB^{(j)} \big(\sqrt{Z_{j-1}}[\psi^{[h,j-1]} + \psi^{(j)}]
\big)}\;,\cr}\Eq(1.38)$$
it is easy to see that $\VV^{(j-1)}$ and $\BB^{(j-1)}$ are of the
same form of $\VV^{(j)}$ and $\BB^{(j)}$ and that the procedure can
be iterated. Note that the above procedure
allows, in particular, to write the running coupling constant $\l_j$,
$0< j\le h$, in terms of $\l_{j'}$, $0\ge j'\ge j+1$:
$$\l_j= \b_j^{(h)}( \l_{j+1},\ldots,\l_0) \virg \l_0=\l,
\Eq(1.39)$$
The function $\b_j^{(h)}(\l_{j+1},...,\l_0)$ is called
the {\it Beta function}. By the remark above on the independence of
scale $j$ propagators of $h$ for $j>h$, it is independent of $h$,
for $j>h$.

\*
\sub(2.2) {\it Tree expansion.}

At the end of the iterative integration procedure, we get
$$\WW(\f,J)=-L\b E_{L,\b} + \sum_{m^\phi+n^J\ge 1}
S_{2m^\phi,n^J}^{(h)}(\phi,J)\;,\Eq(1.40)$$
where $E_{L,\b}$ is the {\it free energy} and
$S_{2m^\phi,n^J}^{(h)}(\phi,J)$ are suitable functional, which can
be expanded, as well as $E_{L,\b}$, the effective potentials and
the various terms in the r.h.s. of \equ(1.16) and \equ(1.15), in
terms of trees (for an updated introduction to trees formalism see
also [GM]). This expansion, which is indeed a finite sum for
finite values of $N,L,\b$, is explained in detail in [BGPS] and
[BM1], which we shall refer to often in the following.

Let us consider the family of all trees which can be constructed
by joining a point $r$, the {\it root}, with an ordered set of
$n\ge 1$ points, the {\it endpoints} of the {\it unlabelled tree},
so that $r$ is not a branching point. Two unlabelled trees are
identified if they can be superposed by a suitable continuous
deformation, so that the endpoints with the same index coincide.

$n$ will be called the {\it order} of the unlabelled tree and the
branching points will be called the {\it non trivial vertices}.
The unlabelled trees are partially ordered from the root to the
endpoints in the natural way; we shall use the symbol $<$ to
denote the partial order.

We shall consider also the {\it labelled trees} (to be called
simply trees in the following), see Fig. \graf(1); they are
defined by associating some labels with the unlabelled trees, as
explained in the following items.

\insertplot{300pt}{150pt}%
{{\ins{30pt}{85pt}{$r$}\ins{50pt}{85pt}{$v_0$}\ins{130pt}{100pt}{$v$}%
\ins{35pt}{-2pt}{$j$}\ins{55pt}{-2pt}{$j+1$}\ins{135pt}{-2pt}{$h_v$}%
\ins{215pt}{-2pt}{$-1$}\ins{235pt}{-2pt}{$0$}\ins{255pt}{-2pt}{$+1$}}%

}%
{treelut}{}

\*\* \centerline{\eqg(1): A tree $\t$ and its labels.} \*

\0 1) We associate a label $j\le 0$ with the root and we denote
$\TT_{j,n}$ the corresponding set of labelled trees with $n$
endpoints. Moreover, we introduce a family of vertical lines,
labelled by an integer taking values in $[j,1]$, and we represent
any tree $\t\in\TT_{j,n}$ so that, if $v$ is an endpoint or a non
trivial vertex, it is contained in a vertical line with index
$h_v>j$, to be called the {\it scale} of $v$, while the root is on
the line with index $j$. There is the constraint that, if $v$ is
an endpoint, $h_v>j+1$.

The tree will intersect in general the vertical lines in set of
points different from the root, the endpoints and the non trivial
vertices; these points will be called {\it trivial vertices}. The
set of the {\it vertices} of $\t$ will be the union of the
endpoints, the trivial vertices and the non trivial vertices. The
definition of $h_v$ is extended in an obvious way to the trivial
vertices and the endpoints.

Note that, if $v_1$ and $v_2$ are two vertices and $v_1<v_2$, then
$h_{v_1}<h_{v_2}$. Moreover, there is only one vertex immediately
following the root, which will be denoted $v_0$ and can not be an
endpoint; its scale is $j+1$.

\0 2) There are two kind of endpoints, {\it normal} and {\it special}.

With each normal endpoint $v$ of scale $h_v$ we associate the
local term $\LL\hat\VV^{(h_v)}(\psi^{[h,h_v-1]})$ of
\equ(1.37) and one space-time point $\xx_v$. We shall say that the endpoint
is of type $\l$.

There are two types of special endpoints, to be called of type
$\phi$ and $J$; the first one is associated with the terms in the
third line of \equ(1.28), the second one with the terms in the
r.h.s. of \equ(1.27). Given $v\in\t$, we shall call $n^\phi_v$ and
$n^J_v$ the number of endpoints of type $\phi$ and $J$ following
$v$ in the tree, while $n_v$ will denote the number of normal
endpoints following $v$. Analogously, given $\t$, we shall call
$n^\phi_\t$ and $n^J_\t$ the number of endpoint of type $\phi$ and
$J$, while $n_\t$ will denote the number of normal endpoints.
Finally, $\TT_{j,n,n^\phi,n^J}$ will denote the set of trees
belonging to $\TT_{j,n}$ with $n$ normal endpoints, $n^\phi$
endpoints of type $\phi$ and $n^J$ endpoints of type $J$. Given a
vertex $v$, which is not an endpoint, $\xx_v$ will denote the
family of all space-time points associated with one of the
endpoints following $v$.

\0 3) There is an important constraint on the scale indices of the
endpoints. In fact, if $v$ is an endpoint normal or of type $J$,
$h_v=h_{v'}+1$, if $v'$ is the non trivial vertex immediately
preceding $v$. This constraint takes into account the fact that at
least one of the $\psi$ fields associated with an endpoint normal
or of type $J$ has to be contracted in a propagator of scale
$h_{v'}$, as a consequence of our definitions.

On the contrary, if $v$ is an endpoint of type $\phi$, we shall
only impose the condition that $h_v\ge h_{v'}+1$. In this case the
only $\psi$ field associated with $v$ is contracted in a
propagator of scale $h_v-1$, instead of $h_{v'}$.

\0 4) If $v$ is not an endpoint, the {\it cluster } $L_v$ with
frequency $h_v$ is the set of endpoints following the vertex $v$;
if $v$ is an endpoint, it is itself a ({\it trivial}) cluster. The
tree provides an organization of endpoints into a hierarchy of
clusters.

\0 5) We associate with any vertex $v$ of the tree a set $P_v$,
the {\it external fields} of $v$. The set $P_v$ includes both the
field variables of type $\psi$ which belong to one of the
endpoints following $v$ and are not yet contracted at scale $h_v$
(in the iterative integration procedure), to be called {\it normal
external fields}, and those which belong to an endpoint normal or
of type $J$ and are contracted with a field variable belonging to
an endpoint $\tilde v$ of type $\phi$ through a propagator
$g^{Q,(h_{\tilde v}-1)}$, to be called {\it special external
fields} of $v$.

These subsets must satisfy various constraints. First of all, if
$v$ is not an endpoint and $v_1,\ldots,v_{s_v}$ are the $s_v$
vertices immediately following it, then $P_v \subset \cup_i
P_{v_i}$. We shall denote $Q_{v_i}$ the intersection of $P_v$ and
$P_{v_i}$; this definition implies that $P_v=\cup_i Q_{v_i}$. The
subsets $P_{v_i}\bs Q_{v_i}$, whose union will be made, by
definition, of the {\it internal fields} of $v$, have to be non
empty, if $s_v>1$, that is if $v$ is a non trivial vertex.

Moreover, if the set $P_{v_0}$ contains only special external
fields, that is if $|P_{v_0}|=n^\phi$, and $\tilde v_0$ is the
vertex immediately following $v_0$, then $|P_{v_0}|< |P_{\tilde
v_0}|$.

\*
\sub(2.3) {\it Dimensional bounds.}

We can write
$$\eqalign{
&\qquad\qquad S_{2m^\phi,n^J}^{(h)}(\phi,J) =\cr
&= \sum_{n=0}^\io \sum_{j_0=h-1}^{-1}
\sum_{\t\in\TT_{j_0,n,2m^\phi,n^J}\atop |P_{v_0}|=2m^\phi} \sum_\oo \int d\xxx
\prod_{i=1}^{2m^\phi} \phi^{\s_i}_{\xx_i,\o_i}
\prod_{r=1}^{n^J} J_{\xx_{2m^\phi+r},\o_{2m^\phi+r}}
S_{2m^\phi,n^J,\t,\oo}(\xxx)\;,\cr}\Eq(1.41)$$
where $\oo=\oo=\{\o_1,\ldots,\o_{2m^\phi+n^J}\}$,
$\xxx=\{\xx_1,\ldots, \xx_{2m^\phi+n^J}\}$ and $\s_i=+$ if $i$ is
odd, $\s_i=-$ if $i$ is even.

Let us define $\bar\l_j=\max_{k\ge j}|\l_k|$; in \S3 of [BM1] it
is proved that the kernels satisfy the following bound:
$$\eqalign{
\int d\xxx &|S_{2m^\phi,n^J,\t,\oo}(\xxx)| \le L\b\
(C\bar\l_{j_0})^n \g^{-j_0(-2+m^\phi+n^J)} \prod_{i=1}^{2m^\phi}
{\g^{-h_i}\over (Z_{h_i})^{1/2}}\;\cdot\cr &\cdot\;
\prod_{r=1}^{n^J }{Z_{\bar h_r}^{(2)}\over Z_{\bar h_r}}
\prod_{\rm v\ not\ e.p} ({Z_{h_v}\over Z_{h_v-1}})^{|P_v|/2}
\g^{-d_v} \;,\cr}\Eq(1.42)$$
where $h_i$ is the scale of the propagator linking the $i$-th
endpoint of type $\phi$ to the tree, $\bar h_r$ is the scale of
the $r$-th endpoint of type $J$ and
$$d_v = -2+|P_v|/2+n_v^J +\tilde z(P_v)\;,\Eq(1.43)$$
with
$$\tilde z(P_v)=\cases{
z(P_v) & if $n_v^\phi\le 1\;, n^J_v=0\;,$\cr 1 & if $n_v^\phi=0\;,
n^J_v=1\;,|P_v|=2\;,$\cr 0 & otherwise\cr} \Eq(1.44)$$
and $z(P_v)=1$ if $|P_v|=4$, $z(P_v)=2$ if $|P_v|=2$ and zero
otherwise. \* As explained in \S 5 of [BM1], one can sum over the
trees $\t$ only if $d_v>0$. While it is not true in general that
$d_v>0$ in \equ(1.42), it is true for the trees contributing to
$\hat G^{2,1}_\o$, $\hat G^{2}_\o$, $\hat G^{4}_+$ with external
momenta computed at the cutoff scale; hence by using the bound
\equ(1.42), one can prove, see [BM2] \S 3.5, the following
theorem.

\*

\theorem(th2) {\it There exists $\e_0$ such that, if $\bar\l_h\le
\e_0$ and $|\bar \kk|=\g^h$, then
$$\hat G^{2,1}_\o(2\bar\kk,\bar\kk) =-{Z_h^{(2)}\over Z_h^2 D_\o(\bar\kk)^2}
[1 +O(\bar\l_h^2)]\;,\Eq(1.45)$$
$$\hat G^{2}_\o(\bar\kk)={1\over Z_h D_\o(\bar\kk)}
[1 +O(\bar\l_h^2)]\;,\Eq(1.46)$$
$$\hat G^{4}_+(\bar\kk,-\bar\kk,-\bar\kk)=
Z_h^{-2} |\bar\kk|^{-4} [-\l_h +O(\bar\l_h^2)]\;.\Eq(1.47)$$ } \*

The expansion for \equ(1.45), \equ(1.46), \equ(1.47) in terms of
the running coupling constant $\l_j$ is convergent {\it if} $\l_j$
is small enough for all $j\ge h$. This property is surely true if
$|h|$ is at most of order $|\l|^{-1}$, but to prove that it is
true for any $|h|$ is quite nontrivial, as it is consequence of
intricate cancellations which are present in the Beta function. In
the following section we will show, by using Ward identities and a
Dyson equation, that indeed $\l_j$ is small enough for all $j\ge
h$ {\it for any $h$}, that is uniformly in the infrared cutoff, so
that the above theorem can be applied.

\vskip1cm

\section(2, Vanishing of Beta function)

\* \sub(3.1a) {\it The main theorem}

The main result of this paper is the following theorem.

\*

\theorem(th3) {\it The model \equ(1.7) is well defined in the
limit $h\to-\io$. In fact there are constants $\e_1$ and $c_2$
such that $|\l|\le \e_1$ implies $\bar\l_j\le c_2\e_1$, for any
$j< 0$.}

\*

\proof The proof of Theorem \thm(th3) is done by contradiction.
Assume that there exists a $h\le 0$ such that
$$\bar\l_{h+1}\le c_2\e_1 < |\l_h|\le 2 c_2\e_1\le\e_0\;,\Eq(2.3)$$
where $\e_0$ is the same as in Theorem \thm(th2). We show that
this is not possible, if $\e_1, c_2$ are suitably chosen.

Let us consider the model with cutoff $\g^h$. In the following
sections we shall prove that, in this model,
$$|\l_h-\l|\le c_3 \bar\l_{h+1}^2\;.\Eq(2.4)$$
However, as a consequence of the remark after \equ(1.35), $\l_j$
is the same, for any $j\ge h$, in the model with or without
cutoff; in fact $g^{(j)}(\kk)$ does not depend on $h$ for $j>h$,
and $\l_h$ only depends on the propagators $g^{(j)}(\kk)$ with
$j>h$ by definition. Hence, from \equ(2.4) we get the bound
$|\l_h|\le \e_1+ c_3 c_2^2 \e_1^2$, which is in contradiction with
\equ(2.3) if, for instance, $c_2=2$ and $\e_1\le
1/(4c_3)$.\Halmos

\*

Theorem \thm(th3) implies, as proved in [BM3], that
$$|\b_{j,\l}(\l_h,..,\l_h)|\le C |\l_h|^2\g^{\t j}\;,\Eq(2.1)$$
a property called {\it vanishing of Beta function}, which implies
that there exists $\lim_{j\to-\io} \l_j$ and that this limit is an
analytic function of $\l$. In its turn, the existence of this
limit implies that there exist the limits
$$\lim_{j\to-\io} \log {Z_{j-1}^{(2)}\over Z_j^{(2)}}=\h_2(\l)
\virg \lim_{k\to-\io} \log {Z_{j-1}\over Z_j}=\h(\l)\;,\Eq(2.2)$$
with $\h(\l)=a_2\l^2+O(\l^3)$ and $\h_1(\l)=a_2\l^2+O(\l^3)$,
$a_2>0$.

\* \sub(3.1aa) {\it The Dyson equation.}

Let us now prove the bound \equ(2.4). We define
$$G^{4,1}_{\o}(\zz,\xx_1,\xx_2,\xx_3,\xx_4)=
<\r_{\zz,\o};\psi^-_{\xx_1,+};\psi^+_{\xx_2,+};
\psi^-_{\xx_3,-};\psi^+_{\xx_4,-}>^T\;,\Eq(2.5)$$
$$G^{4}_+(\xx_1,\xx_2,\xx_3,\xx_4)=
<\psi^-_{\xx_1,+}\psi^+_{\xx_2,+};
\psi^-_{\xx_3,-}\psi^+_{\xx_4,-}>^T\;,\Eq(2.6)$$
where
$$\r_{\xx,\o}=\psi^+_{\xx,\o}\psi^-_{\xx,\o}\;.\Eq(2.10)$$
Moreover, we shall denote by $\hat G^{4,1}_{+\o}(\pp; \kk_1, \kk_2,
\kk_3, \kk_4)$ and $\hat G^{4}_+(\kk_1,\kk_2,\kk_3,\kk_4)$ the
corresponding Fou\-rier transforms, deprived of the momentum
conservation delta. Note that, as a consequence of \equ(1.1), if
the $\psi^+$ momenta are interpreted as ``ingoing momenta'' in the
usual graph pictures, then the $\psi^-$ momenta are ``outgoing
momenta''; our definition of Fourier transform is such that even
$\pp$, the momentum associated with the $\rho$ field, is an
ingoing momentum. Hence, the momentum conservation implies that
$\kk_1 +\kk_3 =\kk_2 +\kk_4+\pp$, in the case of $\hat
G^{4,1}_\o(\pp; \kk_1, \kk_2, \kk_3, \kk_4)$ and $\kk_1 +\kk_3 =
\kk_2 +\kk_4$ in the case of $\hat G^{4}_+(\kk_1, \kk_2, \kk_3,
\kk_4)$.

It is possible to derive a Dyson equation which, combined with the
Ward identity (4.9) of ref. [BM3], gives a relation between $G^4$,
$G^2$ and $G^{2,1}$.

If $Z=\int P(d\psi) \exp \{-V(\psi) \}$ and $<\cdot>$ denotes the
expectation with respect to $Z^{-1} \int P(d\psi)$
$\exp\{-V(\psi)\}$,
$$G^4_+(\xx_1,\xx_2,\xx_3,\xx_4) = <\psi^-_{\xx_1,+} \psi^+_{\xx_2,+}
\psi^-_{\xx_3,-}\psi^+_{\xx_4,-}> -
G^2_+(\xx_1,\xx_2) G^2_-(\xx_3,\xx_4)\;,\Eq(2.7)$$
where we used the fact that $<\psi^-_{\xx,\o}\psi^+_{\yy,-\o}>=0$.

Let $g_\o(\xx)$ be the free propagator, whose Fourier transform is
$g_\o(\kk)=\chi_{h,0}(\kk)/(-ik_0+\o k)$. Then, we
can write the above equation as
$$\eqalign{
&G^4_+(\xx_1,\xx_2,\xx_3,\xx_4)=
-\l \int d\zz\; g_{-}(\zz-\xx_4) <\psi^-_{\xx_1,+} \psi^+_{\xx_2,+}
\psi^-_{\xx_3,-} \psi^+_{\zz,-} \psi^+_{\zz,+}\psi^-_{\zz,+}> +\cr
& +\l\; G^2_+(\xx_1,\xx_2) \int d\zz\; g_{-}(\zz-\xx_4)
<\psi^-_{\xx_3,-} \psi^+_{\zz,-} \psi^+_{\zz,+}\psi^-_{\zz,+}>=\cr
&= -\l \int d\zz g_{-1}(\zz-\xx_4) <[\psi^-_{\xx_1,+}
\psi^+_{\xx_2,+}]\,;[\psi^-_{\xx_3,-}
\psi^+_{\zz,-}\psi^+_{\zz,+}\psi^-_{\zz,+}]>^T\;.\cr}\Eq(2.8)$$
From \equ(2.8) we get
$$\eqalignno{
&-G^4_+(\xx_1,\xx_2,\xx_3,\xx_4)= \l\int d\zz
g_{-}(\zz-\xx_4)<\psi^-_{\xx_1,+}; \psi^+_{\xx_2,+}; \r_{\zz,+}>^T
<\psi^-_{\xx_3,-} \psi^+_{\zz,-}>+\cr &\quad +\l\int d\zz
g_{-}(\zz-\xx_4) <\r_{\zz,+}; \psi^-_{\xx_1,+}; \psi^+_{\xx_2,+};
\psi^-_{\xx_3,-}; \psi^+_{\zz,-}>^T+ &\eq(2.9)\cr &\quad +\l\int
d\zz g_{-}(\zz-\xx_4)<\psi^-_{\xx_1,+}; \psi^+_{\xx_2,+};
\psi^-_{\xx_3,-}; \psi^+_{\zz,-}>^T<\r_{\zz,+}>\;.\cr}$$
The last addend is vanishing, since $<\r_{\zz,\o}>=0$ by the
propagator parity properties. In terms of Fourier transforms, we
get the {\it Dyson equation}
$$\eqalign{
-\hat G^{4}_+(\kk_1,\kk_2,\kk_3,\kk_4) &= \l \hat g_-(\kk_4) \Big[
\hat G^2_-(\kk_3) \hat G^{2,1}_+(\kk_1-\kk_2,\kk_1,\kk_2)+\cr
&+ {1\over L\b} \sum_\pp
G^{4,1}_+(\pp;\kk_1,\kk_2,\kk_3,\kk_4-\pp)\Big]\;,}\Eq(2.11aa)$$
see Fig. 3.

Let us now suppose that $|\kk_4| \le \g^h$; then the support
properties of the propagators imply that $|\pp|\le \g+\g^h\le
2\g$, hence we can freely multiply $G^{4,1}_+$ in the r.h.s. of
\equ(2.11aa) by the compact support function $\c_0(\g^{-j_m}|\pp|)$,
with $j_m= [1+\log_\g 2] + 1$, $\c_0$ being defined as in
\equ(1.4). It follows that \equ(2.11aa) can be written as
$$\eqalignno{
&-\hat G^{4}_+(\kk_1,\kk_2,\kk_3,\kk_4) = \l \hat g_-(\kk_4) \Big[
\hat G^2_-(\kk_3) \hat
G^{2,1}_+(\kk_1-\kk_2,\kk_1,\kk_2)+&\eq(2.11)\cr
& +{1\over L\b} \sum_\pp \chi_M(\pp)
G^{4,1}_+(\pp;\kk_1,\kk_2,\kk_3,\kk_4-\pp)+ {1\over L\b} \sum_\pp
\tilde\chi_M(\pp) G^{4,1}_+(\pp;\kk_1,\kk_2,\kk_3,\kk_4-\pp)
\Big]\;,}$$
where $\chi_M(\pp)$ is a compact support function vanishing for
$|\pp|\ge \g^{h+j_m-1}$ and
$$\tilde\chi_M(\pp) = \sum_{h_p=h+j_m}^{j_m} f_{h_p}(\pp)\;.\Eq(2.11b)$$
Note that the decomposition of the $\pp$ sum is done so that
$\tilde\chi_M(\pp)=0$ if $|\pp|\le 2\g^h$.

\*

{\it Remark.} The l.h.s. of the identity \equ(2.11) is, by
\equ(1.47), of order $\l_h \g^{-4 h}Z_h^{-2}$; if we can prove that the
l.h.s. is proportional to $\l$ with essentially the same
proportionality constants, we get that $\l_h\simeq \l$. This
cannot be achieved if we simply use \equ(1.45), \equ(1.46) and the
analogous bound for $\hat G^{4,1}$, given in Lemma A1.2 of [BM3];
for instance, by using \equ(1.45) and \equ(1.46), we see that the
first addend in the r.h.s. of \equ(2.11) is of size $\g^{- 2h}
Z_h^{(2)} Z_h^{-2} \l [1+O(\bar\l_h^2)]$. We have to take into
account some crucial cancellations in the perturbative expansion,
and this will be done by expressing $\hat G^{2,1}$ and $\hat
G^{4,1}$ in terms of other functional integrals by suitable Ward
identities which at the end will allow us to prove \equ(2.4).

\*

\sub(2.3j) {\it Ward identities.}

By doing in \equ(1.7) the chiral Gauge transformation
$\psi_{\xx,+}^\s\to e^{i\s\a_{\xx,+}} \psi_{\xx,+}^\s$,
$\psi_{\xx,-}^\s\to \psi_{\xx,-}^\s$, one obtains, see [BM2], the
Ward identity, see Fig. \graf(4)
$$D_+(\pp)G^{2,1}_{+}(\pp,\kk,\qq) = G^{2}_{+}(\qq)-
G^{2}_{+}(\kk)+\D^{2,1}_{+}(\pp,\kk,\qq)\;,\Eq(2.12)$$
with
$$\D^{2,1}_{+}(\pp,\kk,\qq)=
{1\over\b L}\sum_\kk C_+(\kk,\kk-\pp)
<\hat\psi^+_{\kk,+}\hat\psi^-_{\kk-\pp,+};
\hat\psi^-_{\kk,+}\hat\psi^+_{\qq,+}>^T\Eq(2.13)$$
and
$$C_\o(\kk^+,\kk^-)=[C_{h,0}(\kk^-)-1]D_\o(\kk^-)
-[C_{h,0}(\kk^+)-1]D_\o(\kk^+)\;.\Eq(2.14)$$


\*

In the same way, we get two other Ward identities
$$\eqalign{
&D_+(\pp) G^{4,1}_{+}(\pp,\kk_1,\kk_2,\kk_3,\kk_4-\pp)
=G^{4}_{+}(\kk_1-\pp,\kk_2,\kk_3,\kk_4-\pp)-\cr &\qquad
-G^{4}_{+}(\kk_1,\kk_2+\pp,\kk_3,\kk_4-\pp)+\D^{4,1}_{+}\;,\cr}
\Eq(2.14a)$$

$$\eqalign{
&D_-(\pp) G^{4,1}_{-}(\pp,\kk_1,\kk_2,\kk_3,\kk_4-\pp)
=G^{4}_{+}(\kk_1,\kk_2,\kk_3-\pp,\kk_4-\pp)-\cr &\qquad -
G^{4}_{+}(\kk_1,\kk_2,\kk_3,\kk_4)+\D^{4,1}_{-}\;,\cr}
\Eq(2.14b)$$
where $\D^{4,1}_\pm $ is the ``correction term''
$$\D^{4,1}_{\pm }(\pp,\kk_1,\kk_2,\kk_3)={1\over\b L}\sum_\kk
C_\pm(\kk,\kk-\pp) <\hat\psi^+_{\kk,\pm}\hat\psi^-_{\kk-\pp,\pm};
\hat\psi^-_{\kk_1,+};\hat\psi^+_{\kk_2,+}; \hat\psi^-_{\kk_3,-};
\hat\psi^+_{\kk_4-\pp,-}>^T\;.\Eq(2.15)$$

\*

\* \sub(3.1b){\it Counterterms}

Eq. \equ(2.14a) can be written, by adding and subtracting suitable
counterterms $\n_\pm$, to be fixed properly later, see Fig. \graf(6)
$$\eqalign{
&(1-\n_+) D_{+}(\pp) \hat G_{+}^{4,1}
(\pp,\kk_1,\kk_2,\kk_3,\kk_4-\pp)-\n_{-} D_{-}(\pp) \hat
G_{-}^{4,1}(\pp,\kk_1,\kk_2,\kk_3,\kk_4-\pp)\cr &= \hat
G_{+}^{4}(\kk_1-\pp,\kk_2,\kk_3,\kk_4-\pp)- \hat
G_{+}^{4}(\kk_1,\kk_2+\pp,\kk_3,\kk_4-\pp) +
H^{4,1}_{+}(\pp,\kk_1,\kk_2,\kk_3,\kk_4-\pp)\;,\cr}\Eq(2.16)$$
where by definition
$$H^{4,1}_{+}(\pp,\kk_1,\kk_2,\kk_3,\kk_4-\pp)=
{1\over\b L}\sum_\kk C_+(\kk,\kk-\pp)
<\hat\psi^+_{\kk,+}\hat\psi^-_{\kk-\pp,+};
\hat\psi^-_{\kk_1,+};\hat\psi^+_{\kk_2,+}; \hat\psi^-_{\kk_3,-};
\hat\psi^+_{\kk_4-\pp,-}>^T -$$
$$- {1\over\b L}\sum_\kk\sum_\o\nu_{\o} D_\o(\pp)
<\hat\psi^+_{\kk,\o}\hat\psi^-_{\kk-\pp,\o};
\hat\psi^-_{\kk_1,+};\hat\psi^+_{\kk_2,+}; \hat\psi^-_{\kk_3,-};
\hat\psi^+_{\kk_4-\pp,-}>^T\;.\Eq(2.16a)$$

In the same way, eq. \equ(2.14b) can be written as
$$\eqalign{
& (1-\n'_-)D_{-}(\pp) \hat G_{-}^{4,1}
(\pp,\kk_1,\kk_2,\kk_3,\kk_4-\pp)-\n'_{+} D_{+}(\pp) \hat
G_{+}^{4,1}(\pp,\kk_1,\kk_2,\kk_3,\kk_4-\pp) =\cr
&= \hat
G_{+}^{4}(\kk_1,\kk_2\kk_3-\pp,\kk_4-\pp)- \hat
G_{+}^{4}(\kk_1,\kk_2,\kk_3,\kk_4) + \hat
H^{4,1}_{-}(\pp,\kk_1,\kk_2,\kk_3,\kk_4-\pp)\;,\cr}\Eq(2.17)$$
where
$$H^{4,1}_{-}(\pp,\kk_1,\kk_2,\kk_3,\kk_4-\pp)={1\over\b L}\sum_\kk
C_-(\kk,\kk-\pp) <\hat\psi^+_{\kk,-}\hat\psi^-_{\kk-\pp,-};
\hat\psi^-_{\kk_1,+};\hat\psi^+_{\kk_2,+}; \hat\psi^-_{\kk_3,-};
\hat\psi^+_{\kk_4-\pp,-}>^T-$$
$$-{1\over\b L}\sum_\kk\sum_\o\nu'_{\o} D_\o(\pp)
<\hat\psi^+_{\kk,\o}\hat\psi^-_{\kk-\pp,\o};
\hat\psi^-_{\kk_1,+};\hat\psi^+_{\kk_2,+}; \hat\psi^-_{\kk_3,-};
\hat\psi^+_{\kk_4-\pp,-}>^T\;.\Eq(2.17a)$$
If we insert in the r.h.s. of \equ(2.16) the value of $\hat
G_-^{4,1}$ taken from \equ(2.17), we get
$$\eqalignno{
&(1+A) D_{+}(\pp) \hat G_{+}^{4,1}
(\pp,\kk_1,\kk_2,\kk_3,\kk_4-\pp)= [\hat
G_{+}^{4}(\kk_1-\pp,\kk_2,\kk_3,\kk_4-\pp)- &\eq(2.18)\cr &\hat
G_{+}^{4}(\kk_1,\kk_2+\pp,\kk_3,\kk_4-\pp)] + B [\hat
G_{+}^{4}(\kk_1,\kk_3-\pp,\kk_4-\pp)- \hat
G_{+}^{4}(\kk_1,\kk_2,\kk_3,\kk_4)] +\hat H^{4,1}_{+}+ B \hat
H^{4,1}_{-}\;,\cr}$$
where
$$A=-\n_+-{\n_{-}\n'_{+}\over 1-\n'_{-}} \virg B={\n_{-}\over
1-\n'_-}\;.\Eq(2.19)$$

If we insert in the last term of the r.h.s. of \equ(2.11) the
value of $\hat G_{+}^{4,1}$ taken from \equ(2.18), we get
$$\eqalignno{
&-\hat G^{4}_+(\kk_1,\kk_2,\kk_3,\kk_4) = &\eq(2.20)\cr
& = \l\hat g_-(\kk_4) \left[\hat G^2_- (\kk_3) \hat
G^{2,1}_+(\kk_1-\kk_2,\kk_1,\kk_2)+{1\over L\b}\sum_\pp\chi_M(\pp)
G^{4,1}_+(\pp;\kk_1,\kk_2,\kk_3,\kk_4-\pp)\right] +\cr
&+ {\l \hat g_-(\kk_4)\over (1+A)} {1\over L\b} \sum_\pp
\tilde\chi_M(\pp) {\hat G_+^{4}(\kk_1-\pp,\kk_2,\kk_3,\kk_4-\pp)-
\hat G_+^{4}(\kk_1,\kk_2+\pp,\kk_3,\kk_4-\pp)\over D_+(\pp)}\;+\cr
&+{\l \hat g_-(\kk_4)\over (1+A)} {1\over L\b} \sum_\pp
\tilde\chi_M(\pp) {\hat G_+^{4}(\kk_1,\kk_2,\kk_3-\pp,\kk_4-\pp)-
\hat G_+^{4}(\kk_1,\kk_2,\kk_3,\kk_4)\over D_+(\pp)}\;+ \cr
&+{\l \hat g_-(\kk_4)\over (1+A)} {1\over L\b} \sum_\pp
\tilde\chi_M(\pp) {\hat
H^{4,1}_{+}(\pp;\kk_1,\kk_2,\kk_3,\kk_4-\pp)+ B \hat
H^{4,1}_{-}(\pp;\kk_1,\kk_2,\kk_3,\kk_4-\pp)\over D_+(\pp)}
\;.\cr}$$
Note that
$${1\over L\b} \sum_\pp \tilde\chi_M(\pp) {\hat
G_+^{4}(\kk_1,\kk_2,\kk_3,\kk_4)\over D_+(\pp)}=0\;,\Eq(2.21)$$
since $D_+(\pp)$ is odd. Then, by using also the Ward identity
\equ(2.12), we get, if
$$\kk_i=\bar\kk_i \virg \bar\kk_1= \bar\kk_4=
-\bar\kk_2= -\bar\kk_3= \bar\kk \virg
|\bar\kk|=\g^h\;,\Eq(2.21a)$$
the identity
$$\eqalignno{
&-\hat G^{4}_+(\bar\kk_1,\bar\kk_2,\bar\kk_3,\bar\kk_4) = \l\hat
g_-(\bar\kk_4)\hat G^2_-(\bar\kk_3) {G^{2}_{+}(\bar\kk_2)-
G^{2}_{+}(\bar\kk_1) \over D_+(2\bar\kk)} + &\eq(2.22)\cr
&+ \l\hat g_-(\bar\kk_4)\hat G^2_-(\bar\kk_3)
{\D^{2,1}_{+}(2\bar\kk,\bar\kk_1,\bar\kk_2)\over D_+(2\bar\kk)} +
\l\hat g_-(\bar\kk_4){1\over L\b}\sum_\pp\chi_M(\pp)
G^{4,1}_+(\pp;\bar\kk_1,\bar\kk_2,\bar\kk_3,\bar\kk_4-\pp)+ \cr
&+ {\l \hat g_-(\bar\kk_4)\over (1+A)} {1\over L\b} \sum_\pp
\tilde\chi_M(\pp) {\hat
G_+^{4}(\bar\kk_1-\pp,\bar\kk_2,\bar\kk_3,\bar\kk_4-\pp)- \hat
G_+^{4}(\bar\kk_1,\bar\kk_2+\pp,\bar\kk_3,\bar\kk_4-\pp)\over
D_+(\pp)}\;+\cr
&+{\l \hat g_-(\bar\kk_4)\over (1+A)} {1\over L\b} \sum_\pp
\tilde\chi_M(\pp) {\hat
G_+^{4}(\bar\kk_1,\bar\kk_2,\bar\kk_3-\pp,\bar\kk_4-\pp)\over
D_+(\pp)}\;+ \cr
&+{\l \hat g_-(\bar\kk_4)\over (1+A)} {1\over L\b} \sum_\pp
\tilde\chi_M(\pp) {\hat
H^{4,1}_{+}(\pp;\bar\kk_1,\bar\kk_2,\bar\kk_3,\bar\kk_4-\pp)+ B
\hat
H^{4,1}_{-}(\pp;\bar\kk_1,\bar\kk_2,\bar\kk_3,\bar\kk_4-\pp)\over
D_+(\pp)} \;.\cr}$$

All the terms appearing in the above equation can be expressed in
terms of convergent tree expansions. The term in the l.h.s. of
\equ(2.22) is given, by \equ(1.47), by
$$\hat G^{4}_+(\bar\kk_1,\bar\kk_2,\bar\kk_3,\bar\kk_4)=\g^{-4 h} Z_h^{-2}
[-\l_h +O(\bar\l_h^2)]\;.\Eq(2.23a)$$
The two terms in the r.h.s., first line, are equal, by \equ(1.46),
to $\g^{-4 h} Z_h^{-2} (\l+O(\bar\l_h^2))$. The first term in the
second line, by \equ(1.46) and eq. (177) of [BM2], can be bounded
as
$$\left| \l\hat g_-(\bar\kk_4)\hat G^2_-(\bar\kk_3)
{\D^{2,1}_{+}(2\bar\kk,\bar\kk_1,\bar\kk_2)\over D_+(2\bar\kk)}
\right| \le C\bar \l_h^2 {\g^{-4 h}\over Z_h^2}\;,\Eq(2.23)$$
while the second term in the second line, by eq. (A1.11) of ref.
[BM3], can be bounded as
$$\left| \l\hat g_-(\bar\kk_4){1\over L\b}\sum_\pp\chi_M(\pp)
G^{4,1}_+(\pp;\bar\kk_1,\bar\kk_2,\bar\kk_3,\bar\kk_4-\pp) \right|
\le C\bar \l_h^3 {\g^{-4 h}\over Z_h^2} \Eq(2.25)$$
Moreover, by using Lemma A1.1 of [BM3], one sees that
$$\eqalign{
&\left| {1\over L\b} \sum_\pp \tilde\chi_M(\pp) {\hat
G_+^{4}(\bar\kk_1-\pp,\bar\kk_2,\bar\kk_3,\bar\kk_4-\pp)- \hat
G_+^{4}(\bar\kk_1,\bar\kk_2+\pp,\bar\kk_3,\bar\kk_4-\pp)\over
D_+(\pp)} \right.\;+\cr
&+ \left. {1\over L\b} \sum_\pp \tilde\chi_M(\pp) {\hat
G_+^{4}(\bar\kk_1,\bar\kk_2,\bar\kk_3-\pp,\bar\kk_4-\pp)\over
D_+(\pp)} \right| \le C \bar\l_{h}  {\g^{-3 h}\over
Z_h^2}\;.}\Eq(2.24)$$
In the following sections we will prove the following Lemma.

\*

\lemma(1) {\it There exists $\e_1\le \e_0$ and four $\l$-functions
$\n_+,\n_-,\n'_+,\n'_-$ of order $\l$ (uniformly in $h$), such
that, if $\bar\l_{h}\le \e_1$,
$$\left| \l\hat  g_-(\bar\kk_4) {1\over L\b} \sum_\pp
\tilde\chi_M(\pp) {\hat
H^{4,1}_{\pm}(\pp;\bar\kk_1,\bar\kk_2,\bar\kk_3,\bar\kk_4-\pp) \over
D_+(\pp)} \right| \le C\bar\l_{h}^2{\g^{-4 h}\over Z_h^2}
\Eq(2.26)$$
}

\*

The above Lemma, together with the identity \equ(2.22), following
from the Dyson equation and the Ward identities, and the previous
bounds, proved in refs. [BM2] and [BM3] and following from the
tree expansion, implies \equ(2.4); this concludes the proof of
Theorem \thm(th3).

\*\*
\section(3, Proof of Lemma {\lm(1)})
\*

\sub(3.1) {\it The corrections.}

We shall prove first the bound \equ(2.26) for $H^{4,1}_+$; the
bound for $H_-^{4,1}$ is done essentially in the same way and will
be briefly discussed later. By using \equ(2.16a), we get
$$\hat g_-(\kk_4) {1\over L\b}\sum_{\pp} \tilde\chi_M(\pp)
D_+^{-1}(\pp) \hat H^{4,1}_{+}(\pp;\kk_1,\kk_2,\kk_3,\kk_4-\pp) =
$$
$$=\hat g_-(\kk_4) {1\over L\b}\sum_{\pp} \tilde\chi_M(\pp)
{1\over L\b}\sum_{\kk} {C_+(\kk,\kk-\pp)\over D_+(\pp)}
<\hat\psi^+_{\kk,+}\hat\psi^-_{\kk-\pp,+};
\hat\psi^-_{\kk_1,+};\hat\psi^+_{\kk_2,+}; \hat\psi^-_{\kk_3,-};
\hat\psi^+_{\kk_4-\pp,-}>^T+$$
$$-\nu_-\hat g_-(\kk_4) {1\over L\b}\sum_{\pp} \tilde\chi_M(\pp)
{1\over L\b}\sum_{\kk} {D_-(\pp)\over D_+(\pp)}
<\hat\psi^+_{\kk,-}\hat\psi^-_{\kk-\pp,-};
\hat\psi^-_{\kk_1,+};\hat\psi^+_{\kk_2,+}; \hat\psi^-_{\kk_3,+};
\hat\psi^+_{\kk_4-\pp,-}>^T-$$
$$-\nu_+\hat g_-(\kk_4) {1\over L\b} \sum_{\pp} \tilde\chi_M(\pp)
 {1\over L\b}\sum_{\kk} <\hat\psi^+_{\kk,+}\hat\psi^-_{\kk-\pp,+};
\hat\psi^-_{\kk_1,+};\hat\psi^+_{\kk_2,+}; \hat\psi^-_{\kk_3,-};
\hat\psi^+_{\kk_4-\pp,-}>^T\;. \Eq(3.1)$$
Let us define
$$\tilde G^4_+(\kk_1, \kk_2,\kk_3, \kk_4) = \left.
{\partial^4\over \partial \phi^+_{\kk_1,+} \partial
\phi^-_{\kk_2,+} \partial \phi^+_{\kk_3,-}
\partial J_{\kk_4}} \tilde W \right|_{\phi=0}\;,\Eq(3.2)$$
where
$$\tilde W = \log \int P(d\hat\psi)e^{-T_1(\psi) + \n_+T_+(\psi)
+ \n_-T_-(\psi)} e^{-V(\hat\psi)+ \sum_\o\int d\xx
[\phi^+_{\xx,\o}\hat\psi^{-}_{\xx,\o}+
\hat\psi^{+}_{\xx,\o}\phi^-_{\xx,\o}]} \;,\Eq(3.3)$$
$$T_1(\psi) = {1\over L\b} \sum_{\pp} \tilde\chi_M(\pp)
{1\over L\b} \sum_{\kk} {C_+(\kk, \kk-\pp) \over D_+(\pp)}
(\hat\psi_{\kk,+}^+ \hat\psi_{\kk-\pp,+}^-)
\hat\psi^+_{\kk_4-\pp,-} \hat J_{\kk_4} \hat
g_-(\kk_4)\;,\Eq(3.4)$$
$$T_+(\psi)= {1\over L\b} \sum_{\pp} \tilde\chi_M(\pp)
{1\over L\b} \sum_{\kk} (\hat\psi_{\kk,+}^+
\hat\psi_{\kk-\pp,+}^-) \hat\psi^+_{\kk_4-\pp,-} \hat J_{\kk_4}
\hat g_-(\kk_4)\;,\Eq(3.5)$$
$$T_-(\psi) = {1\over L\b} \sum_{\pp} \tilde\chi_M(\pp)
{1\over L\b} \sum_{\kk} {D_-(\pp) \over D_+(\pp)}
(\hat\psi_{\kk,-}^+ \hat\psi_{\kk-\pp,-}^-)
\hat\psi^+_{\kk_4-\pp,-} \hat J_{\kk_4} \hat
g_-(\kk_4)\;.\Eq(3.6)$$

\insertplot{300pt}{100pt}%
{\ins{45pt}{0pt}{$T_1$}
\ins{145pt}{0pt}{$T_+$}
\ins{245pt}{0pt}{$T_-$}
}%
{verticiT}{}

\* \vbox{{\ }

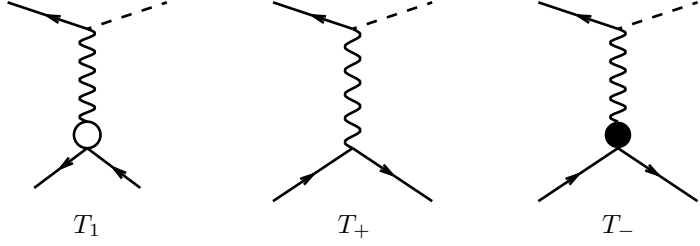
\centerline{\eqg(1v): Graphical representation of $T_1,T_+,T_-$; the dotted
line carries}
\centerline{momentum $\bar\kk_4$, the empty circle represents
$C_+$, the filled one $D_-(\pp)/D_+(\pp)$}}

\* It is easy to see that $\tilde G^4_+$ is related to \equ(3.1)
by an identity similar to \equ(2.8). In fact we can write
$$\eqalignno{& -\tilde G_+^4(\kk_1,\kk_2,\kk_3,\kk_4)= &\eq(3.7)\cr
&= \hat g_-(\kk_4) {1\over L\b} \sum_{\pp} \tilde\chi_M(\pp)
{1\over L\b}\sum_{\kk} {C_+(\kk,\kk-\pp) \over D_+(\pp)}
<[\hat\psi^-_{\kk_1,+}\hat\psi^+_{\kk_2,+}];
[\hat\psi^+_{\kk,+}\hat\psi^-_{\kk-\pp,+}\hat\psi^-_{\kk_3,-}
\hat\psi^+_{\kk_4-\pp,-}]>^T+\cr
&-\nu_-\hat g_-(\kk_4) {1\over L\b} \sum_{\pp} \tilde\chi_M(\pp)
{1\over L\b} \sum_{\kk} {D_-(\pp)\over D_+(\pp)}
<[\hat\psi^-_{\kk_1,+}\hat\psi^+_{\kk_2,+}];
[\hat\psi^+_{\kk,-}\hat\psi^-_{\kk-\pp,-} \hat\psi^-_{\kk_3,+}
\hat\psi^+_{\kk_4-\pp,-}]>^T-\cr
&-\nu_+ \hat g_-(\kk_4) {1\over L\b} \sum_{\pp} \tilde\chi_M(\pp)
{1\over L\b} \sum_{\kk}
<[\hat\psi^-_{\kk_1,+}\hat\psi^+_{\kk_2,+}];
[\hat\psi^+_{\kk,+}\hat\psi^-_{\kk-\pp,+} \hat\psi^-_{\kk_3,-}
\hat\psi^+_{\kk_4-\pp,-}]>^T\;;\cr}$$
moreover, if we introduce the definition
$$\d\r_{\pp,+}={1\over\b L}\sum_\kk{ C_+(\pp,\kk)\over D_+(\pp)}
(\hat\psi_{\kk,+}^+\hat\psi_{\kk-\pp,+}^-)\;,\Eq(3.8)$$
the term in the second line of \equ(3.7) can be rewritten as
$$\eqalignno{
&\hat g_-(\kk_4) {1\over L\b} \sum_{\pp} \tilde\chi_M(\pp)
<[\hat\psi^-_{\kk_1,+} \hat\psi^+_{\kk_2,+}]; [\d\r_{\pp,+}
\hat\psi^-_{\kk_3,-} \hat\psi^+_{\kk_4-\pp,-}]>^T =\cr
&= g_-(\kk_4) {1\over L\b} \sum_{\pp} \tilde\chi_M(\pp) \left[
<\hat\psi^-_{\kk_1,+}; \hat\psi^+_{\kk_2,+}; \d \r_{\pp,+}>^T \;
<\hat\psi^-_{\kk_3,-} \hat\psi^+_{\kk_4-\pp,-}>
+\right.&\eq(3.7a)\cr
&+ \left. <\d \r_{\pp,+}; \hat\psi^-_{\kk_1,+};
\hat\psi^+_{\kk_2,+}; \hat\psi^-_{\kk_3,-};
\hat\psi^+_{\kk_4-\pp,-}>^T \right]\;,\cr}$$
where we used the fact that $\pp\not=0$ in the support of
$\tilde\chi_M(\pp)$ and $<\d \r_{\pp,+}> =0$ for $\pp\not=0$. A
similar decomposition can be done for the other two terms in the
r.h.s. of \equ(3.7); hence, by using \equ(2.16a), we get
$$\eqalignno{
&-\tilde G^4_+(\kk_1,\kk_2,\kk_3,\kk_4)= g_-(\kk_4){1\over L\b}
\sum_{\pp} \tilde\chi_M(\pp)
{H_+^{4,1}(\pp;\kk_1,\kk_2,\kk_3,\kk_4-\pp)\over D_+(\pp)} +\cr
&+ \tilde\chi_M(\kk_1-\kk_2) g_-(\kk_4) G^2_-(\kk_3) \left[
<\hat\psi^-_{\kk_1,+}; \hat\psi^+_{\kk_2,+}; \d
\r_{\kk_1-\kk_2,+}>^T -\right. &\eq(3.7b)\cr
&\left. -\n_+ <\hat\psi^-_{\kk_1,+}; \hat\psi^+_{\kk_2,+};
\r_{\kk_1-\kk_2,+}>^T -\n_- {D_-(\kk_1-\kk_2)\over
D_+(\kk_1-\kk_2)} <\hat\psi^-_{\kk_1,+}; \hat\psi^+_{\kk_2,+};
\r_{\kk_1-\kk_2,-}>^T \right]\cr}$$

We now put $\kk_i=\bar\kk_i$, see \equ(2.21a). Since
$|\bar\kk_1-\bar\kk_2|=2\g^h$, \equ(2.11b) implies that
$\tilde\chi_M(\bar\kk_1- \bar\kk_2)=0$; hence we get
$$-\tilde G^4_+(\kk_1,\kk_2,\kk_3,\kk_4)= g_-(\kk_4){1\over L\b}
\sum_{\pp} \tilde\chi_M(\pp)
{H_+^{4,1}(\pp;\kk_1,\kk_2,\kk_3,\kk_4-\pp)\over
D_+(\pp)}\;.\Eq(3.7c)$$

\* {\it Remark.} \equ(3.7c) says that the last line of the Dyson
equation \equ(2.22) can be written as a functional integral very
similar to the one for $G^4_+$ (we are essentially proceding as in
the derivation of the Dyson equation, in the opposite direction)
except that the interaction $V$
\equ(1.3) is replaced by $V+T_1-\n_+ T_+ -\n_- T_-$; we will
evaluate it via a multiscale integration procedure similar to the
one for $G^4_+$, and in the expansion additional running coupling
constants will appear; the expansion is convergent again if such
new running couplings will remain small uniformly in the infrared
cutoff.

\* \sub(3.2) {\it The properties of $D_\o(\pp)^{-1}
C_\o(\kk,\kk-\pp) $.}

We shall use some properties of the operator $D_\o(\pp)^{-1}
C_\o(\kk,\kk-\pp)$, which were proved in [BM2]. Let us consider
first the effect of contracting both $\hat\psi$ fields of $\d
\r_{\pp,+}$ on the same or two different scales; in the second
case, we also suppose that the regularization procedure (to be
defined later, in agreement with this hypothesis) does not act on
the propagator of higher scale. Hence, we have to study the
quantity
$$\D^{(i,j)}_\o(\kk^+,\kk^-)={C_\o(\kk^+,\kk^-)\over D_\o(\pp)}
\tilde g^{(i)}_\o(\kk^+) \tilde g^{(j)}_\o(\kk^-)\;,\Eq(3.9)$$
where $\pp=\kk^+-\kk^-$. The crucial observation is that
$$\D^{(i,j)}_\o(\kk^+,\kk^-)=0 \virg \hbox{if\ } h<i,j<0\;,\Eq(3.10)$$
since $\c_{h,0}(\kk^\pm)=1$, if $h<i,j<0$. Let us then consider
the cases in which $\D^{(i,j)}_\o(\kk^+,\kk^-)$ is not identically
equal to $0$. Since
$\D^{(i,j)}_\o(\kk^+,\kk^-)=\D^{(j,i)}_\o(\kk^-,\kk^+)$, we can
restrict the analysis to the case $i\ge j$. We define
$$u_0(\kk)=\cases{0 & if $|\kk|\le 1$ \cr 1-f_0(\kk) & if $1\le
|\kk|$ \cr} \virg u_h(\kk)=\cases{0 & if $|\kk|\ge \g^h$ \cr
1-f_h(\kk) & if $|\kk|\le \g^h$ \cr}\;.\Eq(3.10a)$$

Then we get, by using \equ(1.35), the fact that $Z_{-1}=Z_0=1$ and
$\tilde f_j=f_j$ for $j=0,h$,
$$\D^{(0,0)}_\o(\kk^+,\kk^-)={1\over D_\o(\pp)} \left[ {f_0(\kk^+)
\over D_\o(\kk^+)} u_0(\kk^-)- {f_0(\kk^-)\over D_\o(\kk^-)}
u_0(\kk^+) \right]\;,\Eq(3.10b)$$
$$\D^{(h,h)}_\o(\kk^+,\kk^-) = {1\over D_\o(\pp)}{1\over \tilde
Z_{h- 1}(\kk^+) \tilde Z_{h-1}(\kk^-)}
\left[{f_h(\kk^+)u_h(\kk^-)\over D_\o(\kk^+)}-
{u_h(\kk^+)f_h(\kk^-)\over D_\o(\kk^-)}\right]\;,\Eq(3.18)$$
$$\D^{(0,h)}_\o(\kk^+,\kk^-)= {1\over D_\o(\pp)}
{1\over \tilde Z_{h-1}(\kk^-)} \left[ {f_0(\kk^+) u_h(\kk^-)\over
D_\o(\kk^+)} - {f_h(\kk^-) u_0(\kk^+)\over D_\o(\kk^-)}
\right]\;,\Eq(3.10d)$$
$$\D^{(0,j)}_\o(\kk^+,\kk^-)=
-{1\over Z_{j-1}}{\tilde f_j(\kk^-) u_0(\kk^+)\over D_\o(\pp)
D_\o(\kk^-)} \virg h < j< 0 \;,\Eq(3.11)$$
$$\D^{(i,h)}_\o(\kk^+,\kk^-)=
{1\over Z_{h-1}(\kk^-) Z_{i-1}} {\tilde f_i(\kk^+) u_h(\kk^-)\over
D_\o(\pp) D_\o(\kk^+)} \virg j=h < i \le -1 \;.\Eq(3.12)$$

As an easy consequence of the above equations, as shown in \S4.2
of [BM2], one can write, for $0\ge j > h$,
$$\D^{(0,j)}_\o(\kk^+,\kk^-)=
{\pp\over D_\o(\pp)} {\bf S}_\o^{(j)}(\kk^+,\kk^-) \;,\Eq(3.13)$$
where $S^{(j)}_{\o,i}(\kk^+,\kk^-)$ are smooth functions such that
$$|\dpr_{\kk^+}^{m_+} \dpr_{\kk^-}^{m_j} S^{(j)}_{\o,i}(\kk^+,\kk^-)|\le
C_{m_0 +m_j} {\g^{-j(1+m_j)}\over Z_{j-1}}\;.\Eq(3.14)$$

Finally, it is easy to see that, if $0>i\ge h$,
$$|\D^{(i,h)}_\o(\kk^+,\kk^-)|\le C \g^{-(i-h)} {\g^{-h-i}\over
Z_{i-1}} \;.\Eq(3.17)$$
Note that, in the r.h.s. of \equ(3.17), there is apparently a
$Z_{h-1}^{-1}$ factor missing, but the bound can not be improved;
this is a consequence of the fact that $\tilde Z_{h-1}(\kk)=0$ for
$|\kk|\le \g^{h-1}$, see eq. (63) of [BM2], and the support
properties of $u_h(\kk)$. In any case, this is not a problem,
since the dimensional dependence of $\D^{(i,h)}$ on the field
renormalization constants is exactly $Z^{-1}$. Note also the
presence in the bound of the extra factor $\g^{-(i-h)}$, with
respect to the dimensional bound; it will allow us to avoid
renormalization of the marginal terms containing $\D^{(i,h)}$.

\*

\sub(3.3) {\it The multiscale expansion of $\tilde
G^4_+(\bar\kk_1, \bar\kk_2, \bar\kk_3, \bar\kk_4)$: the first
integration step.}

The calculation of $\tilde G_+^4$ is done via a multiscale
expansion; we shall concentrate on the differences with respect to
that described in \sec(1), due to the presence in the potential of
the terms $T_1(\psi)$ and $T_\pm(\psi)$. Moreover we shall suppose
that the momenta $\kk_i$ are put equal to $\bar\kk_i$, defined as
in \equ(2.21a).

Let us consider the first step of the iterative integration
procedure, the integration of the field $\psi^{(0)}$; we shall
describe only the terms linear in the external $J$ field, the only
ones contributing to $\tilde G_+^4$ which were not already
discussed. We call $\bar\VV^{(-1)}(\psi^{[h,-1]})$ the
contribution to the effective potential of such terms and we write
$$\bar\VV^{(-1)}(\psi^{[h,-1]}) = \bar\VV^{(-1)}_{a,1}(\psi^{[h,-1]})
+ \bar\VV^{(-1)}_{a,2}(\psi^{[h,-1]}) +
\bar\VV^{(-1)}_{b,1}(\psi^{[h,-1]}) +
\bar\VV^{(-1)}_{b,2}(\psi^{[h,-1]}) \;,\Eq(3.19)$$
where $\bar\VV^{(-1)}_{a,1} + \bar\VV^{(-1)}_{a,2}$ is the sum of
the terms in which the field $\hat\psi^+_{\bar\kk_4-\pp,-}$
appearing in the definition of $T_1(\psi)$ or $T_\pm(\psi)$ is
contracted, $\bar\VV^{(-1)}_{a,1}$ and $\bar\VV^{(-1)}_{a,2}$
denoting the sum over the terms of this type containing a $T_1$ or
a $T_\pm$ vertex, respectively; $\bar\VV^{(-1)}_{b,1} +
\bar\VV^{(-1)}_{b,2}$ is the sum of the other terms, that is those
where the field $\hat\psi^+_{\bar\kk_4-\pp,-}$ is an external
field, the index $i=1,2$ having the same meaning as before.

Note that the condition \equ(2.21a) on the external momenta
$\kk_i$ forbids the presence of vertices of type $\f$, if $h<0$,
as we shall suppose. Hence, all graphs contributing to
$\bar\VV^{(-1)}$ have, besides the external field of type $J$, an
odd number of external fields of type $\psi$.

\*

Let us consider first $\bar\VV^{(-1)}_{a,1}$; we shall still
distinguish different group of terms, those where both fields
$\hat\psi_{\kk,+}^+$ and $\hat\psi_{\kk-\pp,+}^-$ are contracted,
those where only one among them is contracted and those where no
one is contracted.

If no one of the fields $\hat\psi_{\kk,+}^+$ and
$\hat\psi_{\kk-\pp,+}^-$ is contracted, we can only have terms
with at least four external lines; for the properties of
$\D^{ij}$, at least one of the fields $\hat\psi_{\kk,+}^+$ and
$\hat\psi_{\kk+\pp,+}^-$ must be contracted at scale $h$. If one
of these terms has four external lines, hence it is marginal, it
has the following form
$$\int d\pp \tilde\chi_M(\pp) \hat\psi^+_{-,\bar\kk_4-\pp}
G^{(0)}_2(\bar\kk_4-\pp) \hat g^{(0)}_-(\bar\kk_4-\pp) \hat
g_-(\bar\kk_4) \hat J_{\bar\kk_4} \int d\kk { C(\kk,\pp) \over
D_+(\pp)} \hat\psi^+_{\kk,+}\hat\psi^-_{\kk+\pp,+}\;, \Eq(3.20)$$
where $G^{(0)}_2(\kk)$ is a suitable function which can be
expressed as a sum of graphs with an odd number of propagators,
hence it vanishes at $\kk=0$. This implies that $G^{(0)}_2(0)=0$,
so that we can regularize it without introducing any running
coupling.

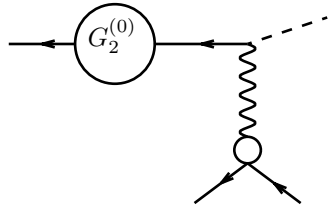
\insertplot{300pt}{100pt}%
{\ins{90pt}{80pt}{$G^{(0)}_2$}
}%
{G20}{}

\centerline{\eqg(2v): Graphical representation of \equ(3.20)}

\*

If both $\hat\psi_{\kk,+}^+$ and $\hat\psi_{\kk-\pp,+}^-$ in
$T_1(\psi)$ are contracted, we get terms of the form
$$
\tilde W_{n+1}^{(-1)} (\bar\kk_4,\kk_1,..,\kk_n) \hat
g_-(\bar\kk_4) \hat J_{\bar\kk_4} \prod_{i=1}^n
\hat\psi^{\e_i}_{\kk_i}\;, \Eq(3.21)$$
where $n$ is an odd integer. We want to define an $\RR$ operation
for such terms. There is apparently a problem, as the $\RR$
operation involves derivatives and in $\tilde W^{(-1)}$ appears
the function $\D^{(0,0)}$ of the form \equ(3.13) and the cutoff
function $\tilde\chi_M(\pp)$, with support on momenta of size
$\g^h$. Hence one can worry about the derivatives of the factor
$\tilde\chi_M(\pp) \pp D_+(\pp)^{-1}$. However, as the line of
momentum $\bar\kk_4-\pp$ is necessarily at scale $0$ (we are
considering terms in which it is contracted), then $|\pp|\ge
\g^{-1} - \g^h\ge \g^{-1}/2$ (for $|h|$ large enough), so that we
can freely multiply by a smooth cutoff function $\bar\chi(\pp)$
restricting $\pp$ to the allowed region; this allows us to pass to
coordinate space and shows that the $\RR$ operation can be defined
in the usual way. We define
$$\LL\tilde W_4^{(-1)} (\bar\kk_4,\kk_1,\kk_2, \kk_3) =
\tilde W_4^{(-1)}(0,..,0)\;,\Eq(3.22)$$
$$\LL\tilde W_2^{(-1)} (\bar\kk_4)=\tilde W_2^{(-1)}(0)+
\bar\kk_4\partial_\kk \tilde W_2^{(-1)}(0)\;. \Eq(3.23)$$
Note that by parity the first term in \equ(3.23) is vanishing;
this means that there are only marginal terms. Note also that the
local term proportional to $\hat J_{\bar\kk_4}
\hat\psi^+_{\bar\kk_4,-}$ is such that the field
$\hat\psi^+_{\bar\kk_4,-}$ can be contracted only at the last
scale $h$; hence it has any influence on the integrations of all
the scales $>h$.

\insertplot{300pt}{100pt}%
{\ins{45pt}{15pt}{$\tilde W_4^{(-1)}$}
\ins{215pt}{15pt}{$\tilde W_2^{(-1)}$}
}%
{W4W2}{}

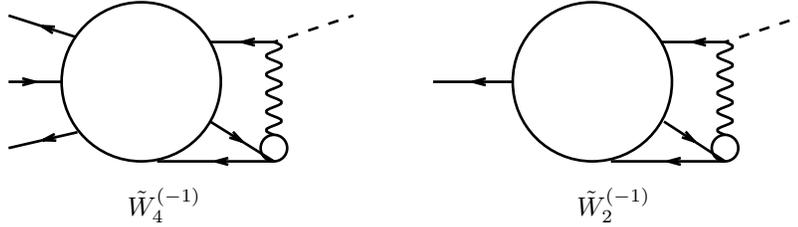
\centerline{\eqg(3v): Graphical representation of $\tilde W_4^{(-1)}$ and
$\tilde W_2^{(-1)}$} \*

\*

If only one among the fields $\hat\psi_{\kk,+}^+$ and
$\hat\psi_{\kk-\pp,+}^-$ in $T_1(\psi)$ is contracted, we note
first that we cannot have terms with two external lines (including
$\hat J_{\kk_4})$; in fact in such a case there is an external
line with momentum $\bar\kk_4$ with $\o=-$ and the other has
$\o=+$; this is however forbidden by global gauge invariance.
Moreover, for the same reasons as before, we do not have to worry
about the derivatives of the factor $\tilde\chi_M(\pp) \pp
D_+(\pp)^{-1}$, related with the regularization procedure of the
terms with four external lines, which have the form
$$\eqalign{
& \int d\kk^+ \hat\psi^+_{\kk_1,+} \hat\psi^-_{\kk^-,+}
\hat\psi^+_{\kk^- +\bar\kk_4- \kk_1,-} \hat g_-(\bar\kk_4) \hat
J_{\bar\kk_4} \tilde\chi_M(\kk^+ -\kk^-) \hat g^{(0)}_-(\bar\kk_4-
\kk^+ +\kk^-)\;\cdot\cr
& \cdot\; G_4^{(0)}(\kk^+, \bar\kk_4, \kk_1) \left\{
{[C_{h,0}(\kk^-)-1] D_{+}(\kk^-) \hat g^{(0)}_+(\kk^+)\over
D_+(\kk^+-\kk^-)} - {u_0(\kk^+) \over D_+(\kk^+-\kk^-)}
\right\}\;, \cr}\Eq(3.24)$$
or the similar one with the roles of $\kk^+$ and $\kk^-$
exchanged.


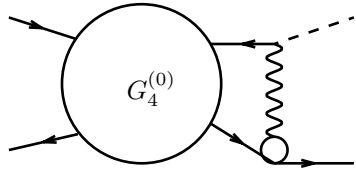
\insertplot{300pt}{100pt}%
{\ins{145pt}{60pt}{$G^{(0)}_4$}
}%
{G40}{}

\centerline{\eqg(4v): Graphical representation of a single addend in
\equ(3.24)}

\* The two terms in \equ(3.24) must be treated differently, as concerns the
regularization procedure. The first term is such that one of the external
lines is associated with the operator $[C_{h,0}(\kk^-)-1] D_+(\kk^-)
D_+(\pp)^{-1}$. We define $\RR=1$ for such terms; in fact, when such external
line is contracted (and this can happen only at scale $h$), the factor
$D_+(\kk^-) D_+(\pp)^{-1}$ produces an extra factor $\g^{h}$ in the bound,
with respect to the dimensional one. This claim simply follows by the
observation that $|D_+(\pp)|\ge 1-\g^{-1}$ as $\pp=\kk^+-\kk^-$ and $\kk^+$
is at scale $0$, while $\kk^-$, as we said, is at scale $h$. This factor has
the effect that all the marginal terms in the tree path connecting $v_0$ with
the end-point to which is associated the $T_1$ vertex acquires negative
dimension.

The second term in \equ(3.24) can be regularized as above, by
subtracting the value of the kernel computed at zero external
momenta, \ie for $\kk^-=\bar\kk_4=\kk_1=0$. Note that such local
part is given by
$$\int d\kk^+ \tilde\chi_M(\kk^+)
\hat g^{(0)}_-(\kk^+) G_2^{(0)}(\kk^+, 0, 0) {u_0(\kk^+) \over
D_+(\kk^+)} \;,\Eq(3.25)$$
and there is no singularity associated with the factor
$D_+(\kk^+)^{-1}$, thanks to the support on scale $0$ of the
propagator $\hat g^{(0)}_-(\kk^+)$.

A similar (but simpler) analysis holds for the terms contributing
to $\bar\VV^{(-1)}_{a,2}$, which contain a vertex of type $T_+$ or
$T_-$ and are of order $\l\n_\pm$. Now, the only thing to analyze
carefully is the possible singularities associated with the
factors $\tilde\chi_M(\pp)$ and  $\pp D_+(\pp)^{-1}$. However,
since in these terms the field $\hat\psi^+_{\bar\kk_4-\pp,-}$ is
contracted, $|\pp|\ge \g^{-1}/2$, for $|h|$ large enough, a
property already used before; hence the regularization procedure
can not produce bad dimensional bounds.

\*

We will define $\tilde z_{-1}$ and $\tilde\l_{-1}$, so that
$$\LL[\bar\VV^{(-1)}_{a,1} +
\bar\VV^{(-1)}_{a,2}](\psi^{[h,-1]}) = \left[ \tilde\l_{-1} \bar
F_\l^{[h,-1]}(\psi^{[h,-1]}) + \tilde z_{-1}
\hat\psi^{[h,-1]+}_{\bar\kk_4,-} D_-(\bar\kk_4) \right] \hat
g_-(\bar\kk_4) \hat J_{\bar\kk_4}\;,\Eq(3.25a)$$
where we used the definition
$$\bar F_\l^{[h,j]}(\psi^{[h,j]}) = {1\over
(L\b)^4}\sum_{\kk_1,\kk_2,\kk_3:C^{-1}_{h,j}(\kk_i)>0}
\hat\psi^{[h,j]+}_{\kk_1,+} \hat\psi^{[h,j]-}_{\kk_2,+}
\hat\psi^{[h,j]+}_{\kk_3,-}
\d(\kk_1-\kk_2+\kk_3-\bar\kk_4)\;.\Eq(1.24a)$$

Note that there is no first order contribution to $\tilde\l_{-1}$,
as follows from a simple calculation, so that $\tilde\l_{-1}$ is
of order $\l^2$ or lower. We expect indeed that it satisfies a non
zero lower bound of order $\l^2$, but this will not play any role
in the following.

\*

Let us consider now the terms contributing to
$\bar\VV^{(-1)}_{b,1}$, that is those where
$\hat\psi^+_{\bar\kk_4-\pp}$ is not contracted and there is a
vertex of type $T_1$.

Besides the term of order $0$ in $\l$ and $\n_\pm$, equal to
$T_1(\psi^{[h,-1]})$, there are the terms containing at least one
vertex $\l$; among these terms, the only marginal ones (those
requiring a regularization) have four external lines (including
$\hat J_{\kk_4}$), since the oddness of the propagator does not
allow tadpoles. These terms are of the form
$$\eqalign{
&\sum_{\tilde \o} \int d\pp \tilde\chi_M(\pp)
\hat\psi^+_{\kk^+,\tilde\o} \int d\kk^+
\hat\psi^+_{\kk^+-\pp,\tilde\o} \hat\psi^+_{\bar\kk_4-\pp,-} \hat
g_-(\bar\kk_4) \hat J_{\bar\kk_4} \;\cdot\cr
&\left[ F^{(-1)}_{2,+,\tilde\o}(\kk^+, \kk^+-\pp) +
F^{(-1)}_{1,+}(\kk^+, \kk^+-\pp) \d_{+,\tilde\o} \right]\;,\cr}
\Eq(3.28)$$
where $F^{(-1)}_{2,+,\tilde\o}$ and $F^{(-1)}_{1,+}$ are defined
as in eq. (132) of [BM2]; they represent the terms in which both
or only one of the fields in $\d \r_{\pp,+}$, respectively, are
contracted. Both contributions to the r.h.s. of \equ(3.28) are
dimensionally marginal; however, the regularization of
$F^{(-1)}_{1,+}$ is trivial, as it is of the form
$$F_{1,+}^{(-1)}(\kk^+,\kk^-)=
[{[C_{h,0}(\kk^-)-1] D_+(\kk^-) \hat g^{(0)}_+(\kk^+)- u_0(\kk^+)
\over D_+(\kk^+-\kk^-)} G^{(2)}(\kk^+)\Eq(3.28a)$$
or the similar one, obtained exchanging $\kk^+$ with $\kk^-$.

\insertplot{300pt}{80pt}%
{\ins{150pt}{30pt}{$+$}
\ins{80pt}{5pt}{$F^{(-1)}_{2,+,\tilde\o}$}
\ins{230pt}{5pt}{$F^{(-1)}_{1,+}$}
}%
{F2F1}{}

\* \centerline{\eqg(5v): Graphical representation of \equ(3.28a)}

\* By the oddness of the propagator in the momentum,
$G^{(2)}(0)=0$, hence we can regularize  such term without
introducing any local term, by simply rewriting it as
$$F_{1,+}^{(-1)}(\kk^+,\kk^-)=
[{[C_{h,0}(\kk^-)-1] D_+(\kk^-) \hat g^{(0)}_+(\kk^+)- u_0(\kk^+)
\over D_+(\kk^+-\kk^-)} [G^{(2)}(\kk^+)-G^{(2)}(0)]\;.\Eq(3.28b)$$

As shown in [BM2], by using the symmetry property
$$\hat g^{(j)}_\o(\kk)=-i\o \hat g^{(j)}_\o(\kk^*) \virg
\kk=(k,k_0),\quad \kk^*=(-k_0,k)\;,\Eq(3.28c)$$
$F^{(-1)}_{2,+,\tilde\o}$ can be written as
$$F^{-1}_{2,+,\tilde\o}(\kk^+, \kk^-) = {1\over D_+(\pp)}
\left[ p_0 A_{0,+,\tilde\o}(\kk^+,\kk^-) + p_1
A_{1,+,\tilde\o}(\kk^+,\kk^-) \right]\;,\Eq(3.29a)$$
where $A_{i,+,\tilde\o}(\kk^+,\kk^-)$ are functions such that, if
we define
$$\LL F^{-1}_{2,+,\tilde\o}={1\over D_+(\pp)}
\left[ p_0 A_{0,+,\tilde\o}(0,0) +p_1 A_{1,+,\tilde\o}(0,0)
\right]\;,\Eq(3.29)$$
then
$$\LL F^{-1}_{2,+,+}=Z_{-1}^{3,+} \virg \LL F^{-1}_{2,+,-}=
{D_{-}(\pp)\over D_+(\pp)} Z_{-1}^{3,-}\;,\Eq(3.30)$$
where  $Z_{-1}^{3,+}$ and $Z_{-1}^{3,-}$ are suitable real
constants. Hence the local part of the marginal term \equ(3.28)
is, by definition, equal to
$$Z_{-1}^{3,+} T_+(\psi^{[h,-1]}) + Z_{-1}^{3,-}
T_-(\psi^{[h,-1]})\;.\Eq(3.30a)$$

\*

Let us finally consider the terms contributing to
$\bar\VV^{(-1)}_{b,2}$, that is those where
$\hat\psi^+_{\bar\kk_4-\pp}$ is not contracted and there is a
vertex of type $T_+$ or $T_-$. If even this vertex is not
contracted, we get a contribution similar to \equ(3.30a), with
$\n_\pm$ in place of $Z_{-1}^{3,\pm}$. Among the terms with at
least one vertex $\l$, there is, as before, no term with two
external lines; hence the only marginal terms have four external
lines and can be written in the form
$$\eqalign{
&\int d\pp \tilde\chi_M(\pp) \hat J_{\kk_4} \hat g_-(\kk_4) \int
d\kk^+ \sum_{\tilde\o} \hat\psi_{\kk^+,\tilde\o}^+ \hat\psi_{\kk^+
-\pp,\tilde\o}^-\;\cdot\cr
&\cdot\; \left[ \n_+ G^{(0)}_{+,\tilde\o}(\kk^+, \kk^+-\pp) + \n_-
{D_-(\pp) \over D_+(\pp)} G^{(0)}_{-,\tilde\o}(\kk^+, \kk^+-\pp)
\right] \;.\cr}\Eq(3.30b)$$
By using the symmetry property \equ(3.28c) of the propagators, it
is easy to show that $G^{(0)}_{\o,-\o}(0,0)=0$. Hence, if we
regularize \equ(3.30b) by subtracting $G^{(0)}_{\o,\tilde\o}(0,0)$
to $G^{(0)}_{\o,\tilde\o}(\kk^+, \kk^+-\pp)$, we still get a local
term of the form \equ(3.30a).

By collecting all the local term, we can write
$$\LL[\bar\VV^{(-1)}_{b,1} + \bar\VV^{(-1)}_{b,2}](\psi^{[h,-1]}) =
\n_{-1,+} T_+(\psi^{[h,-1]}) + \n_{-1,-}
T_-(\psi^{[h,-1]})\;,\Eq(3.31)$$
where $\n_{-1,\o} = \n_\o + Z_{-1}^{3,\o} + G^{(0)}_{\o,\o}(0,0)$.
Hence
$$\eqalign{
&\bar\VV^{(-1)}(\psi^{[h,-1]}) = T_1(\psi^{[h,-1]}) + \n_{-1,+}
T_+(\psi^{[h,-1]}) + \n_{-1,-} T_-(\psi^{[h,-1]}) +\cr
&+ \left[ \tilde\l_{-1} \bar F_\l^{[h,-1]}(\psi^{[h,-1]}) + \tilde
z_{-1} \hat\psi^{[h,-1]+}_{\bar\kk_4,-} D_-(\bar\kk_4) \right]
\hat g_-(\bar\kk_4) \hat J_{\bar\kk_4} +
\bar\VV^{(-1)}_R(\psi^{[h,-1]})\;,\cr}\Eq(3.31a)$$
where $\bar\VV^{(-1)}_R(\psi^{[h,-1]})$ is the sum of all
irrelevant terms linear in the external field $J$.
\*

\sub(3.6) {\it The multiscale expansion of $\tilde
G^4_+(\bar\kk_1, \bar\kk_2, \bar\kk_3, \bar\kk_4)$: the higher
scales integration.}

The integration of the field $\psi^{(-1)}$ is done in a similar
way; we shall call $\bar\VV^{(-2)}(\psi^{[h,-2]})$ the sum over
all terms linear in $J$. As before, the condition
\equ(2.21a) on the external momenta $\kk_i$ forbids the presence
of vertices of type $\f$, if $h<-1$, as we shall suppose.

The main difference is that there is no contribution obtained by
contracting both field variables belonging to $\d\r$ in
$T_1(\psi)$ at scale $-1$, because of \equ(3.10). It is instead
possible to get marginal terms with four external lines (two is
impossible), such that one of these fields is contracted at scale
$-1$. However, in this case, the second field variable will be
necessarily contracted at scale $h$, so that we can put $\RR=1$
for such terms; in fact, the extra factor $\g^{-(-1-h)}$ coming
from the bound \equ(3.17) after the integration of the last scale
field, has the effect of automatically regularize them, and even
the terms containing them as subgraphs.

The terms with a $T_1$ vertex, such that the field variables
belonging to $\d\r$ are not contracted, can be treated as in
\sec(3.3), hence do not need a regularization.

It follows that, if the irrelevant part $\bar\VV^{(-1)}_R$ were
absent in the r.h.s. of \equ(3.31a), then the regularization
procedure would not produce any local term proportional to $\bar
F_\l^{[h,-1]}(\psi^{[h,-2]})$, starting from a graph containing a
$T_1$ vertex.

It is easy to see that all other terms containing a vertex of type
$T_1$ or $T_\pm$ can be treated as in \sec(3.3). Moreover, the
support properties of $\hat g_-(\bar\kk_4)$ immediately implies
that it is not possible to produce a graph contributing to
$\bar\VV^{(-2)}$, containing the $\tilde z_{-1}$ vertex. Hence, in
order to complete the analysis of $\bar\VV^{(-2)}$, we still have
to consider the marginal terms containing the $\tilde\l_{-1}$
vertex, for which we simply apply the localization procedure
defined in \equ(3.22), \equ(3.23). We shall define two new
constants $\tilde\l_{-2}$ and $\tilde z_{-2}$, so that
$\tilde\l_{-2} (Z_{-2})^2$ is the coefficient of the local term
proportional to $\bar F_\l^{[h,-1]}(\psi^{[h,-2]})$, while $\tilde
z_{-2} Z_{-2} \hat\psi^{[h,-2]+}_{\bar\kk_4,-} D_-(\bar\kk_4) \hat
g_-(\bar\kk_4) \hat J_{\bar\kk_4}$ denotes the sum of all local
terms with two external lines produced in the second integration
step.

The above procedure can be iterated up to scale $h+1$, without any
important difference. In particular, for all marginal terms
(necessarily with four external lines) such that one of the field
variables belonging to $\d\r$ in $T_1(\psi)$ is contracted at
scale $i\ge j$, we put $\RR=1$. We can do that, because, in this
case, the second field variable belonging to $\d\r$ has to be
contracted at scale $h$, so that the extra factor $\g^{-(i-h)}$ of
\equ(3.17) has the effect of automatically regularize their
contribution to the tree expansion of $\tilde G^4_+(\bar\kk_1,
\bar\kk_2, \bar\kk_3, \bar\kk_4)$, to be described later.

Note that, as in the case $j=-1$, there is no problem connected
with the presence of the factors $\tilde\chi(\pp)$ and $D_-(\pp)
D_+(\pp)^{-1}$. In fact, if the field
$\hat\psi^+_{\bar\kk_4-\pp,-}$ appearing in the definition of
$T_1(\psi)$ or $T_\pm(\psi)$ is contracted on scale $j$, each
momentum derivative related with the regularization procedure
produces the right $\g^{-j}$ dimensional factor, since $\pp$ is of
order $\g^j$ and the derivatives of $\tilde\chi(\pp)$ are
different from $0$ only for momenta of order $\g^h$. If, on the
contrary, the field $\hat\psi^+_{\bar\kk_4-\pp,-}$ is not
contracted, then the renormalization procedure is tuned so that
$\tilde\chi(\pp)$ and $D_-(\pp) D_+(\pp)^{-1}$ are not affected by
the regularization procedure.

At step $-j$, we get an expression of the form
$$\eqalign{
&\bar\VV^{(j)}(\psi^{[h,j]}) = T_1(\psi^{[h,j]}) + \n_{j,+}
T_+(\psi^{[h,j]}) + \n_{j,-} T_-(\psi^{[h,j]}) +\cr
&+ \left[ \tilde\l_j Z_j^2 \bar F_\l^{[h,j]}(\psi^{[h,j]}) +
\sum_{i=j}^{-1} \tilde z_i Z_i \hat\psi^{[h,j]+}_{\bar\kk_4,-}
D_-(\bar\kk_4) \right] \hat g_-(\bar\kk_4) \hat J_{\bar\kk_4} +
\bar\VV^{j}_R(\psi^{[h,-1]})\;,\cr}\Eq(3.31b)$$
where $\bar\VV^{j}_R(\psi^{[h,-1]})$ is thought as a convergent
tree expansion (under the hypothesis that $\bar\l_h$ is small
enough), to be described in \sec(3.7). Since $Z_{-1}=1$, this
expression is in agreement with \equ(3.31a).

The expansion of $\tilde G^4_+(\bar\kk_1, \bar\kk_2, \bar\kk_3,
\bar\kk_4)$ is obtained by building all possible graphs with four
external lines, which contain one term taken from the expansion of
$\bar\VV^{(h)}(\psi^{(h)})$ and an arbitrary number of terms taken
from the effective potential $\VV^{(h)}(\psi^{(h)})$. One of the
external lines is associated with the free propagator
$g_-(\bar\kk_4)$, the other three are associated with propagators
of scale $h$ and momenta $\bar\kk_i$, $i=1,2,3$.

\* {\it Remark.} With respect to the expansion for $G^{4}_+$,
there are three additional quartic running coupling constants,
$\n_{j,+},\n_{j,-}$ and $\tilde\l_j$. Note that they are all
$O(\l)$, despite of the fact that the interaction $T_1$ has a
coupling $O(1)$; this is a crucial property, which follows from
the properties in \S\secc(3.2), implying that either $T_1$ is
contracted at scale $0$, or it gives no contribution to the
running coupling constants. At a first sight, it seems that now we
have a problem more difficult than the initial one; we started
from the expansion for $G_+^4$, which is convergent if the running
coupling $\l_j$ is small, and we have reduced the problem to that
of controlling the flow of four running coupling constants,
$\n_{+,j}$, $\n_{j,-}$, $\l_j$, $\tilde\l_j$. However, we will see
that, under the hypothesis $\bar\l_h\le \e_1$, also the flow of
$\n_{j,+}$, $\n_{j,-}$, $\tilde\l_j$ is bounded; one uses the
counterterms $\n_+,\n_-$ (this is the reason why we introduced
them in \S\secc(2)) to impose that $\n_{+,j},\n_{j,-}$ are
decreasing and vanishing at $j=h$, and then that the beta
functions for $\tilde\l_j$ and $\l_j$ are identical up to
exponentially decaying $O(\g^{\t j})$ terms.

\*

\sub(3.7) {\it The tree structure of the expansion.}

In order to describe the tree expansion of
$\bar\VV^{(j)}(\psi^{[h,j]})$, $j\in[h+1,-1]$, and $\tilde
G^4_+(\bar\kk_1, \bar\kk_2, \bar\kk_3, \bar\kk_4)$, we have to
modify the tree definition given in \sec(1), after Fig. \graf(1),
in the following way.

\*

\0 1) Instead of two, there are six types of special endpoints.
There are still the endpoints of type $\f$, defined exactly as
before, but there is no endpoint of type $J$. In addition, we have
special endpoints of type $T_1$, $T_+$, $T_-$, $\tilde\l$ and
$\tilde z$, associated in an obvious way with the five local terms
of \equ(3.31b).

\0 2) There are only trees with one and only one special endpoint
of type different from $\f$.

\0 3) The scale index is $\le +1$ for the endpoints of type $T_1$,
$T_+$ or $T_-$ , while it is $\le 0$ for the endpoints of type
$\tilde\l$ or $\tilde z$. Moreover, the scale index of an endpoint
$v$ of type $T_1$, $T_+$, $T_-$ or $\tilde\l$ is equal to
$h_{v'}+1$, if $v'$ is the non trivial vertex immediately
preceding $v$.

\0 4) If the tree has more than one endpoint and one of them is of
type $\tilde z$, the vertex $v_0$ of the tree must have scale $h$
and its scale index is equal to any value between $h+1$ and $0$.

\0 5) Given a tree with one endpoint $v_1$ of type $T_1$ and scale
index $h_{v_1}=+1$, the $\RR$ operation in the non trivial
vertices of the path $\CC$ connecting $v_1$ to $v_0$ depends on
the set $P_1$ of external lines in the vertex $v'_1$ (of scale
$0$) immediately preceding $v_1$.

If $P_1$ contains no one of the two external lines of $v_1$
belonging to the $\d\r$ part of the corresponding $T_1(\psi)$
term, then $\RR$ is defined in agreement with the localization
procedure bringing to \equ(3.30) for all vertices $v\in\CC$, such
that $v> v_2$, $v_2$ being the higher vertex, possibly coinciding
with $v'_1$, whose set of external lines does not contain the
field $\hat\psi^+_{\bar\kk_4-\pp}$ of $T_1(\psi)$. For the
remaining vertices of $\CC$, $\RR$ is defined in the usual way.

If $P_1$ contains both the two external lines of $v_1$ belonging
to $\d\r$ (hence the line of momentum $\bar\kk_4-\pp$ can not
belong to $P_1$), we define $\RR$ in agreement with the remark
following \equ(3.20), up to the higher vertex $v_2<v_1$, where at
least one of the lines of $\d\r$ does not belong to $P_1$ anymore.
For $v\le v_2$, $\RR$ is defined in the usual way.

If $P_1$ contains only one of the two external lines of $v_1$
belonging to $\d\r$ and one defines $v_2$ as before, $\RR$ is
defined along $\CC$ in agreement with the obvious generalization
of \equ(3.28b), for $v>v_2$. In $v_2$ one has to introduce a new
label to distinguish two cases, related with the two different
terms in the braces of the r.h.s. of \equ(3.24). In the first
case, $\RR=1$ for all $v\le v_2$, in the second case $\RR$ is
defined in the usual way.

If $h_{v_1}\le 0$ and we define $P_1$ and $v_2$ as before, the set
$P_1$, as well as the set $P_v$ for all $v<v_1$, must contain at
least one of the two external lines of $v_1$ belonging to $\d\r$.
Moreover, if $P_1$ contains one of these lines, then $\RR=1$ for
all $v\le v_2$.

\0 6) A similar, but simpler, discussion can be done for the trees
containing an endpoint of type $T_\pm$. We do not give the
details, but only stress that there is now no vertex $v$ with
$|P_v|=2$ or $|P_v|=4$, for which $\RR=1$.

\*

\sub(3.8){\it The flow of $\n_{j,\pm}$.}

The definitions of the previous sections imply that there is no
contribution to $\n_{j,\pm}$, coming from trees with a special
endpoint of type $\tilde\l$ or $\tilde z$. Moreover, because of
the symmetry \equ(3.28c) of the propagators (see remark after
\equ(3.30b)), $\n_{j,+}$ gets no contribution from trees with a special
endpoint of type $\n_{j,-}$, and viceversa. Finally, and very
important, if a tree contributing to $\n_{j,\pm}$ has an endpoint
of type $T_1$, this endpoint must have scale index $+1$.

The following Lemma has an important role in the following.

\*

\lemma(2) {\it If $\bar\l_h$ is small enough ( uniformly in $h$),
it is possible to choose $\n_+$ and $\n_-$ so that $\n_{h,\o}=0$
and
$$|\n_{j,\o}| \le c_0 \bar\l_h \g^{\th j} \virg h+1\le j\le 0
\;,\Eq(3.320)$$
where $0<\th<1/4$, $c_0$ is a suitable constant, independent of
$h$, and $\n_{0,\o}=\n_\o$.}

\*

\proof The previous remarks imply that there exists $\bar\e_1 \le
\bar\e_0$, such that, if $\bar\l_h\le \bar\e_1$, we can write
$$\n_{j-1,\o} = \n_{j,\o} + \b^{(j)}_{\n,\o}(\l_j, \n_{j,\o};
\ldots; \l_0,\n_{0,\o}) \virg h+1 \le j \le0\;,\Eq(3.32a)$$
with $\l_0=\l$, $\n_{0,\o}=\n_\o$ and
$$\b^{(j)}_{\n,\o}(\l_j, \n_{j,\o};
\ldots; \l_0,\n_{0,\o}) = \b^{(j,1)}_{\n,\o}(\l_j; \ldots; \l_0) +
\sum_{j'=j}^0 \n_{j',\o} \tilde\b^{(j,j')}_{\n,\o}(\l_j; \ldots;
\l_0)\;.\Eq(3.32)$$
Moreover
given a positive
$\th<1/4$, there are constants $c_1$ and $c_2$ such that
$$|\b^{(j,1)}_{\n,\o}(\l_j; \ldots; \l_0)| \le c_1
\bar\l_h \g^{2\th j} \virg |\tilde\b^{(j,j')}_{\n,\o}(\l_j;
\ldots; \l_0)| \le c_2 \bar\l_h^2 \g^{2\th(j-j')}\;.\Eq(3.33)$$
This follows from the fact that $\b^{(j,1)}_{\n,\o}$ and
$\tilde\b^{(j,j')}_{\n,\o}$ are given by a sum of trees verifying
the bound \equ(1.42) with $d_v>0$, with at least an end-point
respectively at scale $0$ and at scale $j'$, hence one can improve
the bound respectively by a factor $\g^{2\th j}$ and $\g^{2\th
(j-j')}$. In the following we shall call this property the {\it
short memory property}. Note that the bound of
$\tilde\b^{(j,j')}_{\n,\o}$ is of order $\bar\l_h^2$, instead of
$\bar\l_h$, because of the symmetry
\equ(3.28c), but a bound of order $\bar\l_h$ would be sufficient.

By a simple iteration, \equ(3.32a) can also be written in the form
$$\n_{j-1,\o}=\n_{0,\o}+\sum_{j'=j}^{0} \b^{(j')}_{\n,\o}(\l_{j'}, \n_{j',\o};
\ldots; \l_0,\n_{0,\o})\;.\Eq(3.33a)$$
We want to show that it is possible to choose $\n_{0,\o}$, so that
$\n_{0,\o}$ is of order $\bar\l_h$ and $\n_{h,\o}=0$. Since this
last condition, by \equ(3.33a), is equivalent to
$$\n_{0,\o}=-\sum_{j=h+1}^{0} \b^{(j)}_{\n,\o}(\l_{j}, \n_{j,\o};
\ldots; \l_0,\n_{0,\o})\;,\Eq(3.33b)$$
we see, by inserting \equ(3.33b) in the r.h.s. of \equ(3.33a),
that we have to show that there is a sequence $\un=\{\n_j,\, h+1
\le j \le0\}$, such that $\n_{0,\o}$ is of order $\bar\l_h$ and
$$\n_j = -\sum_{j'=h+1}^j \b^{(j')}_{\n,\o}(\l_{j'},
\n_{j'},..,\l_0,\n_0)\;.\Eq(3.34)$$

In order to prove that, we introduce the space $\MMM_\th$ of the
sequences $\un=\{\n_j,\, h+1 \le j \le0\}$ such that $|\n_j|\le c
\bar\l_h\g^{\th j}$, for some $c$; we shall think $\MMM_\th$ as a
Banach space with norm $||\un||_\th = \sup_{h+1\le j\le
0}|\n_j|\g^{-\th j} \bar\l_h^{-1}$. We then look for a fixed point
of the operator $\bT:\MMM_{\th}\to\MMM_{\th}$ defined as:
$$(\bT\un)_j = -\sum_{j'=h+1}^j \b^{(j')}_{\n,\o}(\l_{j'},
\n_{j'},..,\l_0,\n_0)\;.\Eq(3.35)$$

Note that, if $\bar\l_h$ is sufficiently small, then $\bT$ leaves
invariant the ball $\BBB_\th$ of radius $c_0=2c_1$ $\sum_{n=0}^\io
\g^{-n}$ of $\MMM_\th$, $c_1$ being the constant in \equ(3.33). In
fact, by
\equ(3.32) and \equ(3.33), if $||\un||_\th \le c_0$, then
$$|(\bT\un)_j| \le \sum_{j'=h+1}^j
c_1 \bar\l_h \g^{2\th j'} + \sum_{j'=h+1}^j \sum_{i=j'}^0 c_0
\bar\l_h \g^{\th i} c_2 \bar\l_h^2 \g^{2\th (j'-i)} \le c_0
\bar\l_h \g^{\th j}\;,\Eq(3.35a)$$
if $2c_2 \bar\l_h^2 (\sum_{n=0}^\io \g^{-n})^2 \le 1$.

$\bT$ is a also a contraction on $\BBB_\th$, if $\bar\l_h$ is
sufficiently small; in fact, if $\un,\un' \in \MMM_\th$,
$$\eqalign{
&|(\bT \n)_j-(\bT\un')_j|\le \sum_{j'=h+1}^j |
\b^{(j')}_{\n,\o}(\l_{j'}, \n_{j'},..,\l_0,\n_0)
-\b^{(j')}_{\n,\o}(\l_{j'}, \n'_{j'},..,\l_0,\n'_0)| \cr
&\le \sum_{j'=h+1}^j \sum_{i=j'}^0  ||\un-\un'||_\th \bar\l_h
\g^{\th i} c_2 \bar\l_h^2 \g^{2\th (j'-i)} \le {1\over 2}
||\un-\un'||_\th \bar\l_h \g^{\th j}\;, \cr }\Eq(3.36)$$
if $c_2 \bar\l_h^2 (\sum_{n=0}^\io \g^{-n})^2 \le 1/2$. Hence, by
the contraction principle, there is a unique fixed point $\un^*$
of $\bT$ on $\BBB_\th$.\Halmos

\*

\sub(3.9) {\it The constants $\tilde\l_j$ and $\tilde z_j$.}

We shall now analyze the constants $\tilde\l_j$ and $\tilde z_j$,
$h\le j\le -1$, appearing in the expansion of $\tilde
G^4_+(\bar\kk_1, \bar\kk_2, \bar\kk_3, \bar\kk_4)$. We shall do
that by comparing their values with the values of $\l_j$ and
$z_j$. We start noting that the beta function equation for $\l_j$
can be written as
$$\l_{j-1}= \left( {Z_j\over Z_{j-1}} \right)^2 \l_j +
\b_j + \b^{(1)}_{j}\;,\Eq(3.37)$$
where $\b_j$ is the sum over the local parts of the trees with at
least two endpoints and no endpoint of scale index $+1$, while
$\b^{(1)}_{j}$ is the similar sum over the trees with at least one
endpoint of scale index $+1$.

On the other hand we can write
$$\tilde\l_{j-1}= \left( {Z_j\over Z_{j-1}} \right)^2
\tilde\l_j + \tilde\b_j + \tilde\b^{(1)}_{j} + \tilde\b^{(T)}_{j}
+ \tilde\b^{(\n)}_{j}\;,\Eq(3.38)$$
where

\0 1) $\tilde\b_j$ is the sum over the local parts of the trees
with at least two endpoints, no endpoint of scale index $+1$ and
one special endpoint of type $\tilde\l$.

\0 2) $\tilde\b^{(1)}_{j} + \tilde\b^{(T)}_{j}$ is the sum over
the trees with at least one endpoint of scale index $+1$; in this
case, the special endpoint can be of type $\tilde\l$ or $T_1$ and,
if it is of type $T_1$, its scale index must be equal to $+1$.
$\tilde\b^{(1)}_{j}$ and $\tilde\b^{(T)}_{j}$ are, respectively,
the sum over the trees with the endpoint of type $\tilde\l$ or
$T_1$.

\0 3) $\tilde\b^{(\n)}_{j}$ is the sum over the trees with at
least two endpoints, whose special endpoint is of type $T_\pm$.

\*

A crucial role in this paper has the following Lemma.

\*

\lemma(3) {\it Let $\a=\tilde\l_h/\l_h$; then if $\bar\l_h$ is
small enough, there exists a constant $c$, independent of $\l$,
such that $|\a| \le c$ and
$$|\tilde\l_j - \a\l_j| \le c \bar\l_h \g^{\th j} \virg h+1\le j
\le -1\;.\Eq(3.38a)$$ }

\*

\proof The main point is the remark that there is a one to one
correspondence between the trees contributing to $\b_j$ and the
trees contributing to $\tilde\b_j$. In fact the trees contributing
to $\tilde\b_j$ have only endpoints of type $\l$, besides the
special endpoint $v^*$, and the external field with $\o=-$ and
$\s=-$ has to belong to $P_{v^*}$. It follows that we can
associate uniquely with any tree contributing to $\tilde\b_j$ a
tree contributing to $\b_j$, by simply substituting the special
endpoint with a normal endpoint, without changing any label. This
correspondence is surjective, since we have imposed the condition
that the trees contributing to $\tilde\b_j$ and $\b_j$ do not have
endpoints of scale index $+1$. Hence, we can write
$$\left[\left( {Z_j\over Z_{j-1}} \right)^2 -1\right]
(\tilde\l_j - \a\l_j) + \tilde\b_j - \a \b_j = \sum_{i=j}^{-1}
\b_{j,i} (\tilde\l_i - \a\l_i)\;, \Eq(3.43)$$
where, thanks to the ``short memory property'' and the fact that
$Z_j/Z_{j-1}=1 +O(\bar\l_j^2)$, the constants $\b_{j,i}$ satisfy
the bound $|\b_{j,i}|\le C \bar\l_j\g^{2\th(j-i)}$, with $\th$
defined as in Lemma \lm(2).

Among the four last terms in the r.h.s. of \equ(3.38), the only
one depending on the $\tilde\l_j$ is $\tilde\b^{(1)}_{j}$, which
can be written in the form
$$\tilde\b^{(1)}_{j} = \sum_{i=j}^{-1} \b'_{j,i}
\tilde\l_i\;,\Eq(3.43a)$$
the $\b'_{j,i}$ being constants which satisfy the bound
$|\b'_{j,i}| \le C \bar\l_j \g^{2\th j}$, since they are related
to trees with an endpoint of scale index $+1$. For the same
reasons, we have the bounds $|\tilde\b^{(T)}_{j}| \le C \bar\l_j
\g^{2\th j}$, $|\b^{(1)}_{j}| \le C \bar\l_j^2 \g^{2\th j}$.
Finally, by using also Lemma \lm(2), we see that
$|\tilde\b^{(\n)}_{j}| \le C \bar\l_j \bar\l_h \g^{2\th j}$.

We now choose $\a$ so that
$$\tilde\l_h - \a\l_h=0\;,\Eq(3.43b)$$
and we put
$$x_j = \tilde\l_j - \a\l_j \virg h+1 \le j \le -1\;.\Eq(3.43c)$$
We can write
$$x_{j-1} = x_{-1} + \sum_{j'=j}^{-1} \left[ \sum_{i=j'}^{-1}
\b_{j',i} x_i + \sum_{i=j'}^{-1} \b'_{j',i} (x_i + \a\l_i) +
\tilde\b^{(T)}_{j'} + \tilde\b^{(\n)}_{j'} - \a \b^{(1)}_{j}
\right]\;.\Eq(3.39)$$
On the other hand, the condition \equ(3.43b) implies that
$$x_{-1} = - \sum_{j'=h+1}^{-1} \left[ \sum_{i=j'}^{-1}
\b_{j',i} x_i + \sum_{i=j'}^{-1} \b'_{j',i} (x_i + \a\l_i) +
\tilde\b^{(T)}_{j'} + \tilde\b^{(\n)}_{j'} - \a \b^{(1)}_{j}
\right]\;,\Eq(3.40)$$
so that, if $h+1 \le j \le -1$, the $x_j$ satisfy the equation
$$x_j = - \sum_{j'=h+1}^j \left[ \sum_{i=j'}^{-1}
\b_{j',i} x_i + \sum_{i=j'}^{-1} \b'_{j',i} (x_i + \a\l_i) +
\tilde\b^{(T)}_{j'} + \tilde\b^{(\n)}_{j'} - \a \b^{(1)}_{j}
\right]\;.\Eq(3.41)$$

We want to show that equation \equ(3.41) has a unique solution
satisfying the bound
$$|x_j|\le  c_0 (1+|\a|\bar\l_h) \bar\l_h \g^{\th j}\;,\Eq(3.41a)$$
for a suitable constant $c_0$, independent of $h$, if $\bar\l_h$
is small enough. Hence we introduce the Banach space $\MMM_\th$ of
sequences $\ux = \{x_j, h+1\le j \le-1\}$ with norm
$||\ux||_\th\defin \sup_j |x_j|\g^{-\th j} \bar\l_h^{-1}$ and look
for a fixed point of the operator $\bT:\MMM_{\th}\to\MMM_{\th}$
defined by the r.h.s. of
\equ(3.41). By using the bounds on the various constants appearing
in the definition of $\bT$, we can easily prove that there are two
constants $c_1$ and $c_2$, such that
$$|(\bT \ux)_j| \le c_1\bar\l_h (1+|\a|\bar\l_h) \g^{\th j} +
c_2\bar\l_h \sum_{j'=h+1}^j  \sum_{i=j'}^{-1} \g^{2\th (j'-i)}
|x_i|\;.\Eq(3.42)$$
Hence, if we take $c_0=M c_1$, $M\ge 2$, the ball $\BBB_M$ of
radius $c_0 (1+|\a|\bar\l_h)$ in $\MMM_\th$ is invariant under the
action of $\bT$, if $c_2 \bar\l_h (\sum_{n=0}^\io \g^{-n})^2 \le
1/2$, since $1/2 \le (M-1)/M$. On the other hand, under the same
condition, $\bT$ is a contraction in all $\MMM_\th$; in fact, if
$\ux, \ux'\in\MMM_\th$, then
$$|(\bT \ux)_j - (\bT \ux')_j| \le c_2\bar\l_h^2 ||\ux - \ux'||
\sum_{j'=h+1}^j  \sum_{i=j'}^{-1} \g^{2\th (j'-i)} \g^{\th i} \le
{1\over 2} ||\ux - \ux'|| \bar\l_h \g^{\th j}\;,\Eq(3.44)$$
if $c_2 \bar\l_h (\sum_{n=0}^\io \g^{-n})^2 \le 1/2$. It follows,
by the contraction principle, that there is a unique fixed point
in the ball $\BBB_M$, for any $M\ge 2$, hence a unique fixed point
in $\MMM_\th$, satisfying the condition \equ(3.41a) with
$c_0=2c_1$.

To complete the proof, we have to show that $\a$ can be bounded
uniformly in $h$. In order to do that, we insert in the l.h.s. of
\equ(3.40) the definition of $x_{-1}$ and we bound the r.h.s. by
using \equ(3.41a) and \equ(3.42); we get
$$|\tilde\l_{-1}-\a\l_{-1}| \le c_3\bar\l_h + c_4 |\a|
\bar\l_h^2\;,\Eq(3.40a)$$
for some constants $c_3$ and $c_4$. Since $|\l_{-1}|\ge c_5|\l|$,
$\tilde\l_{-1}\le c_6|\l|$ and $\bar\l_h\le 2|\l|$ by the
inductive hypothesis, we have
$$|\a\l_{-1}| \le |\tilde\l_{-1}| + c_3\bar\l_h + c_4 |\a|
\bar\l_h^2 \Rightarrow |\a| \le (c_6 +2c_3 +2c_4|\a|
\bar\l_h)/c_5\;,\Eq(3.40b)$$
so that, $|\a| \le 2(c_6 +2c_3)/c_5$, if $4c_4 \bar\l_h \le
c_5$.\Halmos

\* {\it Remark.} The above Lemma is based on the fact that $\l_j$
and $\tilde\l_j$ have the same Beta function, up to $O(\g^{\th
j})$ terms (note that this is true thanks to our choice of the
counterterms $\n_{\pm}$, which implies that $\n_{j,\pm}$ are
$O(\g^{\th j})$). Hence if $\l_j$ is small, the same is true for
$\tilde\l_j$.

\* We want now to discuss the properties of the constants $\tilde
z_j$, $h\le j\le -1$, by comparing them with the constants $z_j$,
which are involved in the renormalization of the free measure, see
\equ(1.32). There is a tree expansion for the $z_j$, which can be
written as
$$z_j = \b_j + \b^{(1)}_j  \;,\Eq(3.44a)$$
where $\b_j$ is the sum over the trees without endpoints of scale
index $+1$, while $\b^{(1)}_j$ is the sum of the others,
satisfying the bound $|\b^{(1)}_j| \le C\bar\l_h^2 \g^{\th j}$.
The tree expansion of the $\tilde z_j$ can be written as
$$\tilde z_j = \tilde\b_j + \tilde\b^{(\n)}_j + \tilde\b^{(1)}_j\;,
\Eq(3.44b)$$
where $\tilde\b_j$ is the sum over the trees without endpoints of
scale index $+1$, such that the special endpoint is of type
$\tilde\l$, $\tilde\b^{(\n)}_j$ is the sum over the trees whose
special endpoint is of type $T_\pm$, and $\tilde\b^{(1)}_j$ is the
sum over the trees with at least an endpoint of scale index $+1$
(in this case, if the special endpoint is of type $T_1$, its scale
index must be $+1$, see discussion in \sec(3.6)).

Since there is no tree contributing to $\tilde\b^{(1)}_j$ without
at least one $\l$ or $\tilde\l$ endpoint and since all trees
contributing to it satisfy the ``short memory property'', by using
Lemma \lm(3) (which implies that $|\tilde\l_j| \le C\bar\l_h$), we
get the bound $|\tilde\b^{(1)}_j| \le C\bar\l_h\g^{\th j}$. In a
similar manner, by using Lemma \lm(2), we see that
$|\tilde\b^{(\n)}_j| \le C\bar\l_h^2\g^{\th j}$.

Let us now consider $\b_j$ and $\tilde\b_j$. By an argument
similar to that used in the proof of Lemma \lm(3), we can write
$$\tilde\b_j -\a \b_j = \sum_{i=j+1}^{-1} \b_{j,i} (\tilde\l_i -
\a\l_i)\;,\Eq(3.44c)$$
where $\a$ is defined as in Lemma \lm(3) and $|\b_{j,i}|\le
C\bar\l_h\g^{2\th j}$. Hence, Lemma \lm(3) implies that
$$|\tilde z_j - \a z_j|\le C\bar\l_h \g^{\th j}\;.\Eq(3.45)$$

\*

\sub(3.10) {\it The bound of $\tilde G^4_+(\bar\kk_1, \bar\kk_2,
\bar\kk_3, \bar\kk_4)$.}

There are various classes of trees contributing to the tree
expansion of $\tilde G^4_+(\bar\kk_1, \bar\kk_2, \bar\kk_3,
\bar\kk_4)$, depending on the type of the special endpoint. Let us
consider first the family $\TT_{\tilde\l}$ of the trees with an
endpoint of type $\tilde\l$. These trees have the same structure
of those appearing in the expansion of $G^4_+(\bar\kk_1,
\bar\kk_2, \bar\kk_3, \bar\kk_4)$, except for the fact that the
external (renormalized) propagator of scale $h$ and momentum
$\bar\kk_4$ is substituted with the free propagator $\hat
g_-(\bar\kk_4)$. It follows, by using the bound $|\tilde\l_j|\le
C\bar\l_h$, that a tree with $n$ endpoint is bounded by
$(C\bar\l_h)^n Z_h^{-1} \g^{-4h}$, larger for a factor $Z_h$ with
respect to what we need.

Let us now consider the family $\TT_{\tilde z}$ of the trees with
a special endpoint of type $\tilde z$. Given a tree $\t\in
\TT_{\tilde\l}$, we can associate with it the class $\TT_{\tilde
z,\t}$ of all $\t'\in \TT_{\tilde\l}$, obtained by $\t$ in the
following way:

\0 1) we substitute the endpoint $v^*$ of type $\tilde\l$ of $\t$
with an endpoint of type $\l$;

\0 2) we link the endpoint $v^*$ to an endpoint of type $\tilde z$
trough a renormalized propagator of scale $h$.

Note that $\TT_{\tilde z} = \cup_{\t\in \TT_{\tilde\l}}
\TT_{\tilde z,\t}$ and that, if $\t$ has $n$ endpoints, any
$\t'\in \TT_{\tilde z,\t}$ has $n+1$ endpoints. Moreover, since
the value of $\bar\kk_4$ has be chosen so that $f_h(\bar\kk_4)=1$,
$\hat g_-^{(h)}(\bar\kk_4)=Z_{h-1}^{-1} \hat g_-(\bar\kk_4)$;
hence it is easy to show that the sum of the values of a tree
$\t\in \TT_{\tilde\l}$, such the special endpoint has scale index
$j^*+1$, and of all $\t'\in \TT_{\tilde z,\t}$ is obtained from
the value of $\t$, by substituting $\tilde\l_{j^*}$ with
$$\L_{j^*} = \tilde\l_{j^*} - \l_{j^*}
{\sum_{j=h}^{-1} \tilde z_j Z_j\over Z_{h-1}}\;,\Eq(3.45a)$$
see Fig. \graf(5vd).



\*\*

On the other hand, \equ(3.45) and the bound $Z_j\le
\g^{-C\bar\l_h^2 j}$, see [BM1], imply that, if $\bar\l_h$ is
small enough
$$\sum_{j=h}^{-1} \left| \tilde z_j Z_j - \a z_j Z_j
\right|\le \sum_{j=h}^{-1} C \bar\l_h \g^{\th j} Z_j \le
C\bar\l_h\;.\Eq(3.47)$$
It follows, by using also the bound \equ(3.38a), that
$$\L_{j^*} = \a\l_{j^*} \left[1 -
{\sum_{j=h}^{-1} z_j Z_j\over Z_{h-1}} \right] + {O(\bar\l_h)\over
Z_h}\;.\Eq(3.45b)$$
Moreover, since $Z_{j-1}=Z_j (1+z_j)$, for $j\in[-1,h]$, and
$Z_{-1}=1$, it is easy to check that
$$Z_{h-1} - \sum_{j=h}^{-1} z_j Z_j= 1\;.\Eq(3.48)$$
This identity, Lemma \lm(3) and \equ(3.45b) imply the bound
$$|\L_{j^*}| \le C {\bar\l_h\over Z_h}\;,\Eq(3.48a)$$
which gives us the ``missing'' $Z_h^{-1}$ factor for the sum over
the trees whose special endpoint is of type $\tilde\l$ or $\tilde
z$.

\* Let us now consider the family $\TT_{\n}$ of the trees with a
special endpoint of type $T_\pm$. It is easy to see, by using
Lemma \lm(2) and the ``short memory property'', that the sum over
the trees of this class with $n\ge 0$ normal endpoints is bounded,
for $\bar\l_h$ small enough, by $(C\bar\l_h)^{n+1} Z_h^{-1}
\g^{-4h} \sum_{j=h}^{-1} Z_j^{-2} \g^{2\th (h-j)}$ $\g^{\th j} \le
(C\bar\l_h)^{n+1} Z_h^{-3} \g^{-(4-\th)h}$, which is even better
of our needs.

We still have to consider the family $\TT_1$ of the trees with a
special endpoint of type $T_1$. There is first of all the trivial
tree, obtained by contracting all the $\psi$ lines of $T_1$ on
scale $h$, but its value is $0$, because of the support properties
of the function $\tilde\c(\pp)$. Let us now consider a tree $\t\in
\TT_1$ with $n\ge 1$ endpoints of type $\l$, whose structure is
described in item 5) of \sec(3.7), which we shall refer to for
notation. If we put $h_{v_1}=j_1 + 1$ and $h_{v_2}=j_2$, then the
dimensional bound of this tree differs from that of a tree with
$n+1$ normal endpoints contributing to $G^4_+(\bar\kk_1,
\bar\kk_2, \bar\kk_3, \bar\kk_4)$ for the following reasons:

\0 1) there is a factor $Z_h^{-1}$ missing, because the external
(renormalized) propagator of scale $h$ and momentum $\bar\kk_4$ is
substituted with the free propagator $\hat g_-(\bar\kk_4)$;

\0 2) there is a factor $|\l_{j_1}| Z_{j_1}^2$ missing, because
there is no external field renormalization in the
$T_1(\psi^{[h,j]})$ contribution to $\bar\VV^{(j)}(\psi^{[h,j]})$,
see \equ(3.31b);

\0 3) if $P_1$ contains only one of the two external lines of
$v_1$ belonging to $\d\r$, then there is a factor $\g^{-(j_2-h)}$
missing, because the absence of regularization in the vertices
$v\le v_2$, but this is compensated by the same factor arising
because of the bound \equ(3.17), see discussion after \equ(3.24)
and in \sec(3.6), so that the ``short memory property'' is always
satisfied;

\0 4) there is a factor $Z_h^{-1}$ missing, because of the remark
following \equ(3.17).

It follows that the sum of the values of all trees $\t\in \TT_1$
with $n\ge 1$ normal endpoints, if $\bar\l_h$ is small enough, is
bounded by $(C\bar\l_h)^n \g^{-4h} \sum_{j_1=h}^0 Z_{j_1}^{-2}
\g^{2\th(h-j_1)}$ $\le (C\bar\l_h)^n \g^{-4h} Z_h^{-2}$.

By collecting all the previous bounds, we prove that the bound
\equ(2.26) of Lemma \lm(1) is satisfied in the case of
$H_+^{4,1}$.

\* {\it Remark.} In $T_1$ and in the Grassmannian monomials
multiplying $\n_{j,+}, \n_{j,-}$, an external line is always
associated to a free propagator $\hat g_-(\bar\kk_4)$; this is due
to the fact that, in deriving the Dyson equation \equ(2.9), one
extracts a free propagator. Then in the bounds there is a $Z_h$
missing (such propagator is not ``dressed'' in the multiscale
integration procedure), and at the end the crucial identity
\equ(3.48) has to be used to ``dress'' the extracted propagator
carrying momentum $\bar\kk_4$.

\*

\sub(3.11) {\it The bound in the case of $H_-^{4,1}$}.

If we substitute, in the l.h.s. of \equ(3.1) $H_+^{4,1}$ with
$H_-^{4,1}$, we can proceed in a similar way. By using
\equ(2.17a), we get
$$\hat g_-(\kk_4) {1\over L\b}\sum_{\pp} \tilde\chi_M(\pp)
D_+^{-1}(\pp) \hat H^{4,1}_{-}(\pp;\kk_1,\kk_2,\kk_3,\kk_4-\pp) =
$$
$$=\hat g_-(\kk_4) {1\over L\b}\sum_{\pp} \tilde\chi_M(\pp)
{1\over L\b}\sum_{\kk} {C_-(\kk,\kk-\pp)\over D_+(\pp)}
<\hat\psi^+_{\kk,-}\hat\psi^-_{\kk-\pp,-};
\hat\psi^-_{\kk_1,+};\hat\psi^+_{\kk_2,+}; \hat\psi^-_{\kk_3,-};
\hat\psi^+_{\kk_4-\pp,-}>^T+$$
$$-\nu'_-\hat g_-(\kk_4) {1\over L\b}\sum_{\pp} \tilde\chi_M(\pp)
{1\over L\b}\sum_{\kk} {D_-(\pp)\over D_+(\pp)}
<\hat\psi^+_{\kk,-}\hat\psi^-_{\kk-\pp,-};
\hat\psi^-_{\kk_1,+};\hat\psi^+_{\kk_2,+}; \hat\psi^-_{\kk_3,+};
\hat\psi^+_{\kk_4-\pp,-}>^T-$$
$$-\nu'_+\hat g_-(\kk_4) {1\over L\b} \sum_{\pp} \tilde\chi_M(\pp)
 {1\over L\b}\sum_{\kk} <\hat\psi^+_{\kk,+}\hat\psi^-_{\kk-\pp,+};
\hat\psi^-_{\kk_1,+};\hat\psi^+_{\kk_2,+}; \hat\psi^-_{\kk_3,-};
\hat\psi^+_{\kk_4-\pp,-}>^T\;. \Eq(3.1a)$$
We define $\tilde G^4_-(\kk_1, \kk_2,\kk_3, \kk_4)$ as in
\equ(3.2) with $\tilde W$ replaced by $\tilde W_-$ given by
$$\tilde W_- = \log \int P(d\hat\psi)e^{-T_2(\psi) + \n'_+T_+(\psi)
+ \n'_-T_-(\psi)} e^{-V(\hat\psi)+ \sum_\o\int d\xx
[\phi^+_{\xx,\o}\hat\psi^{-}_{\xx,\o}+
\hat\psi^{+}_{\xx,\o}\phi^-_{\xx,\o}]} \;,\Eq(3.3a)$$
$$T_2(\psi) = {1\over L\b} \sum_{\pp} \tilde\chi_M(\pp)
{1\over L\b} \sum_{\kk} {C_-(\kk, \kk-\pp) \over D_+(\pp)}
(\hat\psi_{\kk,-}^+ \hat\psi_{\kk-\pp,-}^-)
\hat\psi^+_{\kk_4-\pp,-} \hat J_{\kk_4} \hat
g(\kk_4)\;,\Eq(3.4aa)$$
$T_+, T_-$ being defined as in \equ(3.5), \equ(3.6). By the
analogues of \equ(3.7a), \equ(3.7b) we obtain
$$-\tilde G^4_-(\kk_1,\kk_2,\kk_3,\kk_4)= g_-(\kk_4){1\over L\b}
\sum_{\pp} \tilde\chi_M(\pp)
{H_-^{4,1}(\pp;\kk_1,\kk_2,\kk_3,\kk_4-\pp)\over
D_+(\pp)}\;.\Eq(3.7ca)$$
The calculation of $\tilde G^4_-(\kk_1,\kk_2,\kk_3,\kk_4)$ is done
via a multiscale expansion essentially identical to the one of
$\tilde G^4_+(\kk_1,\kk_2,\kk_3,\kk_4)$, by taking into account
that $\d\r_{\pp,+}$ has to be substituted with

$$\d\r_{\pp,-}={1\over\b L}\sum_\kk{ C_-(\pp,\kk)\over D_+(\pp)}
(\hat\psi_{\kk,-}^+ \hat\psi_{\kk-\pp,-}^-)\;.\Eq(3.8a)$$
Let us consider the first step of the iterative integration
procedure and let us call again $\bar\VV^{(-1)}(\psi^{[h,-1]})$
the contribution to the effective potential of the terms linear in
$J$. Let us now decompose $\bar\VV^{(-1)}(\psi^{[h,-1]})$ as in
\equ(3.19) and let us consider the terms contributing to
$\bar\VV^{(-1)}_{a,1}(\psi^{[h,-1]})$. The analysis goes exactly
as before when no one or both the fields $\hat\psi_{\kk,-}^+$ and
$\hat\psi_{\kk-\pp,-}^-$ of $\d\r_{\pp,-}$ are contracted. This is
not true if only one among the fields $\hat\psi_{\kk,-}^+$ and
$\hat\psi_{\kk-\pp,-}^-$ in $T_2(\psi)$ is contracted, since in
this case there are marginal terms with two external lines, which
before were absent. The terms with four external lines can be
treated as before; one has just to substitute $D_+(\kk^-) \hat
g_+^{(0)}(\kk^+)$ with $D_-(\kk^-) g_-^{(0)}(\kk^+)$ in the r.h.s.
of \equ(3.24), but this has no relevant consequence. The terms
with two external lines have the form
$$\eqalign{
& \int d\kk^- \hat\psi^+_{\bar\kk_4,-} \hat g_-(\bar\kk_4) \hat
J_{\bar\kk_4} \tilde\chi_M(\bar\kk^4-\kk^-) G_1^{(0)}(\kk_-)
\;\cdot\cr
& \cdot\; \left\{ {[C_{h,0}^\e(\bar\kk_4)-1] D_{-}(\bar\kk_4) \hat
g^{(0)}_-(\kk^-)\over D_+(\bar\kk_4-\kk^-)} - {u_0(\kk^-) \over
D_+(\bar\kk^4-\kk^-)} \right\}\;, \cr}\Eq(3.24b)$$
where $G_1^{(0)}(\kk_-)$ is a smooth function of order $1$ in
$\l$. However, the first term in the braces is equal to $0$, since
$|\bar\kk_4|=\g^h$ implies that $C_{h,0}^\e(\bar\kk_4)-1=0$. Hence
the r.h.s. of \equ(3.24b) is indeed of the form
$$\int d\kk^- \hat\psi^+_{\bar\kk_4,-}
\hat g_-(\bar\kk_4) \hat J_{\bar\kk_4} \tilde\chi_M(\bar\kk_4 -
\kk^-) G_1^{(0)}(\kk_-){u_0(\kk^-) \over D_+(\bar\kk^4-\kk^-)}
\;,\Eq(3.24c)$$
so that it can be regularized in the usual way.

The analysis of $\bar\VV^{(-1)}_{a,2}(\psi^{[h,-1]})$ can be done
exactly as before. Hence, we can define again $\tilde\l_{-1}$ and
$\tilde z_{-1}$ as in \equ(3.25a), with $\tilde\l_{-1}=O(\l)$ and
$\tilde z_{-1}=O(1)$.

Let us consider now the terms contributing to
$\bar\VV^{(-1)}_{b,1}$, that is those where
$\hat\psi^+_{\bar\kk_4-\pp}$ is not contracted and there is a
vertex of type $T_2$. Again the only marginal terms have four
external lines and have the form
$$\eqalign{
&\sum_{\tilde \o} \int d\pp \tilde\chi_M(\pp)
\hat\psi^+_{\kk^+,\tilde\o} \int d\kk^+
\hat\psi^+_{\kk^+-\pp,\tilde\o} \hat\psi^+_{\bar\kk_4-\pp,-} \hat
g_-(\bar\kk_4) \hat J_{\bar\kk_4}\;\cdot\cr
&{D_-(\pp) \over D_+(\pp)} \left[ F^{(-1)}_{2,-,\tilde\o}(\kk^+,
\kk^+-\pp) + F^{(-1)}_{1,-}(\kk^+, \kk^+-\pp) \d_{-,\tilde\o}
\right]\;,\cr} \Eq(3.28aa)$$
where we are using again the definition (132) of [BM2] (hence we
have to introduce in the r.h.s. of \equ(3.28aa) the factor
$D_-(\pp)/ D_+(\pp)$). The analysis of the terms
$F^{(-1)}_{1,-}(\kk^+, \kk^+-\pp)$ is identical to the one in
\sec(3.3), while, as shown in [BM2], the symmetry property
\equ(3.28c) implies now that, if we define
$$F^{-1}_{2,-,\tilde\o}(\kk^+, \kk^-) = {1\over D_-(\pp)}
\left[ p_0 A_{0,-,\tilde\o}(\kk^+,\kk^-) + p_1
A_{1,-,\tilde\o}(\kk^+,\kk^-) \right]\;,\Eq(3.29aa)$$
and
$$\LL F^{-1}_{2,-,\tilde\o}={1\over D_-(\pp)}
\left[ p_0 A_{0,-,\tilde\o}(0,0) +p_1 A_{1,-,\tilde\o}(0,0)
\right]\;,\Eq(3.29ab)$$
then
$$\LL F^{-1}_{2,-,+}= Z_{-1}^{3,-} {D_-(\pp) \over D_+(\pp)}
\virg \LL F^{-1}_{2,-,-}= Z_{-1}^{3,+}\;,\Eq(3.30aa)$$
where $Z_{-1}^{3,+}$ and $Z_{-1}^{3,-}$ are the same real
constants appearing in \equ(3.30). Hence, the local part of the
marginal term \equ(3.28aa) is, by definition, equal to
$$Z_{-1}^{3,+} T_+(\psi^{[h,-1]}) + Z_{-1}^{3,-}
T_-(\psi^{[h,-1]})\;.\Eq(3.30ab)$$
The analysis of $\bar\VV^{(-1)}_{b,2}$ can be done exactly as
before, so that we can write for $\bar\VV^{(-1)}$ an expression
similar to \equ(3.31a), with $T_2(\psi^{[h,-1]})$ in place of
$T_1(\psi^{[h,-1]})$ and $\n'_{-1,\pm}$ in place of $\n_{-1,\pm}$.
One can prove that, for simple symmetry reasons, $\n'_{-1,\pm} =
\n_{-1,\mp}$, if $\n'_\pm = \n_\mp$, but this property will not
play any role, hence we will not prove it.

The integration of higher scales proceed as in \sec(3.6). In fact,
the only real difference we found in the integration of the first
scale was in the calculation of the $O(1)$ terms contributing to
$\tilde z_{-1}$, but these terms are absent in the case of $\tilde
z_j$, $j\le -2$, because the second term in the expression
analogous to \equ(3.24b), obtained by contracting on scale $j<0$
only one of the fields of $\d\r_{\pp,-}$, is exactly zero. It
follows that the tree structure of the expansion is the same as
that described in \sec(3.7) and the constants $\n'_\o$ can be
chosen again so that the bound \equ(3.320) is satisfied even by
the constants $\n'_{j,\o}$.

In the analysis of the constants $\tilde\l_j$ and $\tilde z_j$
there is only one difference, concerning the bound \equ(3.45),
which has to be substituted with $\tilde z_{-1}-\a z_{-1} \le C$,
in the case $j=-1$, but it is easy to see that this has no effect
on the bound \equ(3.48a). It follows that the final considerations
of \sec(3.10) stay unchanged and we get for $\tilde
G^4_-(\kk_1,\kk_2,\kk_3,\kk_4)$ a bound similar to that proved for
$\tilde G^4_+(\kk_1,\kk_2,\kk_3,\kk_4)$, so ending the proof of
Lemma \lm(1).

\* \*\* \appendix(4, The ultraviolet problem and the Thirring
model) \*

Thanks to the linearity of the propagator, the above analysis can
be used with no essential modifications to construct the massless
{\it Thirring model} (see for instance [Z]), by removing the
ultraviolet cutoff. We shall sketch here the main ideas; the
details will be published elsewhere.

The Thirring model describes Dirac fermions in $d=1+1$
interacting with a local current-current interaction;
its action is
$$\int d\xx [-\bar\psi_\xx\not\partial\psi_\xx -
{\l\over 4} J_\m(\xx) J^\m(\xx)]\Eq(1a)$$
where $\not\partial=\g_0\partial_{x_0}+\g_1\partial_{x}$,
$\xx=(x_0,x)$, $\bar\psi_\xx= \psi^+_\xx \g_0$, $\psi_\xx$ is a
two component spinor field (to not be confused with a Grassmannian
field), $J_\m(\xx)=\bar\psi_\xx \g_\m\psi_\xx$ and $\g_0=\s_1$,
$\g_1=\s_2$ are Pauli matrices.

The {\it generating functional} of the Thirring model is the
following Grassmannian integral with infrared cutoff $\g^h$ and
ultraviolet cutoff $\g^N$, with $h,N$ integers and $N>0$
$$\eqalign{
& \WW(\phi,J)= \log \int P_{Z_N}(d\psi^{[h,N]}) \exp\Bigg\{
-V(\psi^{[h,N]})+\cr
&+ \sum_\o \int d\xx \left[ Z_N^{(2)}
J_{\xx,\o}\psi^{[h,N]+}_{\xx,\o}\psi^{[h,N]-}_{\xx,\o}+
\phi^+_{\xx,\o}\psi^{[h,N]-}_{\xx,\o}+
\psi^{[h,N]+}_{\xx,\o}\phi^{[h,N]-}_{\xx,\o}\right]\Bigg\}
\;,\cr}\Eq(1.7a)$$
where $P_{Z_N}(d\psi^{[h,N]})$ is given by
\equ(1.3), with $C_{h,0}(\kk)$ replaced by $C_{h,N}(\kk)=\sum_{j=h}^N
f_j(\kk)$ and $\psi^{[h,0]\s}_{\kk,\o}$ replaced by
$\sqrt{Z_N}\psi^{[h,N]\s}_{\kk,\o}$, and $V(\psi^{[h,N]})$ is
given by
$$V(\psi^{[h,N]})=\tilde\l_N
\int d\xx\; \psi^{[h,N]+}_{\xx,+} \psi^{[h,N]-}_{\xx,+}
\psi^{[h,N]+}_{\xx,-} \psi^{[h,N]-}_{\xx,-}\; ;\Eq(1.2a)$$
$Z_N$ is the (bare )wave function renormalization, $Z^{(2)}_N$ is
the (bare )density renormalization and $\tilde\l_N$ is the (bare)
interaction. In order to get a nontrivial limit as $N\to\io$, it
is convenient to write $\tilde\l_N$ and $Z_N^{(2)}$ in terms of
$Z_N$ and two new bare constants, $\l_N$ and $c_N$, in the
following way:
$$\tilde\l_N=(Z_N)^2\l_N \virg Z_N^{(2)}= c_N Z_N\;.\Eq(aa1)$$
One expects that the model is well defined if $\l_N$ and $c_N$
converge to finite non zero limits and $Z_N\to 0$, as $N\to\io$;
moreover, in order to apply our perturbative procedure, $\l_N$ has
to be small enough, uniformly in $N$.

The proof of this claim is essentially a corollary of the above analysis for
the infrared problem. The RG analysis in \S\secc(1) can be repeated by
allowing the scale index $j$ to be positive or negative. The Ward identity
\equ(2.12) holds with a factor $Z_N/Z_N^{(2)}$ multiplying $G^{2,1}_+$ and
$\D_+^{2,1}$, and we get the identity
$$\left| {Z_h^{(2)}\over c_N Z_h}-1\right|\le C\bar\l_h^2\;.\Eq(aa11)$$
In the same way, from the Dyson equation (identical to
\equ(2.11a), with $\l_N Z_N/ Z_N^{(2)}$ in place of $\l$ in the
r.h.s), and proceeding as in \S\secc(2) and \S\secc(3), we get
that for any $h$ one has $|\l_h-\l_N|\le c_3 \bar\l^2_{h+1}$
(compare with \equ(2.4)), so that the expansion is convergent, if
$\l_N$ is small enough. By \equ(2.2), $Z_N$ must be chosen so that
$Z_N \g^{N \h(\l_N)}$ is convergent for $N\to\io$ to some
constant, which can be fixed by requiring, for instance, that
$Z_0=1$. In the same way we can fix $\lim_{N\to\io} \l_N$ so that,
for instance, $\l_0=\l$, with $\l$ small enough; of course
$\l_N=\l+O(\l^2)$. The choice of $\lim_{N\to\io} c_N$ is a free
parameter, whose value has no special role.

Finally we shall discuss the form taken from Ward identities when
the ultraviolet and the infrared cutoff are removed. The analogous
of \equ(2.16) for the model \equ(1.7a) is
$$\eqalignno{
&{Z_N\over Z_N^{(2)}}(1-\n_+) D_{+}(\pp) \hat G_{+}^{4,1}
(\pp,\kk_1,\kk_2,\kk_3,\kk_4-\pp)-{Z_N\over Z_N^{(2)}}\n_{-}
D_{-}(\pp) \hat G_{-}^{4,1}(\pp,\kk_1,\kk_2,\kk_3,\kk_4-\pp)
&\eq(2.16z)\cr
&= \hat G_{+}^{4}(\kk_1-\pp,\kk_2,\kk_3,\kk_4-\pp)-
\hat G_{+}^{4}(\kk_1,\kk_2+\pp,\kk_3,\kk_4-\pp) + Z_N
H^{4,1}_{+}(\pp,\kk_1,\kk_2,\kk_3,\kk_4-\pp)\;,\cr}$$
with $H^{4,1}_{+}$ given by \equ(2.16a) with the cutoff function
$C_{h,0}^{-1}$ replaced by $C_{h,N}^{-1}$. The counterterms
$\n_\pm$ are found by a fixed point method as in Lemma 4.1, with
the only difference that the ultraviolet scale $0$ is replaced by
the scale $N$; $\n_\pm, \n'_\pm$ tend to a non vanishing well
defined limit as $N\to\io, h\to-\io$.

In the same way the analogous of \equ(2.17) is
$$\eqalignno{
&{Z_N\over Z_N^{(2)}}(1-\n'_-) D_-(\pp) \hat G_-^{4,1}
(\pp,\kk_1,\kk_2,\kk_3,\kk_4-\pp) - {Z_N\over Z_N^{(2)}} \n'_+
D_+(\pp) \hat G_+^{4,1}(\pp,\kk_1,\kk_2,\kk_3,\kk_4-\pp) \cr
&= \hat G_{+}^{4}(\kk_1,\kk_2,\kk_3-\pp,\kk_4-\pp)- \hat
G_{+}^{4}(\kk_1,\kk_2,\kk_3,\kk_4) + Z_N
H^{4,1}_-(\pp,\kk_1,\kk_2,\kk_3,\kk_4-\pp)\;,&\eq(2.17z)\cr}$$
It is a straightforward consequence of our analysis (in particular
of the short memory property we used extensively throughout the
paper) that in \equ(2.16) and \equ(2.17) for the model \equ(1.7a),
if the external momenta have a fixed (\ie independent from $h,N$)
value, then
$$\lim_{N\to\io\atop h\to-\io} Z_N \hat
H^{4,1}_\pm(\pp,\kk_1,\kk_2,\kk_3,\kk_4-\pp)=0\;. \Eq(1.7b)$$
Hence, if we sum the two Ward identities above and remember that $\n_+=\n'_-$
and $\n_-=\n'_+$, we get, in the limit $N\to\io, h\to-\io$
$$\eqalign{
&\sum_{\o=\pm}D_{\o}(\pp) {1-\n_+-\n'_-\over c_N} \hat
G_{\o}^{4,1} (\pp,\kk_1,\kk_2,\kk_3,\kk_4-\pp)=
G_{+}^{4}(\kk_1-\pp,\kk_2,\kk_3,\kk_4-\pp)-\cr
&-\hat G_{+}^{4}(\kk_1,\kk_2+\pp,\kk_3,\kk_4-\pp)+\hat
G_{+}^{4}(\kk_1,\kk_2,\kk_3-\pp,\kk_4-\pp)- \hat
G_{+}^{4}(\kk_1,\kk_2,\kk_3,\kk_4)\cr}\Eq(6.77a)$$

The above Ward identity is identical to the formal one, obtained by a {\it
total} gauge transformation, except for the factor $(1-\n_+-\n'_-)/c_N$
multiplying $G^{4,1}$; in other words, the formal Ward identity holds when
the cutoffs are removed, up to a finite interaction-dependent renormalization
of the density operator. A similar phenomenon appears also in perturbative
QED [H] and is called {\it soft breaking} of gauge invariance. Of course it
is possible to choose $Z^{(2)}_N$ so that the formal Ward identity is
verified, \ie we can choose $c_N=(1-\n_+-\n'_-)$.

On the contrary, the Ward identities \equ(2.16) and \equ(2.17), obtained by a
{\it chiral} gauge transformation, do not tend in the limit to the formal
Ward identities (obtained by \equ(2.16) and \equ(2.17) putting
$\n_\pm=\nu'_\pm=0$); beside the renormalization of the density operator, an
extra factor appears in the identity, namely $G^{4,1}_-$ in \equ(2.16) or
$G^{4,1}_+$ in \equ(2.17). This phenomenon is called {\it chiral anomaly},
see [Z], and is is present also in perturbative QED.

\*\*

\centerline{\titolo References}

\*

\halign{\hbox to 1.2truecm {[#]\hss} &
        \vtop{\advance\hsize by -1.25 truecm \0#}\cr

A& {P.W. Anderson. The theory superconductivity in high $T_c$ cuprates,
Princeton University Press (1997)}\cr
Af& {I. Affleck: Field theory methods and quantum critical
phenomena. Proc. of Les Houches summer school on Critical
phenomena, Random Systems, Gauge theories, North Holland (1984).
}\cr
BG& {G. Benfatto, G. Gallavotti: Perturbation Theory of the Fermi
Surface in a Quantum Liquid. A General Quasiparticle Formalism and
One-Dimensional Systems. {\it J. Stat. Phys.} {\bf 59}, 541--664
(1990). }\cr
BGPS& {G. Benfatto, G. Gallavotti, A. Procacci, B. Scoppola: Beta
Functions and Schwinger Functions for a Many Fermions System in
One Dimension. {\it  Comm. Math. Phys.} {\bf  160}, 93--171
(1994). }\cr
BM1& {G. Benfatto, V. Mastropietro: Renormalization Group, hidden
symmetries and approximate Ward identities in the $XYZ$ model.
{\it Rev. Math. Phys.} {\bf 13}, 1323--1435 (2001). }\cr
BM2& {G. Benfatto, V. Mastropietro: On the density-density
critical indices in interacting Fermi systems. {\it Comm. Math.
Phys.} {\bf 231}, 97--134 (2002).}\cr
BM3& {G. Benfatto, V. Mastropietro: Ward identities and vanishing
of the Beta function for $d=1$ interacting Fermi systems. {\it J.
Stat. Phys.} {\bf 115}, 143--184 (2004).}\cr
BoM& {F. Bonetto, V. Mastropietro: Beta Function and Anomaly of
the Fermi Surface for a $d=1$ System of Interacting Fermions in a
Periodic Potential. {\it Comm. Math. Phys.} {\bf  172}, 57--93
(1995). }\cr
DL&{I.E. Dzyaloshinky, A.I. Larkin: Correlation functions for a
one-dimensional Fermi system with long-range interaction (Tomonaga
model). {\it Soviet Phys. JETP} {\bf 38}, 202--208 (1974). }\cr
G& {T. Giamarchi.: Quantum Physics in one dimension. {\it International Series
of Monographs on Physics} {\bf 121}, Clarendon Press, Oxford (2004).} \cr
GM& {G. Gentile, V. Mastropietro: Renormalization Group for
fermions: a review of mathematical results. {\it Phys. Rep.} {\bf
352}, 273--437 (2001). }\cr
H&{T.R. Hurd: Soft breaking of Gauge
invariance in regularized Quantum Electrodynamics.
{\it Comm. Math. Phys} {\bf 125},
515-526 (1989). }\cr
Le& {A. Lesniewski: Effective action for the Yukawa 2 quantum
field Theory. {\it Commun. Math. Phys.} {\bf 108}, 437--467
(1987). }\cr
M& {V.Mastropietro,
Small denominators and anomalous behaviour in the incommensurate
Hubbard-Holstein model.{\it Comm. Math. Phys.}  201, 81--115 (1999)
}\cr
M1& {V.Mastropietro, Coupled Ising models with quartic interaction
at criticality, {\it Comm. Math. Phys} 244, 595--642 (2004)}\cr
MD&{W. Metzner, C. Di Castro:Conservation laws and correlation
functions in the Luttinger liquid. {\it Phys. Rev. B} {\bf 47},
16107 (1993). }\cr
ML&{D. Mattis, E. Lieb: Exact solution of a many fermion system
and its associated boson field. {\it J. Math. Phys.} {\bf 6},
304--312 (1965). }\cr
S&{J. Solyom: The Fermi gas model of one-dimensional
conductors. {\it Adv. in. Phys.} {\bf 28},
201--303 (1979). }\cr

T&{W. Thirring: {\it Ann. of Phys.} {\bf 3}, 91 (1958). }\cr
Z&{J. Zinn-Justin: Quantum field theory and critical phenomena, Oxford
publications, (1989). }\cr
}

\bye